\documentclass[longauth]{aa}  

\usepackage{graphicx}
\usepackage{txfonts}
\usepackage{amsmath}
\usepackage{booktabs}
\usepackage{gensymb}
\usepackage{caption}
\usepackage{siunitx}
\usepackage{hyperref}
\usepackage{natbib}
 \usepackage{multirow}
 \usepackage[title,page]{appendix}
\usepackage{placeins}
\usepackage{float}
\hypersetup{
    colorlinks,
    citecolor=blue,
    filecolor=magenta,      
    linkcolor=blue,
    }

\newcommand{\twelve}{$^{12}$}
\newcommand{\thirteen}{$^{13}$}
\newcommand{\fourteen}{$^{14}$}
\newcommand{\fifteen}{$^{15}$}
\newcommand{\sixteen}{$^{16}$}
\newcommand{\seventeen}{$^{17}$}
\newcommand{\eighteen}{$^{18}$}
\newcommand{\plus}{$^{+}$}
\newcommand{\kms}{km\,s$^{-1}$ }

\begin{document} 

\titlerunning{Molecular isotopologue measurements of SSCs in NGC~253 with ALCHEMI}

   \title{Molecular isotopologue measurements toward super star clusters and the relation to their ages in NGC~253 with ALCHEMI}

   \subtitle{}

   \author{J. Butterworth
          \inst{\ref{inst.Leiden}} 
          \and S. Viti \inst{\ref{inst.Leiden}}
          \and P. P. Van der Werf\inst{\ref{inst.Leiden}} \and J. G. Mangum \inst{\ref{inst.NRAOCV}}
\and S. Mart\'in \inst{\ref{inst.ESOChile},\ref{inst.JAO}}
          \and N. Harada \inst{\ref{inst.NAOJ},\ref{inst.ASIAA}, \ref{inst.SOKENDAI}} 
          \and K. L. Emig\inst{\ref{inst.NRAOCV}}\thanks{Jansky Fellow of the National Radio Astronomy
          Observatory}   
          \and S. Muller \inst{\ref{inst.ONSALA}}
          \and K. Sakamoto \inst{\ref{inst.ASIAA}
          ,\ref{inst.SOKENDAI}}
          \and Y. Yoshimura \inst{\ref{inst.UTokio}}
          \and K. Tanaka \inst{\ref{inst.KeioUniversity}}
          \and R. Herrero-Illana \inst{\ref{inst.ESOChile},\ref{inst.ICECSIC}}
          \and L. Colzi \inst{\ref{inst.CAB-INTA}}
        \and V. M. Rivilla \inst{\ref{inst.CAB-INTA}}
            \and K. Y. Huang\inst{\ref{inst.Leiden}}
          \and M. Bouvier\inst{\ref{inst.Leiden}}
          \and E. Behrens \inst{\ref{inst.UVA}}
          \and C. Henkel \inst{\ref{inst.MPIfR},\ref{inst.Abdulaziz},\ref{inst.Xinjiang}}         
          \and Y. T. Yan \inst{\ref{inst.MPIfR}}
          \and D. S. Meier \inst{\ref{inst.NMIMT},\ref{inst.NRAOSocorro}}
          \and D. Zhou \inst{\ref{inst.Leiden},\ref{inst.dawn},\ref{inst.dtu},\ref{inst.ubc}}
            }

   \institute{\label{inst.Leiden}Leiden Observatory, Leiden University, PO Box 9513, NL-2300 RA Leiden, the Netherlands \\ \email{butterworth@strw.leidenuniv.nl }
   \and\label{inst.NRAOCV} National Radio Astronomy Observatory, 520 Edgemont Road, Charlottesville, VA 22903, USA
   \and\label{inst.ESOChile}European Southern Observatory, Alonso de C\'ordova, 3107, Vitacura, Santiago 763-0355, Chile  
\and\label{inst.JAO}Joint ALMA Observatory, Alonso de C\'ordova, 3107, Vitacura, Santiago 763-0355, Chile
\and\label{inst.NAOJ}National Astronomical Observatory of Japan, 2-21-1 Osawa, Mitaka, Tokyo 181-8588, Japan
\and\label{inst.ASIAA}Institute of Astronomy and Astrophysics, Academia Sinica, 11F of AS/NTU Astronomy-Mathematics Building, No.1, Sec. 4, Roosevelt Rd, Taipei 10617, Taiwan
\and\label{inst.SOKENDAI}Department of Astronomy, School of Science, The Graduate University for Advanced Studies (SOKENDAI), 2-21-1 Osawa, Mitaka, Tokyo, 181-1855 Japan
\and\label{inst.ONSALA}Department of Space, Earth and Environment, Chalmers University of Technology, Onsala Space Observatory, SE-43992 Onsala, Sweden
\and\label{inst.UTokio}Institute of Astronomy, Graduate School of Science, The University of Tokyo, 2-21-1 Osawa, Mitaka, Tokyo 181-0015, Japan
\and\label{inst.KeioUniversity}Department of Physics, Faculty of Science and Technology, Keio University, 3-14-1 Hiyoshi, Yokohama, Kanagawa 223--8522 Japan
\and\label{inst.ICECSIC}Institute of Space Sciences (ICE, CSIC), Campus UAB, Carrer de Magrans, E-08193 Barcelona, Spain
\and\label{inst.CAB-INTA} Centro de Astrobiología (CAB, CSIC-INTA), Ctra. de Torrej\'on a Ajalvir km 4, 28850, Torrej\'on de Ardoz, Madrid, Spain
\and\label{inst.UVA}Department of Astronomy, University of Virginia, P.~O.~Box 400325, 530 McCormick Road, Charlottesville, VA 22904-4325
\and\label{inst.MPIfR}Max-Planck-Institut f\"ur Radioastronomie, Auf-dem-H\"ugel 69, 53121 Bonn, Germany    
\and\label{inst.Abdulaziz}Astron. Dept., Faculty of Science, King Abdulaziz University, P.O. Box 80203, Jeddah 21589, Saudi Arabia
\and\label{inst.Xinjiang}Xinjiang Astronomical Observatory, Chinese Academy of Sciences, 830011 Urumqi, China
\and\label{inst.NMIMT}New Mexico Institute of Mining and Technology, 801 Leroy Place, Socorro, NM 87801, USA
\and\label{inst.NRAOSocorro}National Radio Astronomy Observatory, PO Box O, 1003 Lopezville Road, Socorro, NM 87801, USA
\and\label{inst.dawn}Cosmic Dawn Center (DAWN)
\and\label{inst.dtu}DTU Space, Technical University of Denmark, Elektrovej 327, DK-2800 Kgs. Lyngby, Denmark
\and\label{inst.ubc}Department of Physics and Astronomy, University of British Columbia, 6225 Agricultural Rd., Vancouver, V6T 1Z1, Canada
}
    
   \date{Received; accepted }

 
  \abstract
   {Determining the evolution of the CNO isotopes in the interstellar medium (ISM) of starburst galaxies can yield important constraints on the ages of superstar clusters (SSCs), or on other aspects and contributing factors of their evolution, such as the Initial Mass Function. Due to the time-dependent nature of the abundances of isotopes within the ISM as they are supplied from processes such as nucleosynthesis or chemical fractionation, this provides the possible opportunity to probe the ability of isotopes ratios to trace the ages of high star forming regions, such as SSCs.}
   {The goal of this study is to investigate whether the isotopic variations in SSC regions within NGC~253 are correlated with their different ages as derived from stellar population modelling.}
   {We have measured abundance ratios of CO, HCN and HCO\plus\ isotopologues in six regions containing SSCs within NGC~253 using high spatial resolution (1.6",$\sim 28$pc) data from the ALCHEMI (ALma Comprehensive High-resolution Extragalactic Molecular Inventory) ALMA Large program. We have then analysed these ratios using \texttt{RADEX} radiative transfer modelling, with the parameter space sampled using the nested sampling Monte Carlo algorithm MLFriends.  These abundance ratios were then compared to ages predicted in each region via the fitting of observed star formation tracers (such as Br$\gamma$) to \texttt{Starburst99} starburst stellar population evolution models.}
   {Through the use of non-LTE radiative transfer modelling, the isotopic column density ratios across multiple regions of SSC activity in NGC~253 are determined. We do not find any significant trend with age for the  CO and HCN isotopologue ratios on the timescales for the ages of the SSC* regions observed. HCO\plus\ may show a correlation with age over these timescales, in \twelve C/\thirteen C.}
   {The driving factors of these ratios within SSCs could be the Initial Mass Function as well as possibly fractionation effects. To further probe these effects in SSCs over time a larger sample of SSCs must be observed spanning a larger age range.}

\maketitle

\section{Introduction}
\label{sec:Intro}
Starburst galaxies are characterized by vigorous star formation activity; this can lead to the emergence of super star clusters (SSCs). SSCs are typically quite young ($<10$~Myr) and have been observed within our Galaxy, with masses of $\sim10^4$~M$_\odot$ \citep{Ginsburg_2018}. Even more massive ($\sim10^5$~M$_\odot$) clusters have been observed in starburst regions of nearby galaxies, such as NGC~4945 and NGC~253 \citep{2018Leroy,Emig_2020,2020_Rico-Villas,2022_Rico_villas}.

SSCs, as compact ($\sim$2-3~pc) regions of high star formation rate (up to 5~M$_\odot$\,yr\textsuperscript{-1} ), serve as ideal laboratories to explore the processes that may influence and transform the interstellar medium (ISM), in particular, the variation of the isotope ratios of carbon (C), nitrogen (N), and oxygen (O). These atoms  and their isotopes are important due to their high abundances and potential to probe stellar nucleosynthesis. This is due to the differing processes under which they typically form:  \twelve C for example is believed to be primarily formed in Helium burning of both low- and high-mass stars. \thirteen C on the other hand is believed to form in the CNO cycle of asymptotic giant branch (AGB) stars \citep{1997Pagal}. Both \fourteen N and \fifteen N are thought to be produced during the CNO cycles of massive stars and in the so-called hot bottom burning of asymptotic giant branch (AGB) stars with a higher proportion of both \fourteen N and \fifteen N formed in intermediate-mass stars relative to massive stars, \fifteen N is also largely produced by novae \citep{Izzard_2004,Romano2017,2019_Romano,2022_Colzi}.
\sixteen O and \eighteen O are believed to be products of the helium burning phase of high-mass stars (with \eighteen O, primarily formed in metal-rich stars) and \seventeen O is formed as a product during the CNO cycle of intermediate mass stars (\citep{2008Wouterloot,2010Wiescher,2014Henkel,Romano_Review_2022}. As such the abundances of these isotopes and their evolution with time in a star-forming region are greatly affected by the initial conditions of star formation, such as the initial mass function (IMF) \citep{Papadopoulos_2010,Papadopoulos_2011}. 

Through the lens of sub-mm measurements these isotopic variations are traced primarily via the observations of isotopologues (isotope-bearing molecules).
Thanks to the Atacama Large Millimeter Array (ALMA), it is possible to observe compact energetic sources, such as SSCs, on a resolvable scale in the starburst regions within nearby galaxies. Within the sub-mm regime probed by ALMA there are multiple rotational transitions of CNO bearing isotopologues available (e.g. CO, HCN, HCO\plus).  
Observations of these isotopologues within the star-forming regions of galaxies (e.g. SSCs) can inform us also on the IMF \citep{Romano2017,Romano_Review_2022}. 
For example, \cite{Romano2017} found that C, N and O isotope ratios imply a Top-Heavy IMF in nearby star-forming galaxies.
Galaxies with a high star formation rate, such as starbursts, have been shown to possess a higher isotopic ratio of \twelve C/\thirteen C than more `normal' galaxies such as the Milky Way.
Chemical fractionation can also influence the observed isotopologue ratios \citep{2015Roueff,2018_Colzi,2019Viti,2020Viti,2020_Colzi,2022_Colzi}. For instance, the presence of certain exothermic chemical reactions in the gas phase of the ISM, under which isotopes are exchanged, may lead to chemical fractionation whereby the enhancement of certain isotopologues is favored over others  \citep{1984Langer,Szucs2014,2019Loison,2019Martin}.

CO, HCN, HCO\plus\ and their isotopologues have been used as tools to investigate the \twelve C/\thirteen C ratio in nearby galaxies \citep{2004Pasquali,Jiang2011,GonzalezAlfonso2012,2013Sliwa,2014Sliwa,2014Henkel,2019Martin,2019_Tang,Martin_ALCHEMI_2021} as well as in higher redshift galaxies (z $\leq 2.5$) \citep{Muller2006,Danielson_2013,2014Spilker,Wallstrom2016}. Meanwhile, \fourteen N/\fifteen N has  been investigated through the use of HCN and its HC\fifteen N isotopologue in extra-galactic sources \citep{1998Henkel,2018Henkel,1999Chin,2009Wang,2014Wang,2016Wang,Jiang2011,Martin_ALCHEMI_2021} and \sixteen O/ \eighteen O has  been investigated towards individual Giant Molecular Clouds (GMCs) previously in NGC~253 using isotopologues of CO \citep{Meier_2015}.

In this work we conduct a study to see whether the observed isotope ratios in starburst regions are determined by age or whether other mechanisms drive the variations in isotope ratio, such as proposed in \cite{Romano2017}. In order to accomplish this, we present ALMA multi-transition observations of CO, HCN, and HCO\plus\ and their \thirteen C, \fifteen N, \seventeen O and \eighteen O isotopologues towards the starburst galaxy NGC~253. These measurements have been extracted from the ALMA large program imaging of the NGC~253 Central Molecular Zone (CMZ), "ALMA Comprehensive High-resolution Extragalactic Molecular Inventory", \citep[ALCHEMI,][]{Martin_ALCHEMI_2021}. 

NGC~253 is a nearby \citep[d $\sim 3.5$~Mpc, ][]{Rekola+2005} galaxy containing a nuclear starburst (Star Formation Rate, SFR $\sim 2$M$_{\rm \odot}$\,yr$^{-1}$, see \citealt{Leroy2015}), which accounts for $\sim 50\%$ of the galaxy's total star-forming activity. A previous study by \citet{2018Leroy} has identified 14 SSCs located within the Central Molecular Zone (CMZ) of NGC~253. These clusters were estimated to have masses of $\sim$10$^5$ M$_{\rm \odot}$ with an upper limit to their ages of $<$10~Myr.

The broad frequency range of ALCHEMI allows for a systematic study of both the physical and chemical properties of NGC~253 as viewed through molecular observations.  
Other studies, made possible thanks to the ALCHEMI data, include: constraining the high cosmic-ray ionization rate (CRIR) nature of the galaxy \citep{Holdship+2021_SpectralRadex,Holdship+2022,Harada+2021,2022_Behrens}, the first detection of a phosphorus-bearing molecule in extragalactic sources \citep{Haasler+2022}, the identification of new methanol maser transitions \citep{Humire+2022}, the use of HOCO\textsuperscript{+} as a chemical tracer of CO$_2$ \citep{Harada+2022}, and also the ability to reconstruct the shock history of the giant molecular clouds located within the CMZ of the galaxy \citep{Huang2023}.

Alongside ALMA observations of the isotopologues of CO, HCN and HCO\plus, the H39$\alpha$ emission was extracted from the ALCHEMI data in order to estimate the ionizing photon rate with a method consistent to that of \cite{Emig_2020}. By combining these ionizing photon rates with other tracers of high star formation such as the Brackett Gamma (Br$\gamma$) line \citep{2020_Pasha}, the ages of star-forming populations can be approximated using stellar population evolution modelling (e.g., \texttt{Starburst99}, \citealt{Starburst99_1999,Starburst99_2014}). Br$\gamma$ is, as a near-infrared hydrogen recombination line, primarily emitted in regions ionized by massive stars (O-, B-type stars).  Br$\gamma$ also possesses relatively low dust attenuation ($\sim14\%$) making it an extremely potent tracer of the age of star-forming regions. We use observations on Br$\gamma$ taken with the SINFONI near-infrared (1.1-2.45~$\mu$m) integral field spectrograph installed on the Very Large Telescope (VLT) \citep{SINFONI_1,SINFONI_2}.

This paper is structured as follows: in Section \ref{sec:Obs}, we present an overview of the ALCHEMI data used  alongside the data from SINFONI. Section \ref{subsec:CubeLineMoment} covers the methods under which the spectral lines were extracted and the moment maps were produced. In Section \ref{sec:LTE} we present an initial LTE analysis  of the isotopologues. This is followed by a Non-LTE Large Velocity Gradient (LVG) modelling  in order to determine the physical conditions within each SSC in Section \ref{sec:non-LTE}. In Section \ref{sec:Ratio_age} we derive the ages of the SSCs and we correlate them with the observed molecular ratios. We summarise our conclusions in Section \ref{sec:Conc}.
\section{Observations}
\label{sec:Obs}

\subsection{ALMA Observations}
\label{subsec:ALMA_Obs}
All of the ALMA observations used within this study are part of ALCHEMI (project code 2017.1.00161.L and 2018.1.00162.S). A detailed explanation of the calibration set-up, data acquisition and imaging of the data of this large program can be found in \cite{Martin_ALCHEMI_2021}. We shall provide a brief summary here. ALCHEMI imaged the Central Molecular Zone (CMZ) of NGC~253 across the ALMA frequency Bands 3 to 7. This provided a near contiguous coverage of the frequencies from 84.2~GHz to 373.2~GHz. The nominal phase center of the observations is $\alpha$ = 00$^h$47$^m$33$^s$.26, $\delta$ = $-25^\circ$17$^\prime$17$^{\prime\prime}.7$ (ICRS). A common rectangular area with size $50^{\prime\prime} \times 20^{\prime\prime}$ ($850\times340$\,pc) at a position angle of $65^\circ$ was imaged to cover the central nuclear region in NGC~253. The final homogeneous angular and spectral resolution for each of the image cubes produced from these measurements was $1.^{\prime\prime}6$ \citep[$\sim28$ pc;][]{Martin_ALCHEMI_2021} and $\sim10$~\kms, respectively. 
The common maximum recoverable angular scale achieved was $15^{\prime\prime}$ after combining the 12~m Array and Atacama Compact Array (ACA) measurements at all frequencies.
The desired species for which the image cubes were created are summarised in Tables \ref{tab:Line_list}-\ref{tab:Line_list_HCOP}, and includes the main most abundant isotopologues of the molecules CO, HCN and HCO\plus.

\begin{table*}[ht!]

  \centering
  \caption{The transitions of CO and its isotopologues used in this work. }
  \label{tab:Line_list}
  \begin{tabular}{ccccccc}
  \hline
  \hline
    {Species} & {Transition} & {Rest Frequency} & $E_{u}$ & $A_{ul}$ & Ref. \\
    {}& {} & {[GHz]}& [K] & [s\textsuperscript{-1}] \\
    \hline
    \hline
    CO & 1-0 & {115.271} & {5.53} &     7.203 $\times 10^{-8}$ & {}  \\
    {} & 2-1 & {230.538} & {16.60} &   6.910 $\times 10^{-7}$ & {1,2,3}  \\
    {} & 3-2 & {345.796} & {33.19} &   2.497 $\times 10^{-6}$ & {}   \\
    \hline
    {\thirteen CO} & 1-0 & {110.201} & {6.25} & 6.333 $\times 10^{-8}$& {}  \\
    {} & 2-1 & {220.399} & {15.87}  & 6.075$\times 10^{-7}$ & 3,4,5 \\
    {} & 3-2 & {330.588} & {31.73}  & 2.194$\times 10^{-6}$ &{}  \\

    \hline
    {C\seventeen O} & 1-0 & {112.359} & {5.39} & 6.697$\times 10^{-8}$ &    \\
    {} & 2-1 & {224.715} & {16.18}  & 6.425$\times 10^{-7}$ & 3,6 \\
    {} & 3-2 & {337.061} & {32.35}  &  1.805$\times 10^{-6}$ &  \\

        \hline
    {C\eighteen O} & 1-0 & {109.782} & {5.27} & 6.266$\times 10^{-8}$ &   \\
    {} & 2-1 & {219.560} & {15.81}  & 6.012$\times 10^{-7}$ & 3,7   \\
    {} & 3-2 & {329.331} & {31.61}  & 2.171$\times 10^{-6}$ &   \\

    \hline
  \end{tabular}\\
  \footnotesize{We use molecular data of CO and its isotopologues from:
  1) \cite{co_moldata_1994}, 2) \cite{co_moldata_1997}, 3) \cite{co_moldata_collis}, 4) \citet{co_iso_moldata_2000}, 5) \citet{13co_moldata_2004}, 6) \citet{c17o_moldata_2003}, 7) \citet{c18o_moldata_2001}. The data was sourced from the LAMDA database \citep{LAMDA_2005} and the Cologne Database for Molecular Spectroscopy (CDMS) catalogue \footnote{https://cdms.astro.uni-koeln.de/} \citep{CDMS_2001,CDMS_2005,CDMS_2016}.}
\end{table*}

\begin{table*}[ht!]
\vspace{1cm}
  \centering
  \caption{The transitions of HCN and its isotopologues used in this work. }
  \label{tab:Line_list_HCN}
  \begin{tabular}{ccccccc}
  \hline
  \hline
    {Species} & {Transition} & {Rest Frequency} & $E_{u}$ & $A_{ul}$ & Ref. \\
    {}& {} & {[GHz]}& [K] & [s\textsuperscript{-1}] \\
    \hline
    \hline
    HCN & 1-0 & {88.632} & {4.25} &     2.407$\times 10^{-5}$ & \multirow{4}{*}{1} \\
    {} & 2-1 & {177.261} & {12.76} &    2.311$\times 10^{-4}$ &  \\
    {} & 3-2 & {265.886} & {25.52} &     8.356$\times 10^{-4}$ & \\
    {} & 4-3 & {354.505} & {42.53} &    2.054$\times 10^{-3}$ &  \\
    \hline
    {H\thirteen CN} & 1-0 & {86.340} & {4.14} & 2.226$\times 10^{-5}$ & \multirow{4}{*}{1,2}   \\
     & 2-1&{172.678} & {12.43} & 2.136$\times 10^{-4}$ & \\
     & 3-2 &{259.012} & {24.86} & 7.725$\times 10^{-4}$ &    \\
     & 4-3 &{345.340} &  {41.44} & 1.899$\times 10^{-3}$ &       \\
        \hline
    {HC\fifteen N} & 1-0 & {86.055} & {4.13} &  2.203$\times 10^{-5}$ & \multirow{4}{*}{1,3} \\
    {} & 2-1 & {172.108} & {12.38}  & 2.115$\times 10^{-4}$ & \\
    {} & 3-2 & {258.157} & {24.78}  &  7.649$\times 10^{-4}$& \\
    {} & 4-3 & {344.200} & {41.30} &     1.880$\times 10^{-3}$ & \\
    \hline
  \end{tabular}\\
  \footnotesize{We use molecular data of HCN and its isotopologues from:
  1) \cite{2023_HCN_rates}, 2) \citet{h13cn_moldata_2004}, 3) ,\citet{hc15n_moldata_2005}. The data was sourced from the Cologne Database for Molecular Spectroscopy (CDMS) catalogue \citep{CDMS_2001,CDMS_2005,CDMS_2016}, or via private communication with authors.}
\end{table*}

\begin{table*}[ht!]
  \centering
  \caption{The transitions of HCO\plus\ and its isotopologues used in this work. }
  \label{tab:Line_list_HCOP}
  \begin{tabular}{ccccccc}
  \hline
  \hline
    {Species} & {Transition} & {Rest Frequency} & $E_{u}$ & $A_{ul}$ & Ref. \\
    {}& {} & {[GHz]}& [K] & [s\textsuperscript{-1}] \\
    \hline
    \hline
    HCO\plus & 1-0 & {89.188} & {4.28} &     4.187$\times 10^{-5}$ & \multirow{4}{*}{1,2,3,4} \\
    {} & 2-1 & {178.375} & {12.84} &    4.019$\times 10^{-4}$ &   \\
    {} & 3-2 & {267.558} & {25.68} &     1.453$\times 10^{-3}$ &  \\
    {} & 4-3 & {356.734} & {42.80} &    3.572$\times 10^{-3}$ &   \\
    \hline
    {H\thirteen CO\plus} & 1-0 & {86.754} & {4.16} & 3.853$\times 10^{-5}$ & \multirow{4}{*}{1,2,5} \\
     & 2-1&{173.507} & {12.46} & 3.699$\times 10^{-4}$ &   \\
     & 3-2 &{260.255} & {24.98} & 1.337$\times 10^{-3}$ &    \\
     & 4-3 &{346.998} &  {41.63} & 3.287$\times 10^{-3}$ &       \\
        \hline
    {HC\eighteen O\plus} & 1-0 & {85.162} & {4.09} & 3.6449$\times 10^{-5}$ & \multirow{3}{*}{1,2,6} \\
    {} & 2-1 & {170.322} & {12.26}  & 3.498$\times 10^{-4}$ & \\
    {} & 3-2 & {255.479} & {24.52}  &  1.265$\times 10^{-3}$ & \\
    \hline
  \end{tabular}\\
  \footnotesize{We use molecular data of HCO\plus and its isotopologues from:
  1) \cite{1999_hcop_collis}, 2) \cite{2020_hcop_collis}, 3) \citet{hcop_moldata_2007}, 4) \citet{hcop_tinti_moldata_2007}, 5) \citet{h13cop_moldata_2001}, 6) \citet{hc18op_moldata_2004}. The data was sourced from the LAMDA database \citep{LAMDA_2005} and the Cologne Database for Molecular Spectroscopy (CDMS) catalogue \citep{CDMS_2001,CDMS_2005,CDMS_2016}.}
\end{table*}

\begin{table}[ht!]
  \centering
  \caption{All the selected NGC~253 positions of the beam sized SSC* regions described in Sect. \ref{subsec:Region_Selection}. The SSCs defined in \cite{2018Leroy} that are contained within each SSC group have been provided for clarity, using the nomenclature of that paper. }
  \label{tab:SSC_locations}
  \begin{tabular}{cccc}
  \hline
    {SSC} & {R.A.(ICRS)} & {Dec.(ICRS)} & Leroy\\
    {Group}& {(00$^h$ 47$^m$)} & {($-25^\circ$ 17$^\prime$)} & SSCs\\
    \hline
    \hline
    SSC-1* & 32$^s$.8108 & 21$^{\prime\prime}$.293 & 1,2,3 \\
    SSC-4* & 32$^s$.9178 & 20$^{\prime\prime}$.426 & 4 \\
    SSC-7* & 33$^s$.0147 & 19$^{\prime\prime}$.341 & 5,6,7 \\
    SSC-9* & 33$^s$.0970 & 18$^{\prime\prime}$.053 & 8, 9 \\
    SSC-13* & 33$^s$.1927 & 16$^{\prime\prime}$.928 & 10, 11, 12, 13 \\
    SSC-14* & 33$^s$.2971 & 15$^{\prime\prime}$.583 & 14 \\
    \hline
  \end{tabular}
\end{table}

\subsection{SINFONI Observations}
\label{subsec:SINFONI_Obs}

The Spectrograph for INtegral Field Observations in the Near-Infrared (SINFONI) at the ESO VLT was used to obtain infrared J-, H- and K-band images toward NGC~253. The SINFONI instrument is mounted at the Cassegrain focus of the Unit Telescope 4 at the Very Large Telescope (VLT). This paper uses the SINFONI K band observations first published in \cite{MullerSanchez2010}, and a more detailed description of these observations may be found in that paper. In short, the observations of NGC~253 were made in visitor mode on August 28th, 2005. The K-band observations were taken using a spatial pixel scale of 0.25'' corresponding to a field of view of 8'' $\times$ 8'' per frame and a spectral resolution of 68~\kms. These observations were taken primarily to observe H$_2$ emission across the CMZ; thus multiple frames were taken along the CMZ until no H$_2$ was observed, resulting in 10 pointings altogether. This mosaic covers most of the CMZ of NGC~253; further details of these observations available in \cite{MullerSanchez2010}.

\section{Moment Maps and Region Selection}

\subsection{Spectral-Line Extraction}
\label{subsec:CubeLineMoment}
The molecules we focus on in this study are the \thirteen C, \fifteen N, \seventeen O (of CO) and \eighteen O isotopologues of CO, HCN and HCO\plus. A full summary of the investigated lines are shown in Tables \ref{tab:Line_list} - \ref{tab:Line_list_HCOP}.
The integrated spectral line intensities from our data cubes were extracted using \texttt{CubeLineMoment}\footnote{\url{https://github.com/keflavich/cube-line-extractor}} \citep{2019_Mangum_Heart}. 
\texttt{CubeLineMoment} works by extracting integrated intensities for a given list of targeted spectral frequencies by applying a set of spectral and spatial masks (defined by the user).
  The \texttt{CubeLineMoment} masking process uses a brighter spectral line (typically the main isotopologue line of the same rotational transition), whose velocity structure over the galaxy is most representative of the science target line, which is used as a tracer of the velocity of the gas component inspected. The products of \texttt{CubeLineMoment} are moment 0, 1 and 2 maps in the desired units of the user, masked below a chosen $\sigma$ threshold (channel-based). \cite{2019_Mangum_Heart} provides a more detailed overview of this process.

For use in the age approximation fitting (see Section \ref{sec:Age}), we have also extracted H39$\alpha$ (rest freq: 106.73738 GHz) radio recombination line emission from the ALCHEMI dataset in the six SSC apertures in order to estimate a dust-unobscured ionizing photon rate, elaborated upon in Section \ref{sec:Ratio_age}. In order to completely image the line, we used the spectra over barycentric velocity ranges [$-$200, 750]~\kms to fit and measure the line emission. The Br$\gamma$ line was also extracted from the SINFONI data described, and is consistent with the map shown in Figure 11 of \cite{2013Rosenberg}.

\subsection{Region Selection}
\label{subsec:Region_Selection}

We have selected six SSC regions using the aperture size of the beam of the ALCHEMI observations. Due to the lower resolution of the ALCHEMI observations ($\sim28$ pc) with respect to the continuum observations of \cite{2018Leroy} ($\sim2$ pc), we define new SSC regions, see Table \ref{tab:SSC_locations} for details. These regions were selected and identified such that they covered all of the SSC regions defined in \cite{2018Leroy} at the ALCHEMI beam size resolution, with no overlap.
Each of these regions contains within them between 1 to 4 of the SSC regions approximately resolved in \cite{2018Leroy}.
Due to the fact that our newly defined regions do not correspond to those of \cite{2018Leroy}, a new nomenclature is needed. Thus from now on each region has been designated as a SSC* region with its corresponding number chosen from one of the SSCs that it contains. 

\begin{figure}
  \centering
  \begin{tabular}[b]{@{}p{0.50\textwidth}@{}}
    \centering\includegraphics[width=1.0\linewidth]{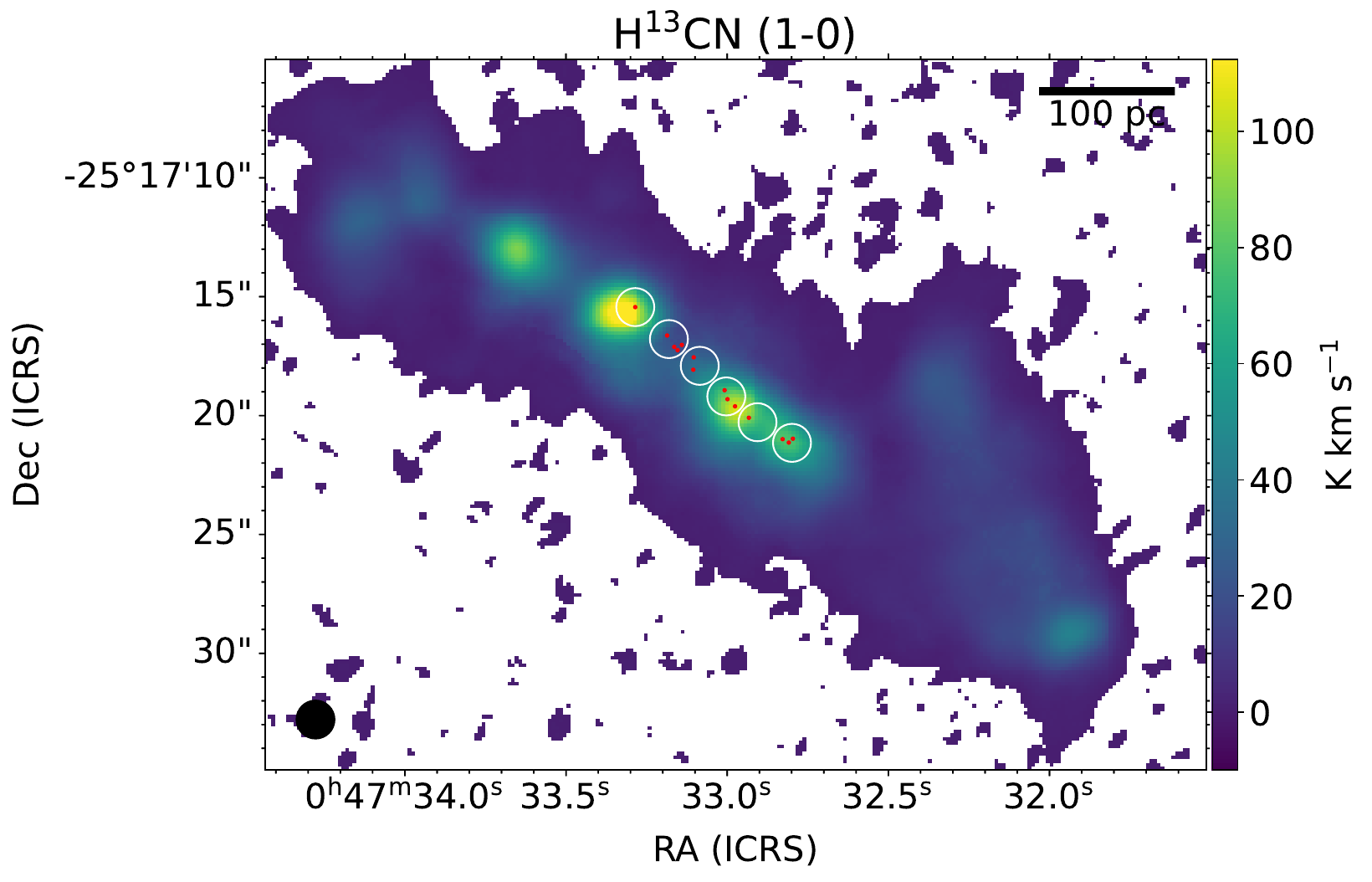} \\
    \centering\small (a) 
  \end{tabular}%
  \quad
 \begin{tabular}[b]{@{}p{0.50\textwidth}@{}}
    \centering\includegraphics[width=1.0\linewidth]{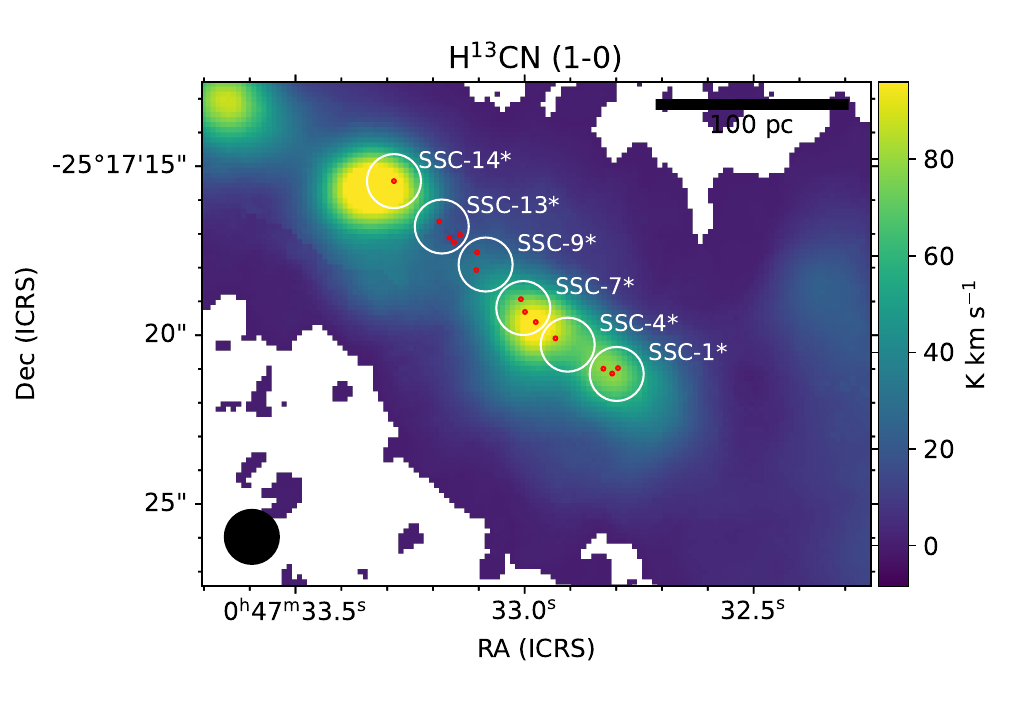} \\
    \centering\small (b) 
  \end{tabular}

 
  \caption{Velocity-integrated line intensities in units of K\,\kms  H\thirteen CN (1-0). Panel (a) shows a region covering the entire CMZ of NGC~253 and Panel (b) covers a zoomed in region of the SSC* regions studied in this paper. The studied SSC regions as listed in Tab.~\ref{tab:SSC_locations} are labeled in white texts on the map. The original SSC locations with appropriate beam sizes from \cite{2018Leroy} are shown by the red regions. The ALCHEMI $1''.6 \times 1''.6$ beam is displayed in the lower-left corner of the map. 
  }
  \label{fig:mom0_H13CN_10_21}
\end{figure}

\begin{figure}[ht!]
  \centering
  \begin{tabular}[b]{@{}p{0.5\textwidth}@{}}
    \centering\includegraphics[width=1.0\linewidth]{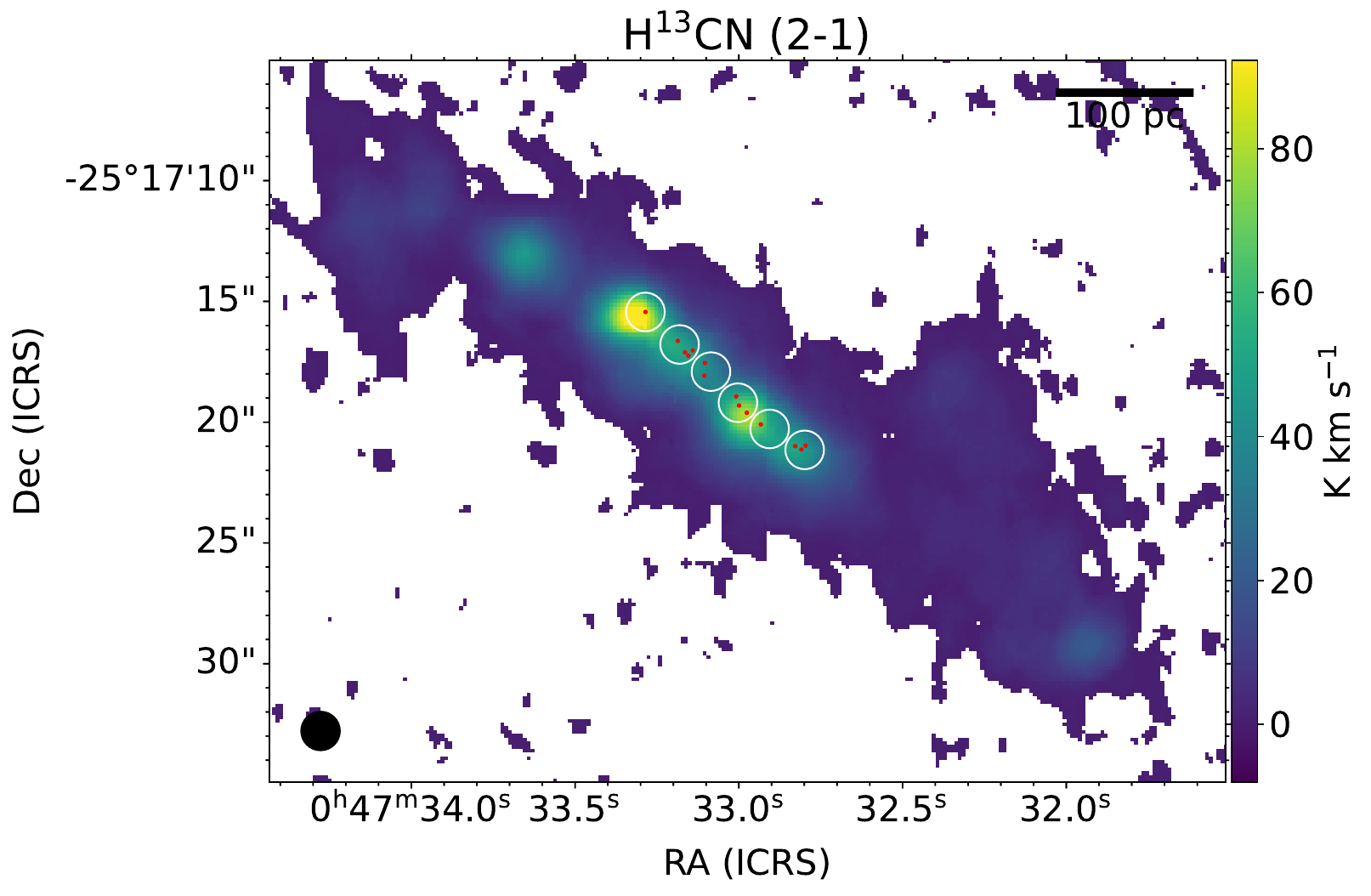} \\
    \centering\small (a)
      \end{tabular}%
  \quad
  \begin{tabular}[b]{@{}p{0.5\textwidth}@{}}
    \centering\includegraphics[width=1.0\linewidth]{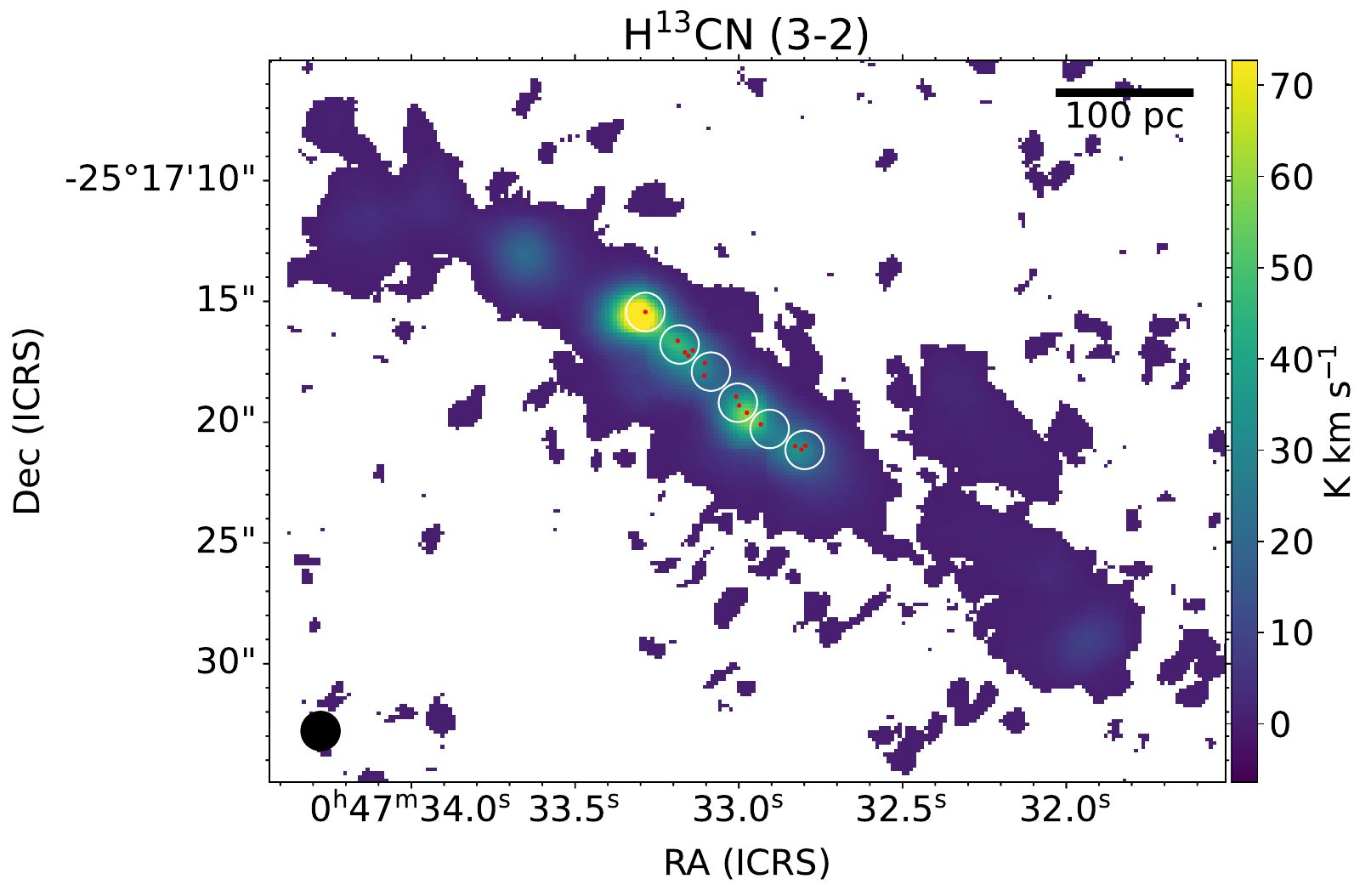} \\
    \centering\small (b) 
  \end{tabular}%
  \quad
\begin{tabular}[b]{@{}p{0.5\textwidth}@{}}
    \centering\includegraphics[width=1.0\linewidth]{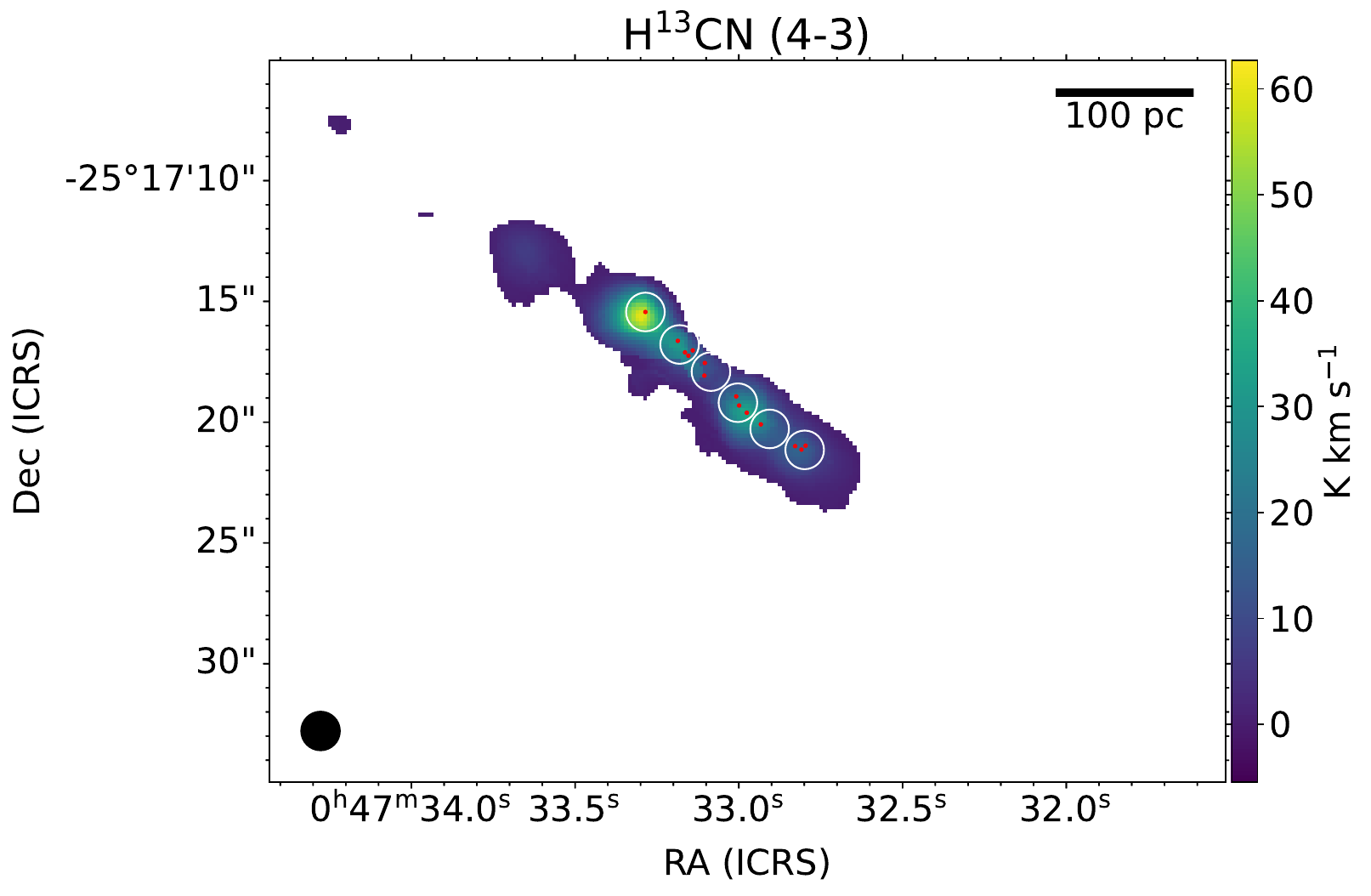} \\
    \centering\small (c) 
  \end{tabular}
  \caption{Velocity-integrated line intensities (moment 0) maps in units of K\,\kms for H\thirteen CN (2-1), (3-2) and (4-3). Each of the maps shown have been generated using a signal-to-noise cutoff of 3. The studied SSC regions as listed in Tab.~\ref{tab:SSC_locations} are labeled in white texts on the map. The original SSC locations with appropriate beam sizes are from \cite{2018Leroy}. The ALCHEMI $1''.6 \times 1''.6$ beam is displayed in the lower-left corner of the map. 
  }
  \label{fig:mom0_H13CN_32_43}
\end{figure}

Within each SSC* region, we have obtained the line intensities extracted over a beam-sized aperture of the available J transitions for each isotopologue.
In Figs.~\ref{fig:mom0_H13CN_10_21} and \ref{fig:mom0_H13CN_32_43} we present the velocity-integrated line intensity moment 0 maps from H\thirteen CN (1-0) to (4-3) as an example of the lines we have imaged (Tables \ref{tab:Line_list}-\ref{tab:Line_list_HCOP}).  The remaining moment maps are shown in Appendix \ref{app:Mom0}. 
In each moment 0 map, the SSC* regions listed in Table \ref{tab:SSC_locations} are shown with the ALCHEMI beam size plotted in the lower left corner in each map. 
The line intensities in each of these maps are shown in [K\,\kms] units, the conversion to brightness temperature units from Jy/beam was accomplished using the \texttt{astropy} Python package \citep{2022_Astropy}.
It must be noted that SSC-9* and SSC-13* tend to show weaker emission in each of our molecules, when compared to the other regions in the CMZ; this is seen towards the center of the moment maps shown in Figures \ref{fig:mom0_H13CN_10_21} and \ref{fig:mom0_H13CN_32_43}. 
In fact, we identify an absorption feature at these locations, within certain J lines of certain isotopologues, which particularly affects the weaker emission species such as HC\fifteen N, and H\thirteen CO\plus\ and affects lower J lines most prominently. This absorption feature affecting both SSC-9* and SSC-13* is consistent with the position of the nuclear continuum source designated as `TH2' \citep{1985_TH2,2010_TH2,2019_Mangum_Heart}. `TH2' is a strong continuum source at the dynamical center of NGC~253 first identified in \cite{1985_TH2} and long thought to possibly contain a Low Luminosity AGN (LLAGN), however it is now believed to be a supernova/supernova remnant source \citep{2019_Mangum_Heart}. This absorption primarily affects weaker lines, and as such no transitions of HC\fifteen N were used for either SSC-9* or SSC-13*. An example of the effect of this absorption upon HC\fifteen(1-0) in SSC-13* compared to SSC-1* is shown in Appendix \ref{app:absorption}.

\section{Gas Properties Inference}
\subsection{LTE Analysis}
\label{sec:LTE}
 
 In order to aid in informing the priors of our non-LTE analysis, shown in the next sub-section \ref{sec:non-LTE}, we obtain estimates of the molecular column densities assuming local thermodynamic equilibrium (LTE).
 The upper state column density, $N_u$,  of a molecular transition can be approximated as 
\begin{equation}
    N_u = \frac{8\pi k \nu^2W}{hc^3A_{ul}}\left(\frac{1}{\eta_{ff}}\right) \frac{\tau}{1 - e^{- \tau}}, 
    \label{eq:uppercol}
\end{equation}
where $W$ is the integrated line intensity, $_u$ is signifying the upper level of a rotational transition, $A_{ul}$ is the  Einstein A-coefficient, and $\tau$ is the optical depth. $\eta_{ff}$ is the beam filling factor. Given the variation between the beam size of this project ($1''.6 \times 1''.6$) and that of \cite{2018Leroy} where these regions were first defined ($0''.11 \times 0''.11$) we accounted for the beam filling factor, $\eta_{ff}$ by defining it as:

\begin{equation}
\eta_{ff}=\frac{\theta^{2}_{S}}{\theta^{2}_{beam}+\theta^{2}_{S}}
\end{equation} 

where $\theta_{S}$ and $\theta_{beam}$ are the respective resolutions of the source and the beam. For the purposes of the LTE estimation, $\theta_{S}$ was defined as the size of the regions observed in \cite{2018Leroy}.
We could then use these transition column densities, $N_u$, to derive the total column density of the emitting molecule, $N$, by using
\begin{equation}
    N=\frac{N_uZ(T)}{g_ue^{\frac{E_u}{kT}}}
    \label{eq:coldens},
\end{equation}
where $Z(T)$ is the partition function, $g_u$ is the statistical weight of level $u$, and $E_u$ is the excitation energy of level $u$. To compute the partition function, we make use of molecular data from the Cologne Database for Molecular Spectroscopy (CDMS) \citep{2005_Muller,CDMS_2016}. 
We used a combination of Eqs.~\ref{eq:uppercol} and~\ref{eq:coldens} following the methodology of \cite{Goldsmith_1999} to compute an estimate of the total column density for each molecule. These total column densities were estimated over a range of temperatures and optical depth. The temperature range was left free up to the maximum range chosen to be consistent with the \texttt{RADEX} analysis (shown in Section \ref{sec:non-LTE}), with the upper limit being 500 K. We used a range of values for optical depth ($\tau$) from 0 to 10 in increments of 0.1 in the optically thin regime, $\tau < 1$ and 0.5 in the optically thick regime $\tau > 1$. These column densities, shown in Table \ref{tab:cols}, are now used to define the prior distribution column density ranges studied during the \texttt{RADEX} modelling in the following section. 
Thanks to the fact that we have multiple transitions across each region for most of the species (the exceptions being weaker isotopologues (e.g. HC\fifteen N and HC\eighteen O\plus) in the absorption affected regions) the temperature can also be constrained via the use of rotational diagrams. Most species across the regions, show a behaviour that can be explained by either multiple temperature components or the optical depth effects upon the higher J lines relatively to low-J. A notable exception to this is HCN which can consistently be fit with a single temperature component. This analysis drove the initial temperature ranges for each species seen in the non-LTE analysis (see Section \ref{sec:non-LTE}).

\begin{table}[]
\caption{The minimum and maximum LTE column densities generated for each species directly from the observed intensities using the equations \ref{eq:uppercol} and \ref{eq:coldens}.}
\label{tab:cols}
\centering
\begin{tabular}{ccc}
Species & $N_{\rm min}$ & $N_{\rm max}$ \\ \hline\hline
CO       & 1.8e+18  & 8.9e+20  \\
$^{13}$CO     & 6.1e+17  & 2.0e+20  \\
C$^{17}$O     & 5.3e+15  & 3.9e+17  \\
C$^{18}$O     & 9.1e+16  & 3.2e+19  \\
HCN      & 3.1e+15  & 3.5e+17  \\
H$^{13}$CN    & 7.7e+12  & 4.3e+16  \\
HCO$^+$     & 1.4e+14  & 1.8e+17 \\
H$^{13}$CO\plus   & 5.5e+12  & 9.8e+16  \\
HC$^{15}$N    & 8.4e+11  & 1.0e+15  \\
HC$^{18}$O\plus   & 2.0e+12  & 8.5e+15  \\\hline
\end{tabular}
\end{table}

\subsection{Non-LTE Radex Analysis}
\label{sec:non-LTE}

We use a non-LTE analysis in order to infer the physical characteristics, as well as optical depth (per transition), of each region.
We employ the radiative transfer code \texttt{RADEX} \citep{Van_der_Tak_2007}.  We reference the sources of the collisional data in the subtext of Tables \ref{tab:Line_list}-\ref{tab:Line_list_HCOP}.
In order to determine these physical conditions we followed a similar methodology, combining \texttt{SpectralRadex} and \texttt{UltraNest}, as \cite{Holdship+2021_SpectralRadex,2022_Behrens,2022Huang,Huang2023}, with the difference being that in our study each group of isotopologues (e.g. HCN, H\thirteen CN and HC\fifteen N) were modelled together for the process. The physical conditions constrained within our modelling are: gas density ($n_{\rm H2}$), gas temperature ($T_{\rm kin}$),  the total column density of each isotopologue (i.e. N(HCN), N(H$^{13}$CN), and N(HC$^{15}$N)), and the beam filling factor ($\eta_{\rm ff}$). The beam filling factor has been allowed to be a free parameter between 0 and 1, this has been done as the number of lines that we possess allow it to be constrained despite its degeneracy with N and $n_H$.

We inferred the gas properties by combining the \texttt{RADEX} modelling with a Bayesian inference process in order to properly sample the parameter space and obtain reliable uncertainties. While collisional data were available for HCO\plus, HCN, H\thirteen CN, HC\fifteen N, CO, \thirteen CO, C\eighteen O and C\seventeen O, this was not the case for the isotopologues of HCO\plus. 
Collisional data of HCO\plus\ thus were used for its respective isotopologues, recommended by the LAMDA database \citep{LAMDA_2020} (See Table \ref{tab:Line_list_HCOP}) . In order to test this, \texttt{RADEX} models were computed under the same physical parameters of the different isotopologues of CO, whose data are available, and the results were then compared. They were shown to be consistently similar and thus the assumption we make regarding the other isotopologues is valid.

The posterior probability distributions are derived with the nested sampling Monte Carlo algorithm MLFriends \citep{ultranest16,ultranest19}, which was integrated into our analysis via the python package
\texttt{UltraNest} \footnote{\url{https://johannesbuchner.github.io/UltraNest/}} \citep{ultranest21}. 
Similarly to the approach adopted by \citet{2022Huang, Huang2023}, for each modelling run we assume prior distributions of our parameters to be uniform or log-uniform, where the column density priors are based upon the $N_u$ results obtained during the LTE analysis (see Section \ref{sec:LTE}; these are given for each species group in Tables \ref{tab:table_prior_CO}-\ref{tab:table_prior_HCOP}). We assume that the uncertainty on our measured intensities is Gaussian so that our likelihood is given by $P(\theta | d) \sim \exp(-\frac{1}{2}\chi^2),$ where $\chi^2$ is the chi-squared statistic between our measured intensities and the \texttt{RADEX} output for a set of physical parameters $\theta$ (summarised earlier).

\begin{table}[ht!]
  \centering
  \caption{The parameter space allowed as a prior for the nested sampling of the CO isotopologues \texttt{RADEX} models.}
  \label{tab:table_prior_CO}
  \begin{tabular}{c|cc}
  \hline 
    Variable  & Range & Distribution type\\
    \hline\hline
    Gas density, $n_{\rm H2}$ [cm\textsuperscript{-3}] & $10^{3}-10^{7}$ & Log-uniform\\
    Gas temperature, $T_{\rm kin}$ [K] & $40-500$ & Uniform\\
    $N$(CO) [cm\textsuperscript{-2}] & $10^{18}-10^{21}$ & Log-uniform\\
    $N$(\thirteen CO) [cm\textsuperscript{-2}] & $10^{17}-10^{20}$ & Log-uniform\\
    $N$(C\eighteen O) [cm\textsuperscript{-2}] & $10^{15}-10^{19}$ & Log-uniform\\
        $N$(C\seventeen O) [cm\textsuperscript{-2}] &$10^{16}-10^{19}$ & Log-uniform\\
    Beam filling factor, $\eta_{ff}$ & $0.0-1.0$ & Uniform\\
    \hline
  \end{tabular}
\end{table}
\begin{table}[ht!]
  \centering
  \caption{The parameter space allowed as a prior for the nested sampling of the HCN isotopologues \texttt{RADEX} models.}
  \label{tab:table_prior_HCN}
  \begin{tabular}{c|cc}
  \hline
    Variable  & Range & Distribution type\\
    \hline \hline
    Gas density, $n_{\rm H2}$ [cm\textsuperscript{-3}] & $10^{3}-10^{7}$ & Log-uniform\\
    Gas temperature, $T_{\rm kin}$ [K] & $40-500$ & Uniform\\
    $N$(HCN) [cm\textsuperscript{-2}] & $10^{14}-10^{18}$ & Log-uniform\\
    $N$(H\thirteen CN) [cm\textsuperscript{-2}] & $10^{12}-10^{16}$ & Log-uniform\\
    $N$(HC\fifteen N) [cm\textsuperscript{-2}] & $10^{11}-10^{15}$ & Log-uniform\\
    Beam filling factor, $\eta_{ff}$ & $0.0-1.0$ & Uniform\\
    \hline
  \end{tabular}
\end{table}
\begin{table}[ht!]
  \centering
  \caption{The parameter space allowed as a prior for the nested sampling of the HCO\plus\ isotopologues \texttt{RADEX} models.}
  \label{tab:table_prior_HCOP}
  \begin{tabular}{c|cc}
  \hline
    Variable  & Range & Distribution type\\
    \hline \hline
    Gas density, $n_{\rm H2}$ [cm\textsuperscript{-3}] & $10^{3}-10^{7}$ & Log-uniform\\
    Gas temperature, $T_{\rm kin}$ [K] & $40-500$ & Uniform\\
    $N$(HCO\plus) [cm\textsuperscript{-2}] & $10^{14}-10^{18}$ & Log-uniform\\
    $N$(H\thirteen CO\plus) [cm\textsuperscript{-2}] & $10^{12}-10^{17}$ & Log-uniform\\
    $N$(HC\eighteen O\plus) [cm\textsuperscript{-2}] & $10^{12}-10^{16}$ & Log-uniform\\
    Beam filling factor, $\eta_{ff}$ & $0.0-1.0$ & Uniform\\
    \hline
  \end{tabular}
\end{table}

This analysis provides an indication of the mean gas properties of our SSC* regions, under the assumption that all of the fitted transitions in each group arise from the same gas component.
Thus the results provided here are more of a mean property of the gas present in each SSC* region as predicted by each molecule and its isotopologues. Along with the physical parameters described, this \texttt{RADEX} fitting also provides us with an indication of the optical depths, $\tau$, of the fitted lines for each region.

\begin{figure*}
  \centering
  \includegraphics[width=\textwidth]{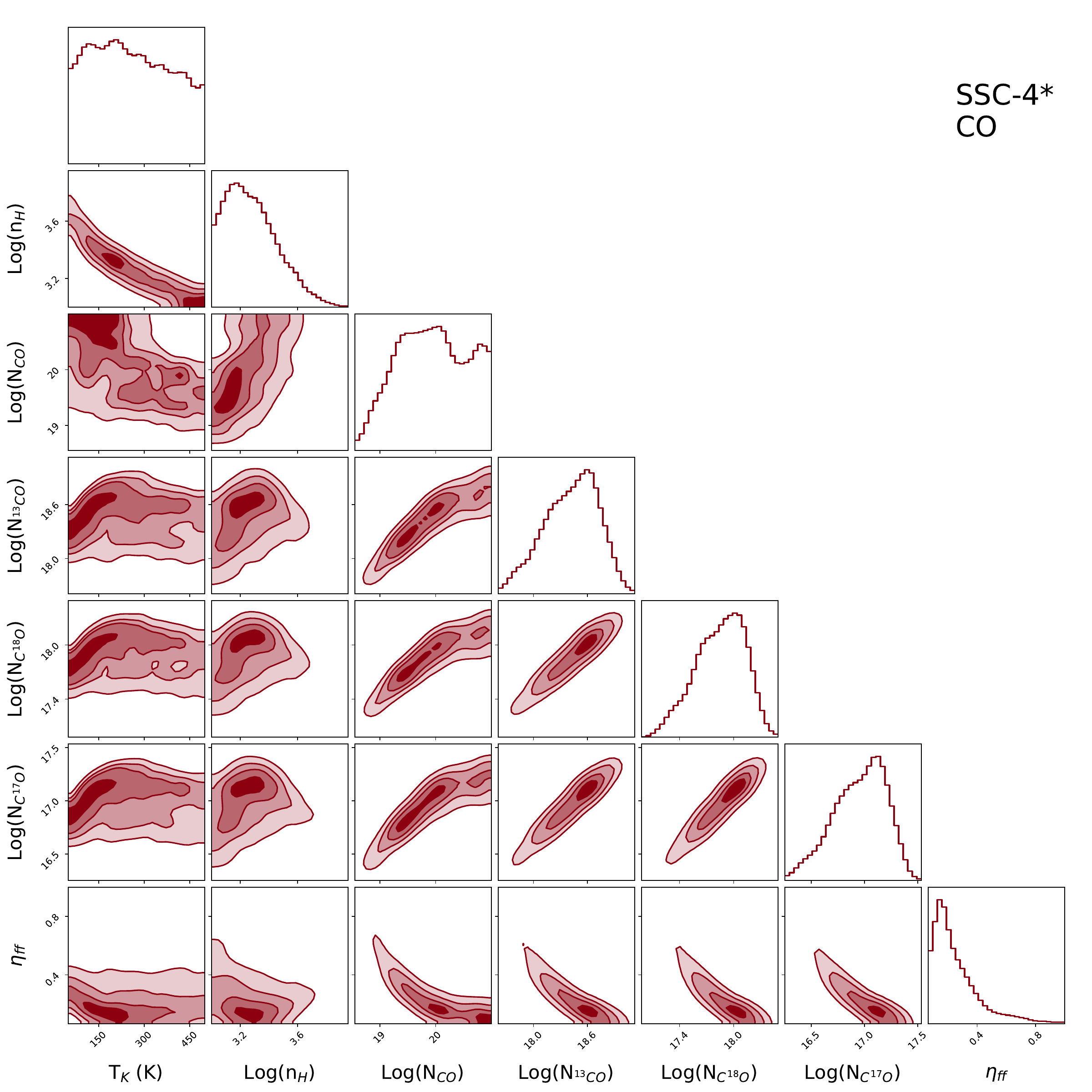}
  \caption{The posterior distributions for SSC-4* of the kinetic temperature, neutral H\textsubscript{2} number density, beam filling factor and column densities of CO, \thirteen CO, C\eighteen O, and C\seventeen O, as predicted by \texttt{RADEX}. }
  \label{fig:RADEX_corner_SSC4_CO}
\end{figure*}

\begin{figure*}
  \centering
  \includegraphics[width=\textwidth]{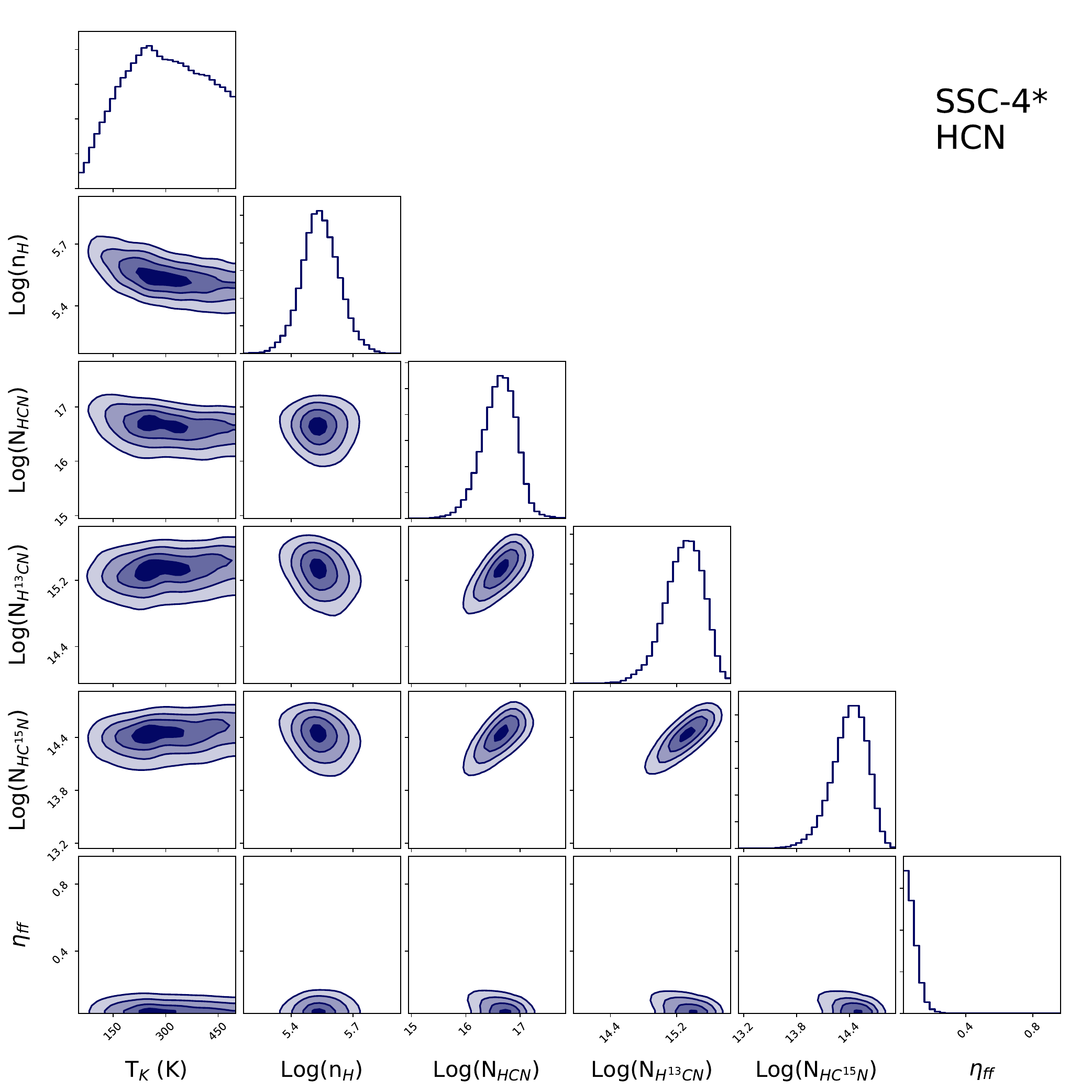}
  \caption{The posterior distributions for SSC-4* of the kinetic temperature, neutral H\textsubscript{2} number density, beam filling factor and column densities of HCN, H\thirteen CN and HC\fifteen N, as predicted by \texttt{RADEX}. }
  \label{fig:RADEX_corner_SSC4_HCN}
\end{figure*}

\begin{figure*}
  \centering
  \includegraphics[width=\textwidth]{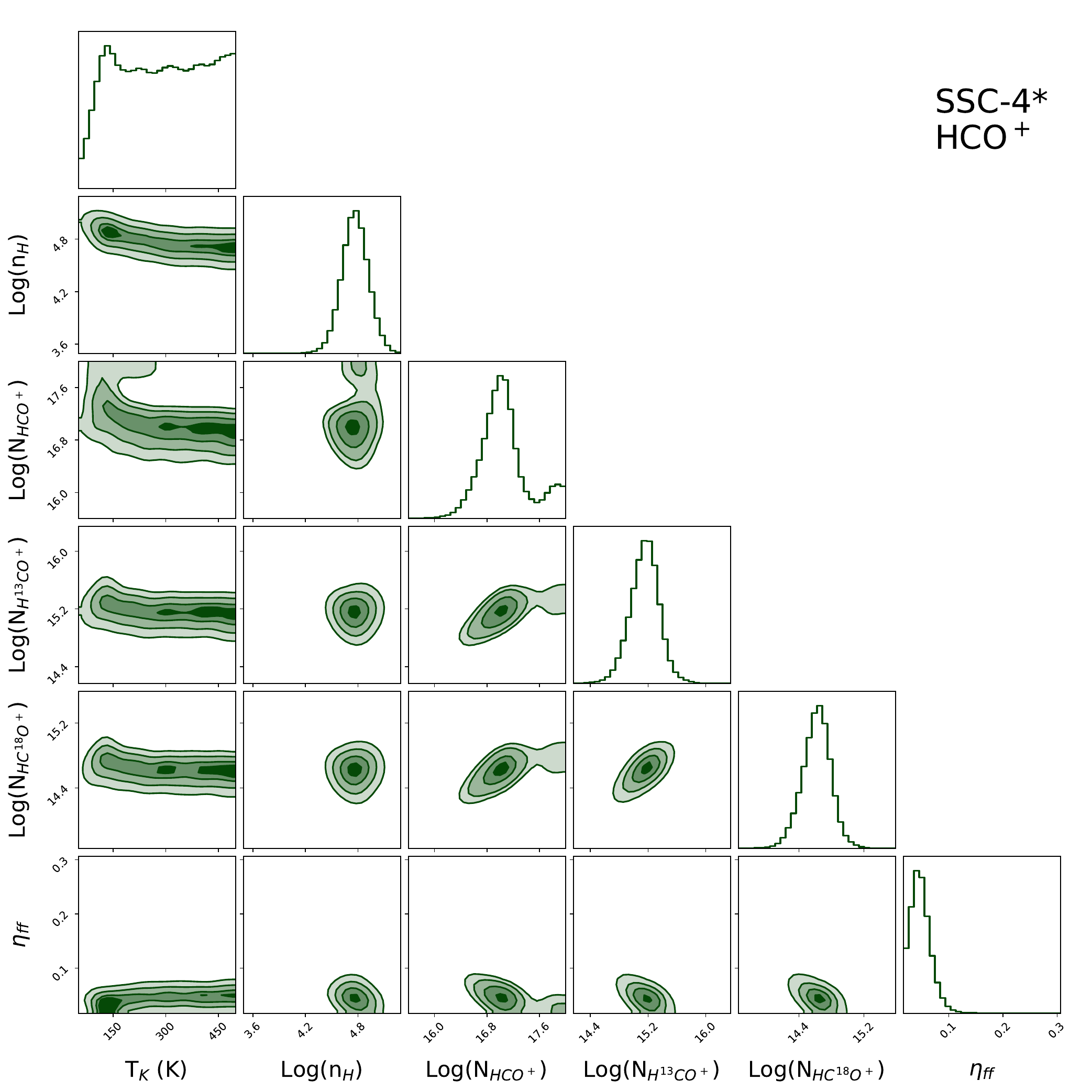}
  \caption{The posterior distributions for SSC-4* of the kinetic temperature, neutral H\textsubscript{2} number density, beam filling factor and column densities of HCO\plus, H\thirteen CO\plus\ and HC\eighteen O\plus, as predicted by \texttt{RADEX}. }
  \label{fig:RADEX_corner_SSC4_HCOP}
\end{figure*}

Figures \ref{fig:RADEX_corner_SSC4_CO}-\ref{fig:RADEX_corner_SSC4_HCOP} show the posterior distributions of the sampling for the CO, HCN, and HCO$^+$ isotopologues toward the SSC-4* region.  The remaining posterior distributions for the other regions are shown in Appendix \ref{app:Corners}. A summary of the results of the \texttt{Ultranest} \texttt{RADEX} model fitting can be found in Tables \ref{tab:CO_RADEX_Results}-\ref{tab:HCOP_RADEX_Results}. As can be seen from the figures, the majority of the parameters are constrained for  each isotopologue group sampling, thanks to the large number of lines available in the ALCHEMI survey. The kinetic temperature, $T_{\rm K}$, however, remains largely unconstrained regardless of the region or isotopologue group. As already hinted by the LTE analysis, this may be (at least partly) due to the fact that there are multiple temperature components of the gas, a cold one being traced by the lower J transitions and another hot one being traced by higher J transitions. Another contributing factor is that the transitions being observed have an excitation temperature of $<\sim40$K, and since we are probing a temperature regime of $>100$K \citep[see ][]{2019_Mangum_Heart} in these regions, these lines are not suited to constrain $T_{\rm K}$.

As one would expect, the HCN and HCO\plus\ isotopologues trace  denser regions of the gas ($\text{log}(n_H) > 4.3$) compared to the CO isotopologues ($\text{log}(n_H) \lesssim 3.5$). Also, as may be expected, CO isotopologues appear to be tracing more extended regions of the gas as shown by the larger beam filling factors, $\eta_{ff}$ ($\sim 0.3$, $\sim 5$~pc), compared to those of the HCN and HCO\plus\ isotopologues ($\lesssim 0.1$, $\lesssim 2$~pc, with the exception the 2 absorption affected regions for HCN). These lower values of $\eta_{ff}$ correspond to the scale of the constituent SSCs, as defined by \cite{2018Leroy} (2-3~pc), located within each beam, thus reassuring us that the emission we are tracing and modelling is sourced from the SSCs we are observing rather than the larger scale GMCs defined in NGC~253 \citep{Martin_ALCHEMI_2021,Huang2023}.  

A particular region of note for HCN and HCO\plus\ and their isotopologues is SSC-14* as it shows significantly higher-intensity emission than the other regions. This can be associated with the increased predicted column density of each of these molecules by $\leq0.5$~dex for some isotopologues, compared to the next most abundant region. In fact these high column densities approach the parameter space under which \texttt{RADEX} struggles to converge as a result of line saturation at high opacities, particularly for HCO\plus\ (N$_{\text{HCO}^+}>10^{17}$ \text{cm}$^{-2}$).

The predicted column densities of both HCO\plus\ and its corresponding \thirteen C isotopologue are shown to consistently be higher than their corresponding HCN species counterparts. This enhancement is nearly of an order of magnitude for a couple of the regions (SSC-1*, SSC-14*). This supports the conclusion made in previous studies that HCO\plus\ (and thus its isotopologues) are relatively more abundant than HCN in regions of high star formation, usually when compared to other active regions, such as AGN-dominated regions \citep{Krips_2008,Izumi2016,2022Butterworth}.

\begin{table*}[]
\centering
\caption{The median, 16th, and 84th percentile parameter value results from the \texttt{Ultranest} \texttt{RADEX} model fitting of the CO, \thirteen CO, C\eighteen O and C\seventeen O observed transition lines for each SSC* region.}
\label{tab:CO_RADEX_Results}
\begin{tabular}{cccccccc}
\hline
Location & T$_K$                     & Log(n$_{\rm H}$ )               & Log(N$_{\rm CO}$)             & Log(N$_{\rm ^{13}CO}$)        & Log(N$_{\rm C^{18}O}$)        & Log(N$_{\rm C^{17}O}$)        & Beam Filling      \\ & (K) & (cm$^{-3}$) & (cm$^{-2}$) & (cm$^{-2}$) & (cm$^{-2}$) & (cm$^{-2}$) & Factor \\ \hline \hline
SSC-1*    & $253_{-120}^{123}$ & $3.19_{-0.12}^{0.18}$ & $20.0_{-0.7}^{0.6}$ & $18.5_{-0.4}^{0.2}$ & $18.0_{-0.3}^{0.2}$ & $17.1_{-0.3}^{0.2}$ & $0.167_{-0.056}^{0.164}$ \\ \hline
SSC-4*   & $252_{-133}^{158}$ & $3.26_{-0.16}^{0.20}$ & $19.9_{-0.6}^{0.7}$ & $18.5_{-0.3}^{0.3}$ & $17.9_{-0.3}^{0.2}$ & $17.0_{-0.3}^{0.2}$ & $0.206_{-0.077}^{0.185}$ \\ \hline
SSC-7*    & $256_{-129}^{150}$ & $3.26_{-0.15}^{0.21}$ & $19.9_{-0.7}^{0.7}$ & $18.4_{-0.4}^{0.3}$ & $17.8_{-0.3}^{0.2}$ & $16.9_{-0.3}^{0.2}$ & $0.211_{-0.077}^{0.201}$ \\ \hline
SSC-9*    & $239_{-134}^{170}$ & $3.29_{-0.17}^{0.24}$ & $19.8_{-0.7}^{0.8}$ & $18.3_{-0.4}^{0.3}$ & $17.7_{-0.3}^{0.3}$ & $16.8_{-0.3}^{0.2}$ & $0.233_{-0.098}^{0.245}$ \\ \hline
SSC-13*   & $203_{-117}^{183}$ & $3.44_{-0.21}^{0.28}$ & $19.7_{-0.6}^{0.8}$ & $18.3_{-0.3}^{0.3}$ & $17.7_{-0.3}^{0.3}$ & $16.8_{-0.3}^{0.2}$ & $0.241_{-0.103}^{0.199}$ \\ \hline
SSC-14*   & $227_{-123}^{158}$ & $3.39_{-0.18}^{0.21}$ & $20.2_{-0.6}^{0.6}$ & $18.7_{-0.3}^{0.2}$ & $18.1_{-0.3}^{0.2}$ & $17.2_{-0.2}^{0.2}$ & $0.137_{-0.045}^{0.103}$\\ \hline

\end{tabular}
\end{table*}

\begin{table*}[]
\centering
\caption{The median, 16th, and 84th percentile parameter value results from the \texttt{Ultranest} \texttt{RADEX} model fitting of the HCN, H\thirteen CN and HC\fifteen N observed transition lines for each SSC* region.}
\label{tab:HCN_RADEX_Results}
\begin{tabular}{ccccccc}
\hline
Location & T$_K$              & Log(n$_{\rm H}$ )               & Log(N$_{\rm HCN}$)            & Log(N$_{\rm H^{13}CN}$)       & Log(N$_{\rm HC^{15}N}$)       & Beam Filling       \\ & (K) & (cm$^{-3}$) & (cm$^{-2}$) & (cm$^{-2}$) & (cm$^{-2}$) & Factor\\\hline \hline
SSC-1*    & $271_{-109}^{134}$ & $5.48_{-0.09}^{0.10}$ & $16.5_{-0.3}^{0.3}$ & $15.2_{-0.3}^{0.3}$ & $14.4_{-0.2}^{0.2}$ & $0.069_{-0.022}^{0.041}$ \\ \hline
SSC-4*    & $294_{-120}^{131}$ & $5.54_{-0.08}^{0.09}$ & $16.6_{-0.3}^{0.3}$ & $15.3_{-0.2}^{0.2}$ & $14.4_{-0.2}^{0.2}$ & $0.063_{-0.017}^{0.031}$ \\ \hline
SSC-7*    & $325_{-147}^{120}$ & $5.56_{-0.08}^{0.09}$ & $16.7_{-0.3}^{0.2}$ & $15.5_{-0.2}^{0.2}$ & $14.5_{-0.2}^{0.1}$ & $0.055_{-0.011}^{0.023}$ \\ \hline
SSC-9*    & $310_{-144}^{127}$ & $5.78_{-0.26}^{0.15}$ & $15.4_{-0.5}^{1.2}$ & $14.2_{-0.5}^{0.9}$ & -                   & $0.260_{-0.187}^{0.472}$ \\ \hline
SSC-13*   & $211_{-99}^{163}$  & $6.05_{-0.14}^{0.17}$ & $15.3_{-0.5}^{1.0}$ & $14.2_{-0.4}^{0.7}$ & -                   & $0.293_{-0.222}^{0.437}$ \\ \hline
SSC-14*   & $264_{-130}^{138}$ & $5.78_{-0.09}^{0.09}$ & $16.8_{-0.2}^{0.2}$ & $15.7_{-0.2}^{0.1}$ & $14.9_{-0.1}^{0.1}$ & $0.047_{-0.007}^{0.012}$ \\ \hline
\end{tabular}
\end{table*}
\begin{table*}[]
\centering
\caption{The median, 16th, and 84th percentile parameter value results from the \texttt{Ultranest} \texttt{RADEX} model fitting of the HCO\plus, H\thirteen CO\plus\ and HC\eighteen O\plus\ observed transition lines for each SSC* region.}
\label{tab:HCOP_RADEX_Results}
\begin{tabular}{ccccccc}
\hline
Location & T$_K$                     & Log(n$_{\rm H}$)               & Log(N$_{\rm HCO^+}$)          & Log(N$_{\rm H^{13}CO^+}$ )     & Log(N$_{\rm HC^{18}O^+}$ )     & Beam Filling      \\
& (K) & (cm$^{-3}$) & (cm$^{-2}$) & (cm$^{-2}$) & (cm$^{-2}$) & Factor\\
\hline \hline
SSC-1*    & $297_{-137}^{143}$ & $4.44_{-0.57}^{0.22}$ & $16.9_{-0.5}^{0.6}$ & $15.3_{-0.4}^{0.6}$   & $14.7_{-0.3}^{0.6}$   & $0.068_{-0.018}^{0.032}$ \\ \hline
SSC-4*    & $291_{-151}^{147}$ & $4.77_{-0.13}^{0.13}$ & $17.0_{-0.3}^{0.4}$ & $15.2_{-0.2}^{0.2}$   & $14.6_{-0.2}^{0.1}$   & $0.045_{-0.010}^{0.017}$ \\ \hline
SSC-7*    & $180_{-67}^{208}$  & $4.93_{-0.14}^{0.14}$ & $17.2_{-0.3}^{0.6}$ & $15.3_{-0.2}^{0.2}$   & $14.7_{-0.1}^{0.1}$   & $0.033_{-0.008}^{0.011}$ \\ \hline
SSC-9*    & $270_{-154}^{154}$ &  $4.36_{-0.73}^{0.62}$ & $17.1_{-0.8}^{0.6}$ & $15.4_{-0.7}^{0.6}$   & $14.5_{-0.6}^{0.6}$   & $0.081_{-0.030}^{0.066}$ \\ \hline 
SSC-13*   & $263_{-148}^{165}$ & $4.38_{-0.66}^{0.57}$ & $17.1_{-0.6}^{0.6}$ & $15.4_{-0.6}^{0.6}$   & $14.6_{-0.5}^{0.6}$   & $0.091_{-0.032}^{0.064}$ \\ \hline
SSC-14*   & $125_{-36}^{212}$  & $4.80_{-0.34}^{0.21}$ & $17.6_{-0.5}^{0.3}$ & $15.7_{-0.3}^{0.4}$   & $15.0_{-0.2}^{0.3}$   & $0.039_{-0.010}^{0.014}$ \\ \hline
\end{tabular}
\end{table*}

Moreover, the optical depth conditions of each investigated species and its respective transitions for each SSC* region were obtained. These results are shown in Appendix \ref{app:tau}. We observe that the majority of non-main isotopologues are indeed optically thin ($\tau < 1$) for the majority of our SSC* regions, with a couple of notable exceptions. Specifically, \thirteen CO, H\thirteen CO\plus\ and HC\eighteen O\plus\ have multiple lines across the different regions that are optically thick, primarily driven by the significant column densities that we observe of these species. The optical depth for each species is also predicted to be lower for the J=1-0 transitions than for the J=2-1 transitions ($\tau_{10}<\tau_{21}$). The physical reason for this occurrence can be explained by the dependence of $\tau$ on the $N_u$ and $B_{ul}$ parameters \citep[see Eq.~(6), ][]{Goldsmith_1999} where $B_{ul}$ is the Einstein coefficient for induced emission. In the case of our lines the increase of $B_{ul}$ with increasing transitions is larger than the decrease of $N_u$, this thus results in increasing optical depth with transition level.

\section{Isotopic Ratios and their Relationship to SSC Ages}
\label{sec:Ratio_age}

\subsection{SSC Age Determination}
\label{sec:Age}

Since isotopes are generated and supplied to the ISM over varying timescales, dependent on the evolutionary stages of intermediate-mass or more massive stars, by examining and comparing how these ages correlate (or not) to abundance ratios we can more reliably test the hypothesis of that factors such as the IMF are the primary driving forces of observed isotopic ratios \citep{Romano2017}. In order to determine ages for each SSC* region we used the stellar population modelling software \texttt{Starburst99} to compute the evolution of how the equivalent width of the Br$\gamma$, EW(Br$\gamma$), as well as the ionizing photon rate estimated from the H39$\alpha$ line evolved with the age of the star-forming region \citep{Starburst99_1999,Starburst99_2014}. How the ionizing photon rate is derived  has been summarised in Appendix \ref{app:H39}. Ionizing photon rate and EW(Br$\gamma$) are both inversely proportional to the age of a high star-forming region such as an SSC, as they are primarily driven by the presence of the most massive stars. \texttt{Starburst99} operates by populating a main sequence of stars under specified initial conditions and allowing the evolution of these populations to occur over a user-defined timescale (in this study, $10^4-10^9$ yrs). The evolution of these populations results in various effects upon observable properties, such as ionizing photon rate and EW(Br$\gamma$), as a result of how its constituent stars' progress along their evolutionary track. \texttt{Starburst99} (S99) allows for the specification of the IMF and of upper mass boundary of formed stars. Means of formation of new stars in this model can be conducted in two primary ways: either all of the population can be created initially (known as Instantaneous formation) or continuously over the entire period of the model at intermittent time steps.

Before we can obtain ages for the SSC* regions, we have to verify which combination of parameters reliably reproduces the observed conditions in these regions. S99 allows for various input parameters, but the ones most prudent to this study are IMF power law, stellar mass cutoff and formation type.
To do this we fit the ionizing photon rates of H39$\alpha$ and the equivalent width of Br$\gamma$, as both are available outputs of S99. The range of initial conditions can be found in Table \ref{tab:S99_Params}. The different formation types (Instantaneous and Continuous) determine whether S99 is entirely populated with stars initially or if there is a constant rate of new stars formed, respectively. The masses used as a part of these models were taken from a previous study of the SSCs in NGC~253 ($\sim 10^5$~M$_\odot$, \citealt{2018Leroy}). We found that, when setting appropriate masses to be achieved $10^5$~M$_\odot$ the Continuous models were not able to reproduce the observed values of Br$\gamma$ EW or ionizing photon rate over the entire time range and thus were discarded in favour of Instantaneous models. The remaining parameters of both IMF and Star Mass Boundary were then fit using the ionizing rate and EW(Br$\gamma$). The IMF power-law values were chosen to represent typical IMF scenarios, 1.5, 2.35 and 3.3 representing top-heavy, a typical Salpeter IMF and a bottom-heavy scenario respectively \citep{1955_Salpeter}.

\begin{table}[ht!]
  \centering
  \caption{The investigated initial parameters of \texttt{Starburst99}.}
  \label{tab:S99_Params}
  \begin{tabular}{c|c}
  \hline
    Variable  & Range \\
    \hline \hline
    Formation Type  & Instantaneous, Continuous\\
    IMF (Power-law) & 1.5, 2.35, 3.3 \\
    Star Mass Boundary (M$_\odot$) & 100, 150, 250 \\
    \hline
  \end{tabular}
\end{table}

We extracted H39$\alpha$ radio recombination line emission from the ALCHEMI dataset in the six SSC* apertures in order to estimate an ionizing photon rate free of dust obscuration. In all SSCs* a component consistent with He39$\alpha$ which has a velocity offset from H39$\alpha$ of $-$122.15~\kms, was also present. In SSC-1*, 4*, 7*, and 14*, we fit one Gaussian component to hydrogen and one Gaussian component to helium. In SSCs-9* and 13*, we fit two Gaussian components to the two observed velocity components of hydrogen and one Gaussian component to helium. The spectra, shown with fitting, for each region are supplied in Appendix \ref{app:H39}. We use the total hydrogen H39$\alpha$ integrated emission, summed over two components where applicable, to estimate the ionizing photon rate, as described in \cite{Emig_2020} and briefly in Appendix \ref{app:H39}. 

The equivalent width of the Br$\gamma$ line was obtained by measuring the line emission over the SSC* apertures using the Br$\gamma$ image from \cite{2013Rosenberg}. The equivalent width was obtained by using this line emission in combination with the K continuum.

A top-heavy IMF power-law of 1.5 and a upper stellar mass cutoff of 150~M$_{\odot}$ with instantaneous star formation was shown to be the best fit to both the H39$\alpha$ and Br$\gamma$ data. We were thus able to obtain ages for each of the SSC* regions as predicted by \texttt{Starburst99}.
We find an average best fit for the age of the order of $10^6$~years, which implies that nucleosynthesis effects are unlikely to lead to isotopic differentiation across the SSC regions due to their young ages, on the order of $10^6$~years,  relative to the necessary timescales required for the production of the isotopes ($\sim10^8$~yrs, \citealt{Milam_2005}). For example, \thirteen C is primarily produced from \twelve C during the CNO cycle of intermediate mass stars ($1.5-6$~M$_\odot$) and ejected during their AGB phase \citep{13C}. Similarly, \fifteen N can be produced and supplied to the ISM via nova explosions, this can lead to variations in the \fourteen N/\fifteen N across the galactic disk, however $10^9$~yrs are typically required for significant enough nova to occur \citep{Romano2017,2019_Romano,2022_Colzi}. We investigate the relationship between the SSCs* ages and isotopic ratios in the next section to see whether a correlation indeed exists. 


\subsection{Is There a Correlation Between SSC Age and Isotopic Ratio?}

From the derived column densities we have the possibility to investigate whether there is a correlation between the age of each star-forming region and the isotopic ratio. Table \ref{tab:Ratio_Ranges} contains the ranges and means of the isotopologue column ratios produced from the \texttt{RADEX} modelling.
To this end we shall summarise the primary isotopic ratios investigated in this study and how their predicted column densities correlate to the modelled ages of each region. The plots for the column densities ratios versus the predicted ages will be shown for each isotopologue ratio, but the line intensity ratio versus age will only be shown for the J(1-0) transition of \twelve CO/\thirteen CO. The \texttt{RADEX} column density predictions give the abundance of a molecule and \texttt{RADEX} takes into account the optical depth effects, as such they give typically a more reliable result than individual line intensity ratios.  For optically-thin LTE gas, both results should be comparable. Given the low sample size of 4 to 6 regions (depending on the ratio) quantified statistics of any possible correlation between age and column density ratio are not given. Instead a qualitative assessment of each ratio vs age plot has been completed.

\begin{table}[]
\caption{The ranges and means of the isotopologue column density ratios from across all the SSC* regions.}
\label{tab:Ratio_Ranges}
\centering
\begin{tabular}{ccc}
\hline
Ratio                      & X$_{Ratio}$ & $\bar{X}_{Ratio}$ \\ \hline
CO/$^{13}$CO               & 26.7-35.9         & 28.3              \\
HCO$^{+}$/H$^{13}$CO$^{+}$ & 36.3-89.9         & 58.1              \\
HCN/H$^{13}$CN             & 11.1-22.4         & 15.4              \\
HCO$^{+}$/HC$^{18}$O$^{+}$ & 144.6-409.6       & 262.7             \\
CO/C$^{18}$O               & 100.0-145.1       & 106.6             \\
HCN/HC$^{15}$N             & 75.7-183.4        & 128.4             \\
CO/C$^{17}$O               & 778.8-1147.0      & 847.2 \\ \hline
\end{tabular}
\end{table}
\subsubsection{\twelve C/\thirteen C}
\label{subsubsec:12C/13C}

\begin{figure}
  \centering
  \begin{tabular}[b]{@{}p{0.48\textwidth}@{}}
    \centering\includegraphics[width=1.0\linewidth]{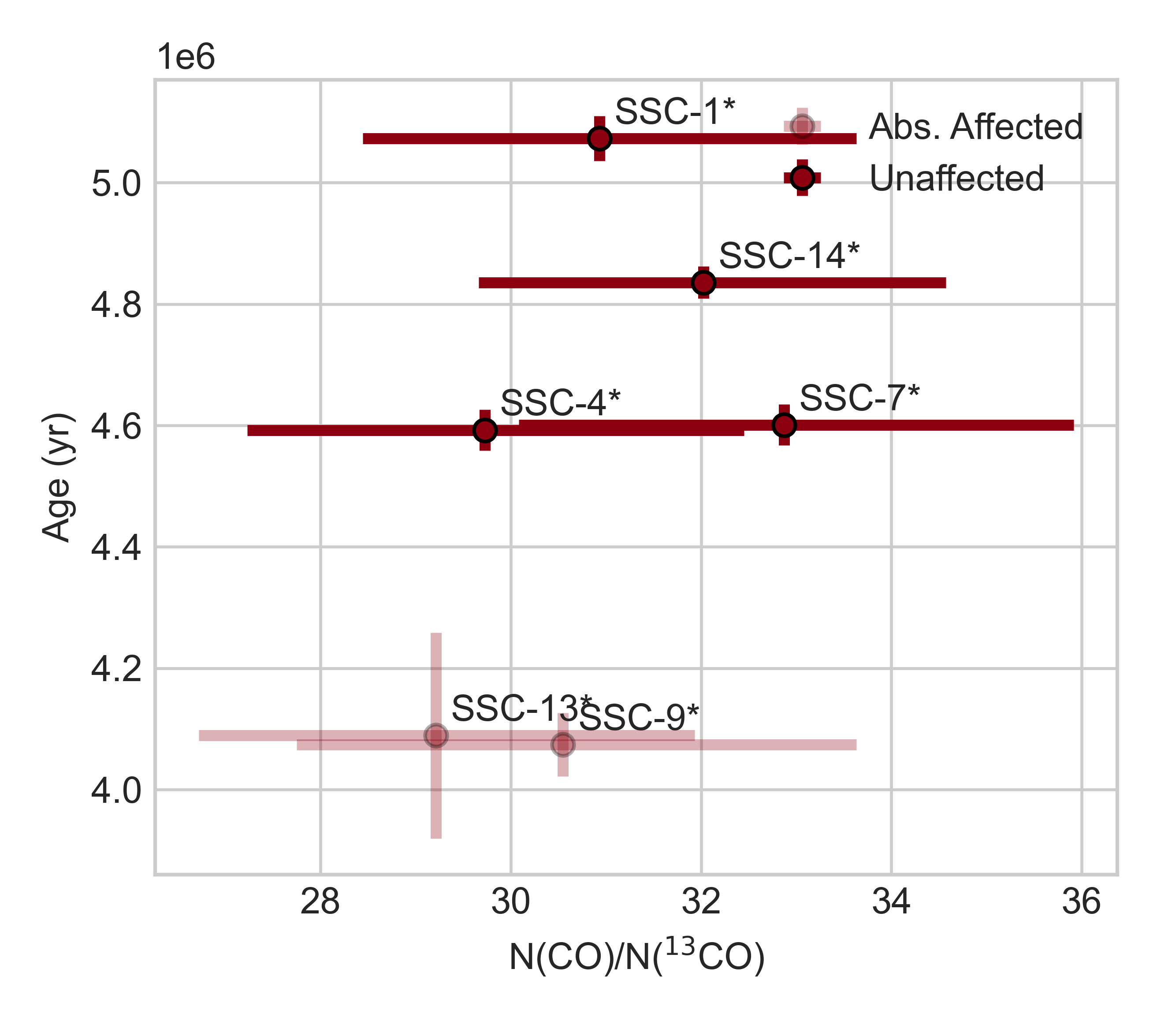} \\
    \centering\small (a) 
  \end{tabular}%
  \quad
\begin{tabular}[b]{@{}p{0.48\textwidth}@{}}
    \centering\includegraphics[width=1.0\linewidth]{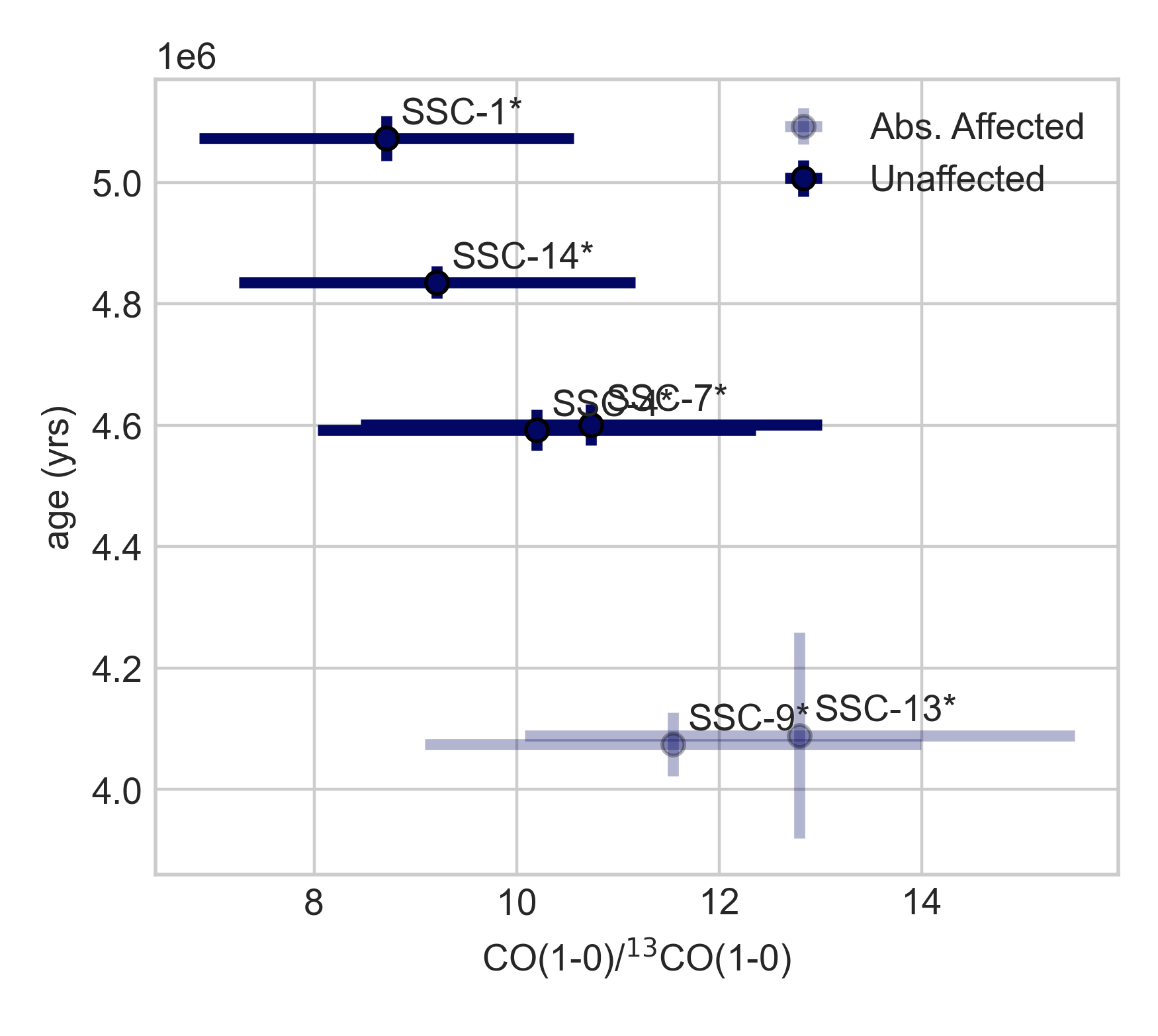} \\
    \centering\small (b) 
  \end{tabular}
  \caption{The \twelve CO/\thirteen CO column density ratio derived from \texttt{RADEX} is shown in (a) while the line intensity ratio for the J=$1-0$ transition from the ALCHEMI observations is shown in (b), of the ratio \twelve CO/\thirteen CO. The regions unaffected by the absorption effects observed near GMC-5 in NGC~253 are shown in full colour, whereas those regions that are affected are shown shaded.
  }
  \label{fig:CO_13CO_Ratios}
\end{figure}

Figure \ref{fig:CO_13CO_Ratios} shows the intensity and column density ratios of \twelve CO/\thirteen CO. It is clear that, especially in column densities, this ratio shows no correlation with regards to the predicted ages of each region. It is indeed possible that this lack of correlation is a result of the fact that these regions are too young ($<<10^{8}$~yrs) to observe a meaningful effect as a result of nucleosynthesis from any but the most massive of stars. There is also no relation with regards to spatial location of these regions column density ratios. The regions away from the `TH2' region do show lower \twelve C(1-0)/\thirteen C(1-0) intensities ratios, though all fall within error bars. 

\begin{figure}
  \centering
  \includegraphics[width=0.48\textwidth]{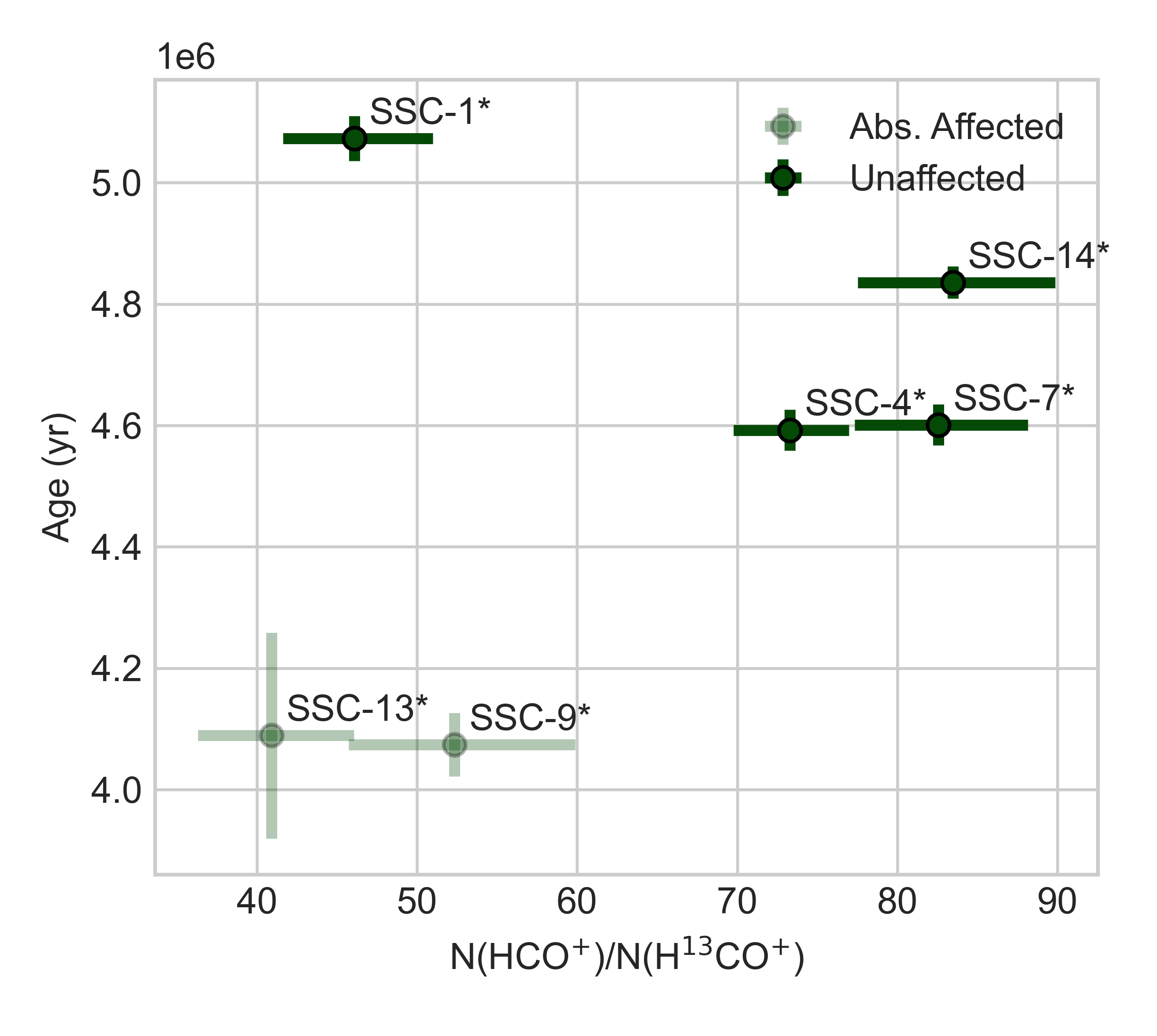}
  \caption{The H\twelve CO\plus/H\thirteen CO\plus\ column density ratio derived from \texttt{RADEX}. The regions unaffected by the absorption effects observed near GMC-5 in NGC~253 are shown in full colour, whereas those regions that are effected are shown shaded. }
  \label{fig:HCOP_H13OP}
\end{figure}

Figure \ref{fig:HCOP_H13OP} shows the column density ratio of the H\twelve CO\plus/H\thirteen CO\plus\ ratio (range, $\sim36-90$) versus age. This ratio does indeed show a negative correlation with age, with the exception of SSC*-13. However, we note that SSC*-13 is the region most affected by the absorption source observed near the dynamical center of NGC~253, as discussed earlier. If we disregard this region it would imply that there are some processes, other than nucleosynthesis, on this timescale that can affect this ratio. In fact, \cite{2015Roueff} found that H\twelve CO\plus/H\thirteen CO\plus\ experiences a turbulent variation due to fractionation processes up until $\sim10^6$ yr, which is of the same order as the observed ages of these regions \citep{2023_Sipila}. This is a tentative hypothesis however as we have only a few observed regions over a relatively small timescale, as well as assuming that SSC-13* is significantly affected by absorption features.

 If fractionation processes are indeed responsible for the trend seen in H\twelve CO\plus/H\thirteen CO\plus\ then it is possible that ISM fractionation processes have already approached an equilibrium with regards to the \twelve CO/\thirteen CO ratio by the time frame covered by our regions. Indeed this would be consistent with the chemical models of \cite{2015Roueff} who found that the \twelve CO/\thirteen CO ratio approached equilibrium swiftly and with less variation than H\twelve CO\plus/H\thirteen CO\plus. 

\begin{figure}
  \centering
  \includegraphics[width=0.48\textwidth]{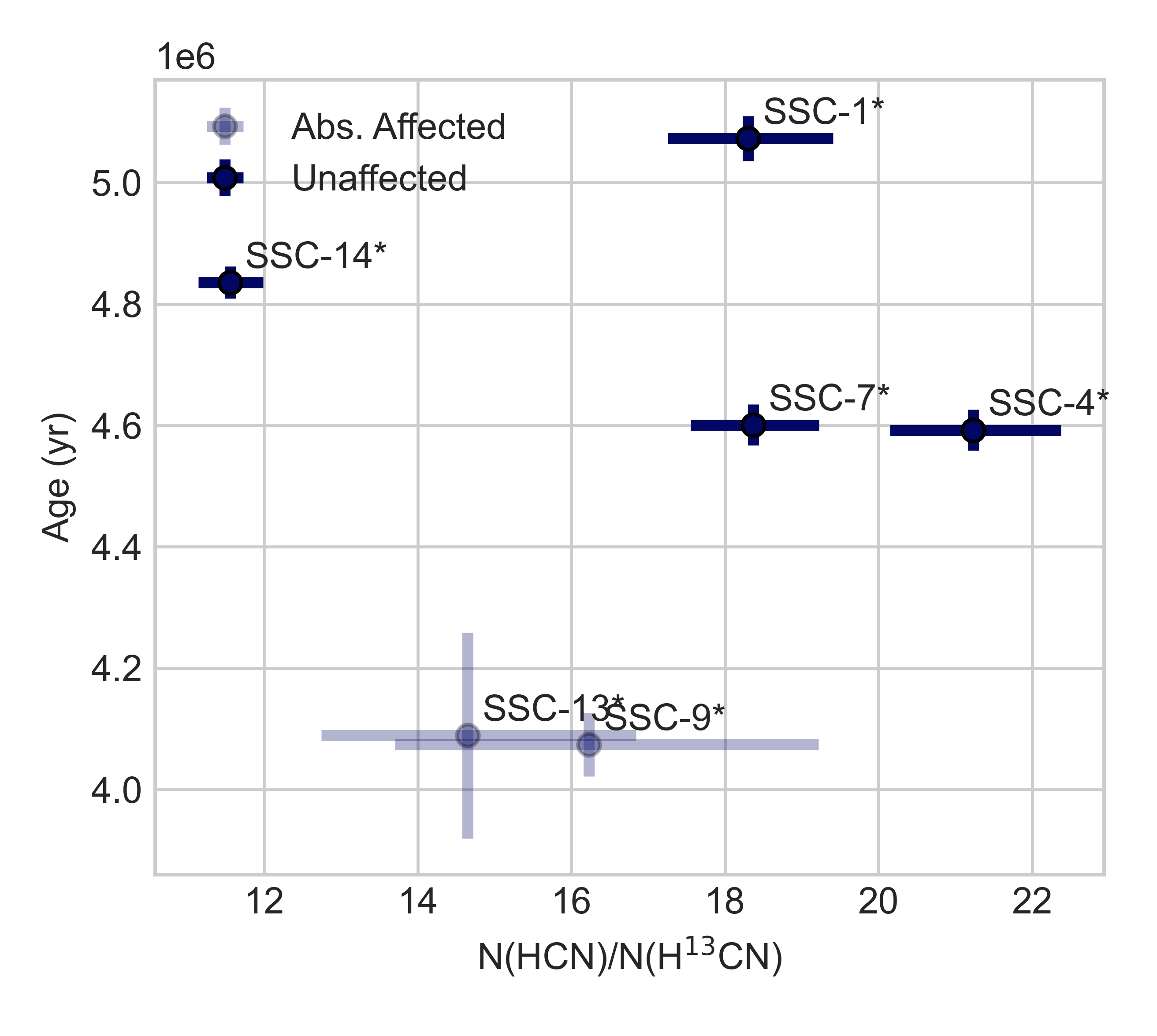}
  \caption{The H\twelve CN/H\thirteen CN column density ratio derived from \texttt{RADEX}. The regions unaffected by the absorption effects observed near GMC-5 in NGC~253 are shown in full colour, whereas those regions that are effected are shown shaded.}
  \label{fig:HCN_H13CN}
\end{figure}

Similarly to the \twelve CO/\thirteen CO ratio, the H\twelve CN/H\thirteen CN ratio, shown in Figure \ref{fig:HCN_H13CN}, shows little correlation with age, though the ratio shows significantly more variation between the regions than its CO counterpart. The H\twelve CN/H\thirteen CN ratio has been predicted to undergo a similarly turbulent variation as H\twelve CO\plus/H\thirteen CO\plus\ approaching the timescales observed across these SSC* regions; this may explain the significant variation observed \citep{2015Roueff,2023_Sipila}.

\begin{figure}[ht!]
    \centering\includegraphics[width=1.0\linewidth]{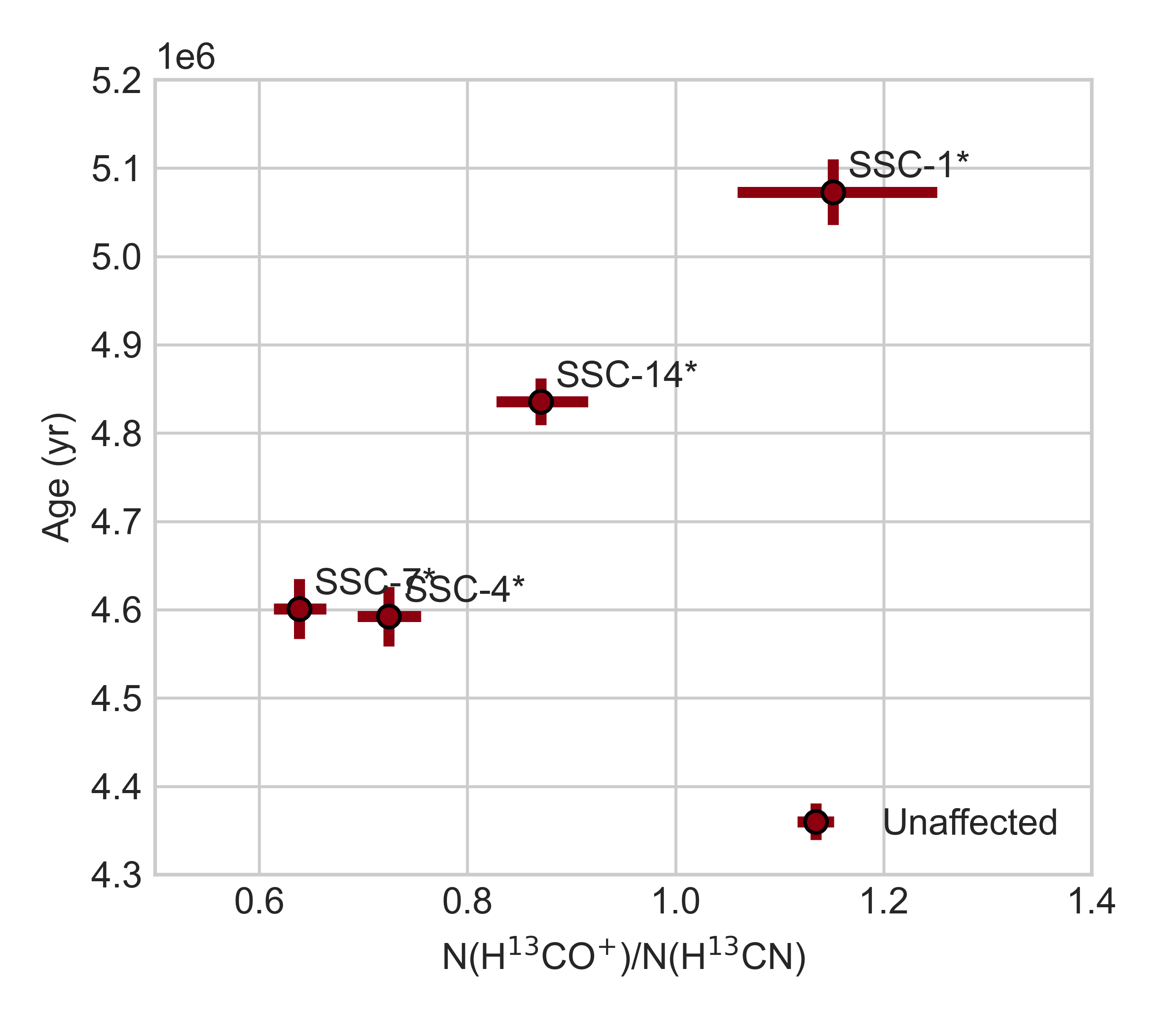} \\

      \caption{The H\thirteen CO\plus/H\thirteen CN column density ratio derived from \texttt{RADEX}. Due to the significant effect of the absorption on this ratio only the unaffected region are displayed here, the equivalent plot including the absorption affected regions is shown in Appendix \ref{app:more:ratio_age_plots}.}
  \label{fig:H13COP_H13CN}
\end{figure}

\begin{figure}[ht!]
    \centering\includegraphics[width=1.0\linewidth]{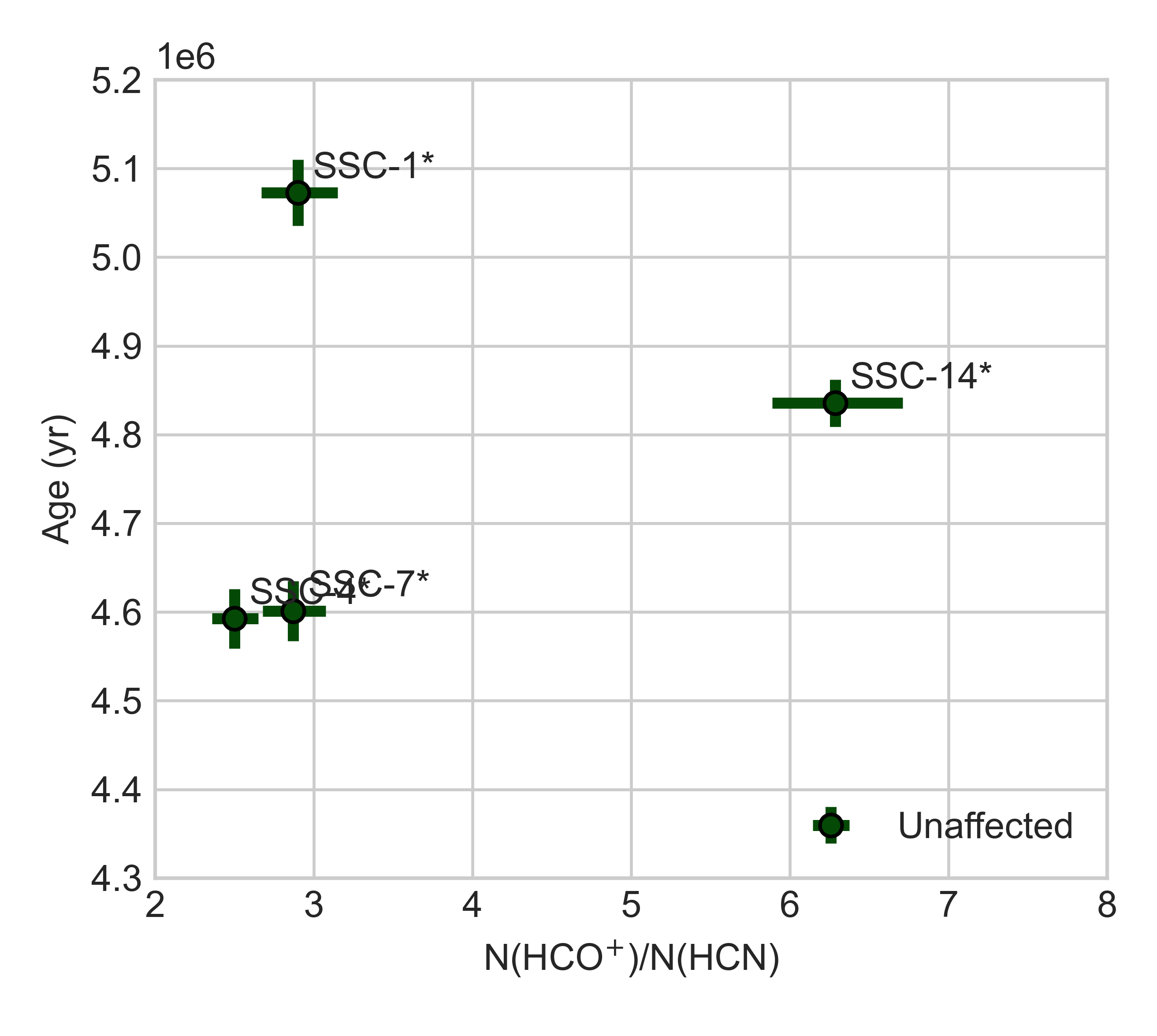} \\

      \caption{The HCO\plus/HCN column density ratio derived from \texttt{RADEX}. Due to the significant effect of the absorption on this ratio only the unaffected region are displayed here, the equivalent plot including the absorption affected regions is shown in Appendix \ref{app:more:ratio_age_plots}}
  \label{fig:HCOP_HCN}
\end{figure}

H\thirteen CO\plus\ and H\thirteen CN are the \thirteen C-bearing isotopologues of the dense gas tracers HCN and HCO\plus. These isotopologues are optically thin unlike their \twelve C counterparts.
An interesting relation was observed in the column density ratio of these isotopologues. As shown in Figure \ref{fig:H13COP_H13CN}, once we disregard the absorption affected regions of SSC-9* and SSC-13* (for the reasons stated earlier, the equivalent plot including the absorption affected region can be seen in Appendix \ref{app:more:ratio_age_plots}) there appears to be a positive relation between age and this ratio. This is in contrast to the equivalent ratio of HCO\plus/HCN shown in Figure \ref{fig:HCOP_HCN} in which no correlation with age can be seen.
This result perhaps supports the fractionation hypothesis for the correlation seen in H\twelve CO\plus/H\thirteen CO\plus. 
This ratio may also be a useful tool as a tracer of the age of SSC regions in the future as a result.

\cite{2019Martin} observed isotope ratios using spatially resolved optically thin isotopologues of CO in NGC~253. They observed \twelve C/ \thirteen C $\sim 21\pm 6$ using C\eighteen O and \thirteen C\eighteen O. In comparison we observed a larger ratio than this when viewed with CO/\thirteen CO ($\sim27-36$) and HCO\plus/H\thirteen CO\plus\ ($\sim36-80$) in our SSC* regions. A smaller ratio is observed with HCN/H\thirteen CN ($\sim11-22$), this range of values is consistent with observations of \cite{Meier_2015} who observed HCN/H\thirteen CN at 2'' scale with a range of $10 - 20$. \cite{Meier_2015} also observed HCO\plus/H\thirteen CO\plus\ at this scale and found a range of $10 - 20$.\cite{Martin_ALCHEMI_2021} observed these isotopologue ratios too, but at 15'' ($\sim 255$pc) resolution, where they observed values for CO/\thirteen CO of $13.3 \pm 1.4$, HCN/H\thirteen CN of $26.1 \pm 1.3$ and HCO\plus/H\thirteen CO\plus\ of $21.2 \pm 1.5$. It should be noted that the assumed source size was 10'' ($\sim$ 170 pc) for the CO ratio and 5'' for the HCN and HCO\plus\ ratios. Thus, the resolution and region in which these ratios are observed significantly affect the result. \cite{2019_Tang} observed CN and \thirteen CN at 3.6" $\times$ 1.7" ($\sim68 \times 32$pc) resolution in NGC~253 and observed an average \twelve C/\thirteen C ratio of $41.6 \pm 0.2$.

A similar study at a comparable resolution observing regions of intense star formation, like SSCs, has not been completed previously in nearby galaxies, as such a comparison to observations of the \twelve C/\thirteen C ratio would not be appropriate. The \twelve C/\thirteen C ratio has been observed over many regions over many galactic regions.
\cite{2023_Yan} provides an overview (In Table 7 of that paper) of \twelve C/\thirteen C ratios observed across different regions of the Milky Way. On average the \twelve C/\thirteen C varies between $\sim25 \pm 2$ in the CMZ up to $\sim69 \pm 12$ in the Outer Galaxy \citep{1985_Henkel,1990_Langer,1993_Langer,1996_Wouterloot,1998_Keene,2002_Savage, Milam_2005, 2019_Yan, 2023_Yan}.


In summary, the \twelve C/\thirteen C column density ratio from our \texttt{RADEX} modelling shows little correlation with the ages of these star-forming regions. A possible exception being the H\twelve CO\plus/H\thirteen CO\plus\ ratio, which could be a result of chemical fractionation affecting this ratio on the observed timescales  (rather than nucleosynthesis). The lack of correlation we observe supports the conclusion of \cite{Romano2017} that the IMF is a primary driving factor in the isotopic ratio observed in star- forming regions. Though it must be noted that this is only the age range covered by these SSCs.

\subsubsection{\sixteen O/\eighteen O}

\begin{figure}
  \centering
  \includegraphics[width=0.48\textwidth]{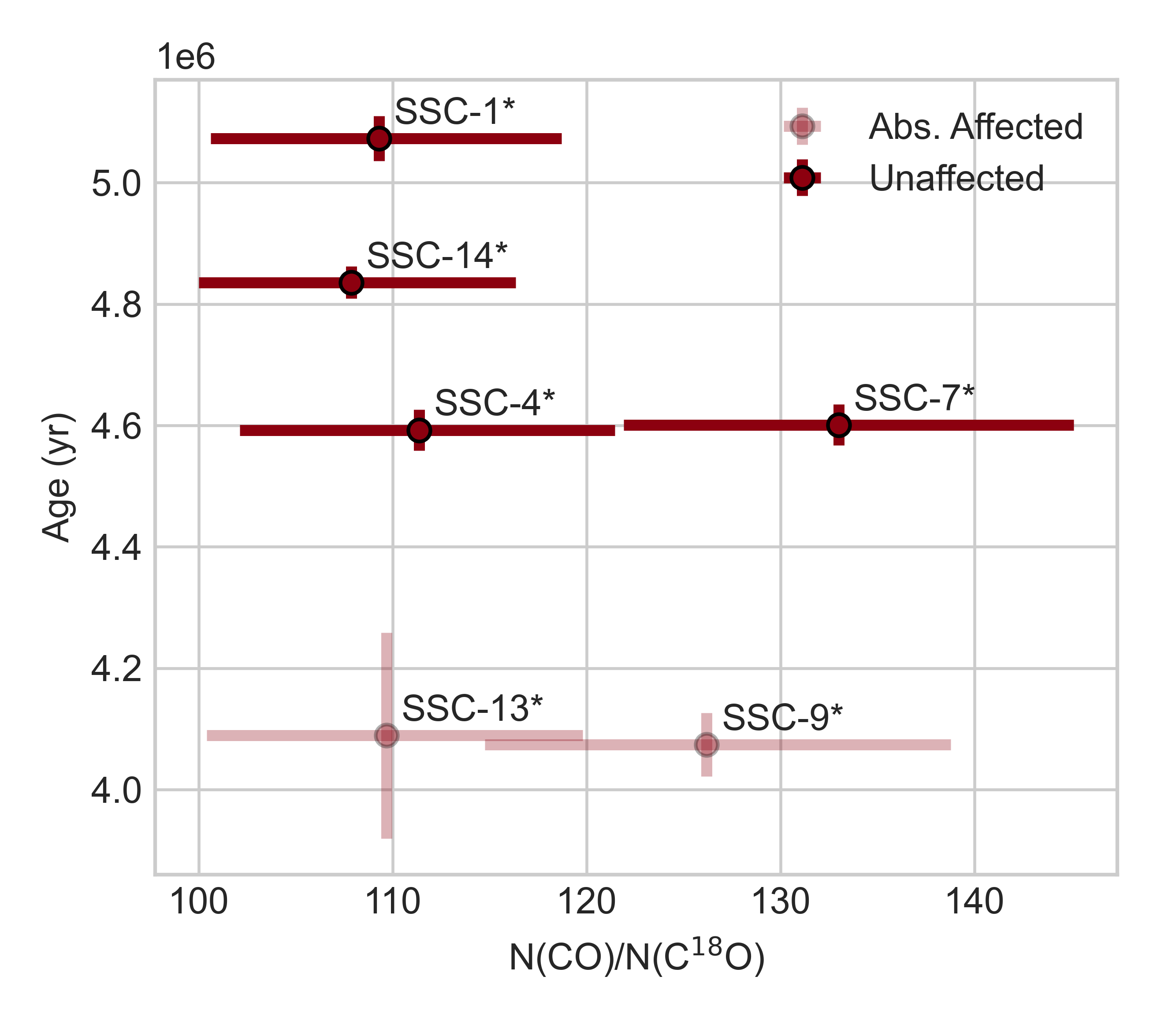}
  \caption{The C\sixteen O/C\eighteen O column density ratio derived from \texttt{RADEX}. The regions unaffected by the absorption effects observed near GMC-5 in NGC~253 are shown in full colour, whereas those regions that are effected are shown shaded.}
  \label{fig:CO_C18O}
\end{figure}

\begin{figure}
  \centering
  \includegraphics[width=0.48\textwidth]{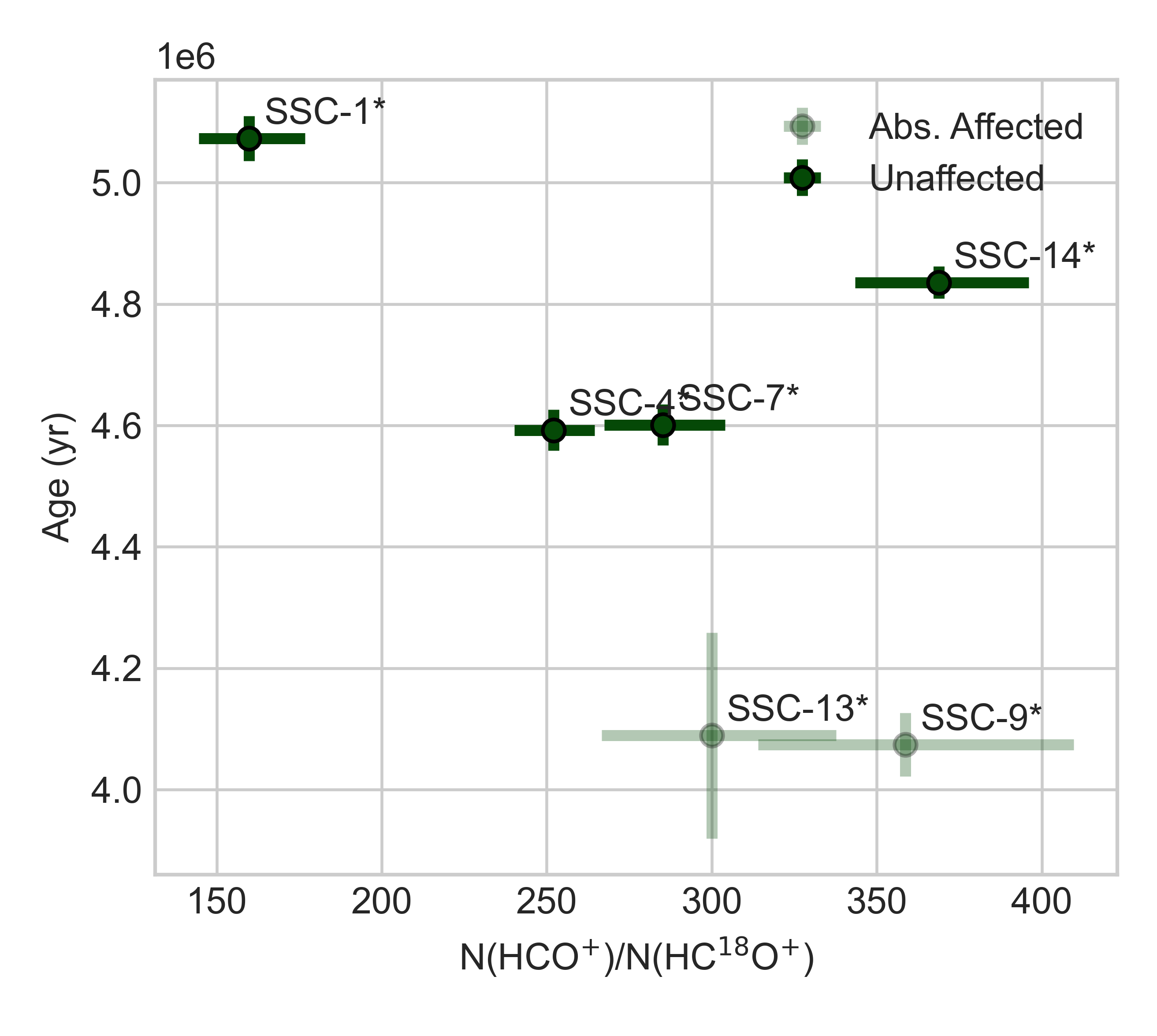}
  \caption{The HC\sixteen O\plus/HC\eighteen O\plus\ column density ratio derived from \texttt{RADEX}. The regions unaffected by the absorption effects observed near GMC-5 in NGC~253 are shown in full colour, whereas those regions that are effected are shown shaded.}
  \label{fig:HCOP_HCO18P}
\end{figure}

Figures \ref{fig:CO_C18O} and \ref{fig:HCOP_HCO18P} show the column density ratio of the C\sixteen O/C\eighteen O and HC\sixteen O\plus/HC\eighteen O\plus\ ratio, respectively. C\sixteen O/C\eighteen O clearly shows no correlation between age and column density ratio. The HC\sixteen O\plus/HC\eighteen O\plus\ ratio generally does not show a clear trend, though the oldest predicted region SSC-1* does show a significantly smaller ratio than the rest. To explore whether this is a trend or not would require an investigation into more regions and a larger range of timescales. When compared to previous observations of \sixteen O/\eighteen O, CO/C\eighteen O ($\sim100-145$) is largely consistent with those derived at 3'' ($\sim51$~pc) resolution in \cite{2019Martin} in NGC~253, using \thirteen CO/\thirteen C\eighteen O (130 $\pm 40$), however our results for the ratio of HCO\plus/HC\eighteen O\plus\ are significantly higher ($140-410$). \cite{Martin_ALCHEMI_2021} provides ALMA Compact Array observations of \sixteen O/\eighteen O ratios in NGC~253 at 25'' ($\sim 255$~pc) resolution calculated from CO/C\eighteen O ($48 \pm 5$), \thirteen CO/\thirteen C\eighteen O (520 $\pm 60$) and  HCO\plus/HC\eighteen O\plus\ ($100 \pm 20$). \cite{Aladro2015} observed the HCO\plus/HC\eighteen O\plus\ at an even lower resolution of 28'' found it to be $69 \pm 2$. 

As was the case for \twelve C/\thirteen C an appropriate comparison study in nearby galaxies has not yet been completed. \sixteen O/\eighteen O ratios, however, have been observed across different regions of the Milky Way. As viewed with H$_2$CO and its respective \eighteen O isotopologue, the \sixteen O/\eighteen O varies from $\sim263 \pm 45$ in the CMZ up to $\sim625 \pm 144$ in the Outer Galaxy \citep{1981_Gardner,1994_Wilson,2023_Yan}. 

The variations across all these observations and resulting ratios within the same galaxy, show that resolution and the regions being traced play a key role upon observed isotope ratio. As stated earlier, nucleosynthesis effects upon these ratios can most likely be discarded due to the young ages of these regions. 

\subsubsection{\sixteen O/\seventeen O}

\begin{figure}
  \centering
  \includegraphics[width=0.48\textwidth]{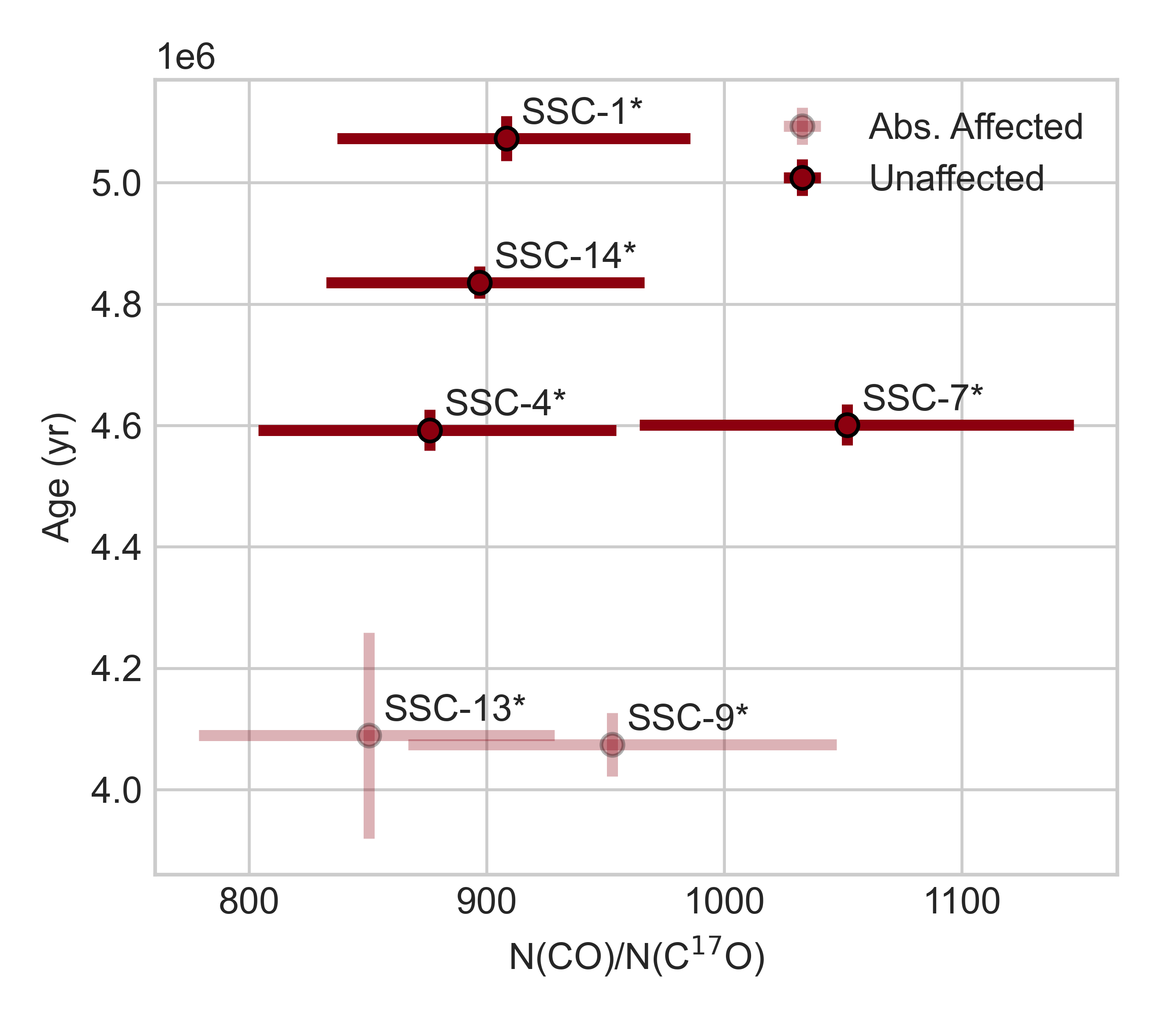}
  \caption{The C\sixteen O/C\seventeen O column density ratio derived from \texttt{RADEX}. The regions unaffected by the absorption effects observed near GMC-5 in NGC~253 are shown in full colour, whereas those regions that are effected are shown shaded.}
  \label{fig:CO_C17O}
\end{figure}

The column density ratio of C\sixteen O/C\seventeen O is shown in Figure \ref{fig:CO_C17O}. This ratio is largely unaffected by the SSC* age, similar to what is observed  in the other ratios. \seventeen O, unlike \eighteen O, can be produced during nucleosynthesis in the hydrogen burning of less massive stars, and so its enrichment in the ISM would likely be observed at even larger timescales than those expected by \eighteen O.
From the limited sample of this study, on the time scale probed by these regions, this ratio appears to be independent of SSC age. This ratio was observed at an observed source size of 10'' as $400 \pm 40$ in NGC253 in \cite{Martin_ALCHEMI_2021}, this is significantly lower than the what we observe in these SSC* regions $\sim 780-1150$.

Again, due to a lack of an appropriate comparison study in nearby galaxies, we supply instead a comparison to galactic observations. \sixteen O/\seventeen O ratios have been observed across different regions of the Milky Way. As viewed with H$_2$CO and its respective \seventeen O isotopologue, the \sixteen O/\seventeen O ratio varies from $\sim894 \pm 155$ in the CMZ up to $\sim3000 \pm 786$ in the Outer Galaxy according to \cite{2023_Yan} using data from \citep{1981_Gardner,1994_Wilson,2020_Zhang}. 

\subsubsection{\fourteen N/\fifteen N}

\begin{figure}
  \centering
  \includegraphics[width=0.48\textwidth]{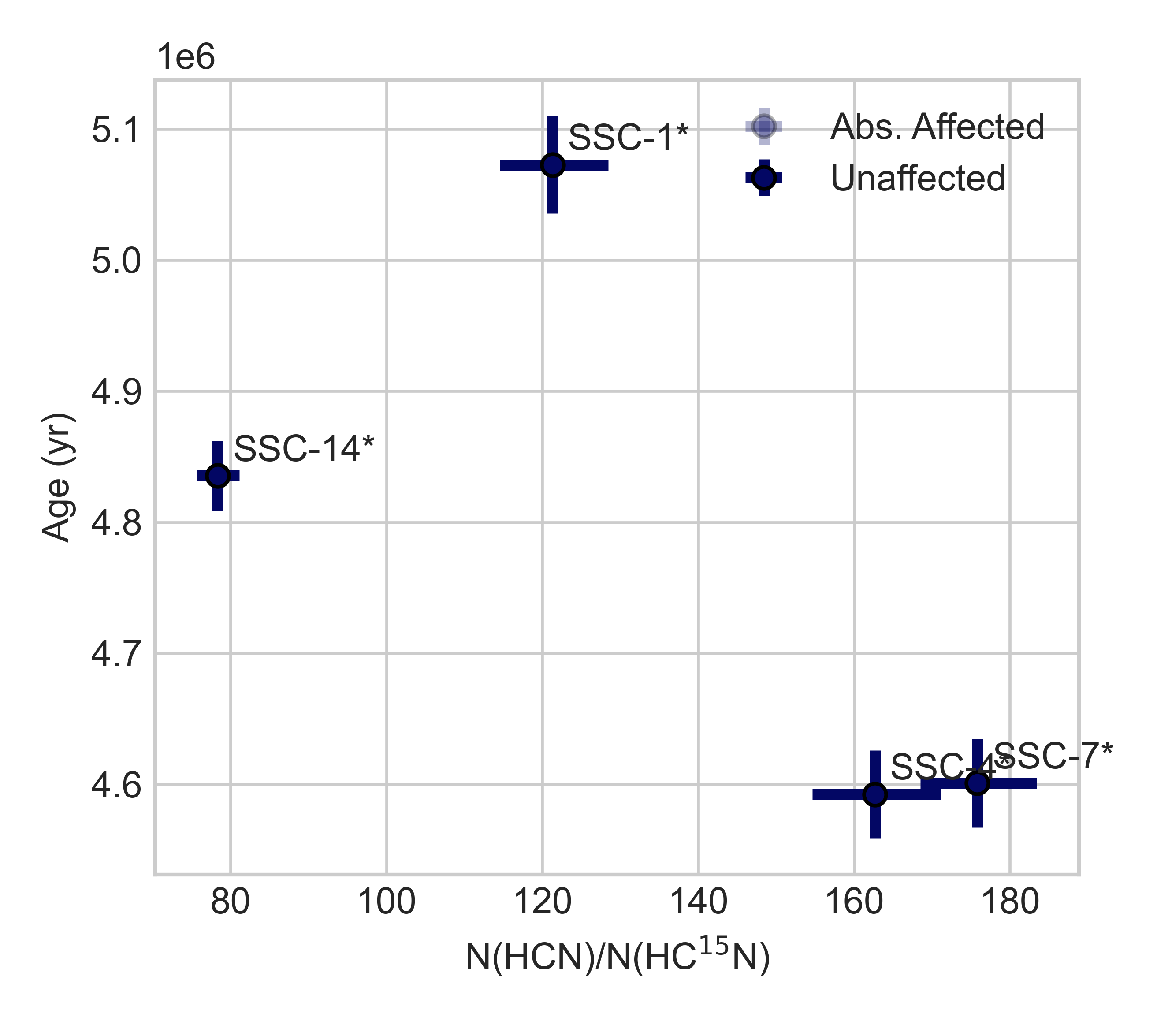}
  \caption{The column density ratio produced from \texttt{RADEX} of the ratio HC\fourteen N/HC\fifteen N. }
  \label{fig:HCN_HC15N}
\end{figure}

Figure \ref{fig:HCN_HC15N} shows the column density ratio of the HC\fourteen N/HC\fifteen N ratio. The analysis of this ratio is affected greatly by the necessity to discount the two regions, SSC*-9 and SSC*-13 where absorption effects become noticeable in the HC\fifteen N line profile. From the four remaining regions however we find again no sign of a correlation between age and the column density ratio. An interesting distinction is that the two regions located closest to the edge of the CMZ (SSC*-1 and SSC*-14) both have significantly lower HC\fourteen N/HC\fifteen N ratios than their more central counterparts (SSC*-4 and SSC*-7). Though without  observing more regions we can not draw any conclusions from this. 
Galactic observations of \fourteen N/\fifteen N have been made in previous studies, including with HC\fourteen N/HC\fifteen N as was observed in this paper. The values of \fourteen N/\fifteen N viewed across the Milky Way, when observed with HC\fourteen N/HC\fifteen N, range from $284 \pm 63$ within the Inner Disk up to $388 \pm 32$ in the Outer Galaxy \citep{2012_Adande,2018_Colzi}.

\subsubsection{Isotopic Ratios and ISM Fractionation}

Table \ref{tab:Ratio_Ranges} lists the observed ranges of isotopic ratios for the various species. Focusing initially on \twelve C/\thirteen C, \cite{2020Viti} conducted a theoretical chemical modelling study on carbon fractionation across different physical conditions that may represent different types of galaxies. Taking the most suitable model from this study, representing starburst galaxies (see Figure 10 in \cite{2020Viti}), H\twelve CN/H\thirteen CN is predicted ($\sim 70$) to be higher than what we observe ($\sim11$-$22$). The models agree relatively well for the younger regions with regards to H\twelve CO\plus/H\thirteen CO\plus\ ($\sim 55$). \twelve CO/\thirteen CO is also relatively well predicted by these models ($\sim 30$) relative to our observations ($\sim26$-$\sim36$).

\cite{2019Viti} completed a similar study on nitrogen fractionation. 
They found that, depending on the physical parameters, the resulting HC\fourteen N/HC\fifteen N abundance ratio could vary from $\sim 10$ to $\sim 1000$. The scenario for the most elevated ratios ($\sim 1000$) are where fractionation processes were suppressed; these were typically for cases of relatively low density $n_H \sim 10^4$~cm\textsuperscript{-3} and high cosmic-ray ionisation rate, visual extinctions or radiation fields. The significantly lower range of HC\fourteen N/HC\fifteen N seen in this study ($\sim 75-180$) as well as the higher densities predicted by the non-LTE results ($n_H > 10^5$~cm\textsuperscript{-3}) would imply that  fractionation processes are occurring within these regions.


\section{Conclusions}
\label{sec:Conc}

By using the plethora of data made available to us thanks to the ALCHEMI collaboration, we made use of lines from 10 isotopologues, collectively, of CO, HCN and HCO\plus\ in order to investigate the possibility of a correlation between the age of star-forming regions in NGC~253 and the observed isotopic ratios.  Our findings are summarised below:
\begin{itemize}

\item From our \texttt{RADEX} results HCN and its isotopologues appear to be tracing a slightly denser component of the gas ($n_H>10^5$~cm\textsuperscript{-3}) than HCO\plus\ and its isotopologues.
\item HCO\plus\ and its isotopologues  have larger column densities than their HCN counterparts. This is consistent with previous studies  that have found HCO\plus\ to be more abundant than HCN in regions of high star formation. CO and its isotopologues, as one might expect, trace a more diffuse and extended component of the gas in these regions, as shown by their lower densities ($n_H<10^4$~cm\textsuperscript{-3}).
\item Our results regarding the \twelve CO/\thirteen CO and H\twelve CN/H\thirteen CN seem to discount a clear relation between the age of a star forming region and the isotopic ratio \twelve C/\thirteen C.
\item H\twelve CO\plus/H\thirteen CO\plus\ may show a tentative negative correlation, as long as one makes the assumption that the results for SSC*-13 are being greatly perturbed by the absorption feature. In order to verify this trend, a follow-up study of these regions at higher resolution to differentiate the individual SSCs in HCO\plus\ and its isotopologues would be required. This would allow a more complete understanding of the possible fractionation occurring in these regions to be obtained.
\item The \sixteen O/\eighteen O ratios generally show little correlation with age, with the exception of the oldest region (SSC*-1) appearing to have a significantly lower HC\sixteen O\plus/HC\eighteen O\plus\ ratio than the younger regions.
\item The ratios HC\fourteen N/HC\fifteen N and HC\sixteen O\plus\ and HC\seventeen O\plus\ also show no correlation with the age of the SSCs.
\item The ratios observed in this study generally support the conclusion that age has a minor effect upon isotopic ratios, at least over the limited age range observed in these regions, thus supporting the hypothesis that the IMF is a primary driving factor of the observed isotopic ratios in these regions \citep{Romano2017} or possibly fractionation effects.


\end{itemize}


To more confidently make conclusive statements regarding the use of C, N, or O isotopic ratios, a future study either  observing the SSC regions in NGC~253 at higher spatial resolution, or perhaps a multi-object study observing a wider range of star-forming regions spanning a larger range of ages should be completed.

\section*{Acknowledgements}
J.B., S.V., K.Y.H. and M.B. have received funding from the European Research Council (ERC) under the European Union’s Horizon 2020 research and innovation programme MOPPEX 833460.  This paper makes use of the following ALMA data: ADS/JAO.ALMA\#2017.1.00161.L and ADS/JAO.ALMA\#2018.1.00162.S. ALMA is a partnership of ESO (representing its member states), NSF (USA) and NINS (Japan), together with NRC (Canada), MOST and ASIAA (Taiwan), and KASI (Republic of Korea), in cooperation with the Republic of Chile. The Joint ALMA Observatory is operated by ESO, AUI/NRAO and NAOJ. V.M.R. has received support from the project RYC2020-029387-I funded by MCIN/AEI /10.13039/501100011033, and from the the Consejo Superior de Investigaciones Cient{\'i}ficas (CSIC) and the Centro de Astrobiolog{\'i}a (CAB) through the project 20225AT015 (Proyectos intramurales especiales del CSIC).
 L.C. acknowledges financial support through the Spanish grant PID2019-105552RB-C41 funded by MCIN/AEI/10.13039/501100011033.

\bibliographystyle{aa}
\bibliography{refs}

\appendix

\section{Additional Moment 0 Maps}
\label{app:Mom0}

 This appendix includes the remaining moment 0 maps generated using \texttt{CubeLineMoment}. The maps are provided in temperature units and cover the CMZ of NGC~253.
 
\begin{figure}
  \centering
  \begin{tabular}[b]{@{}p{0.48\textwidth}@{}}
    \centering\includegraphics[width=1.0\linewidth]{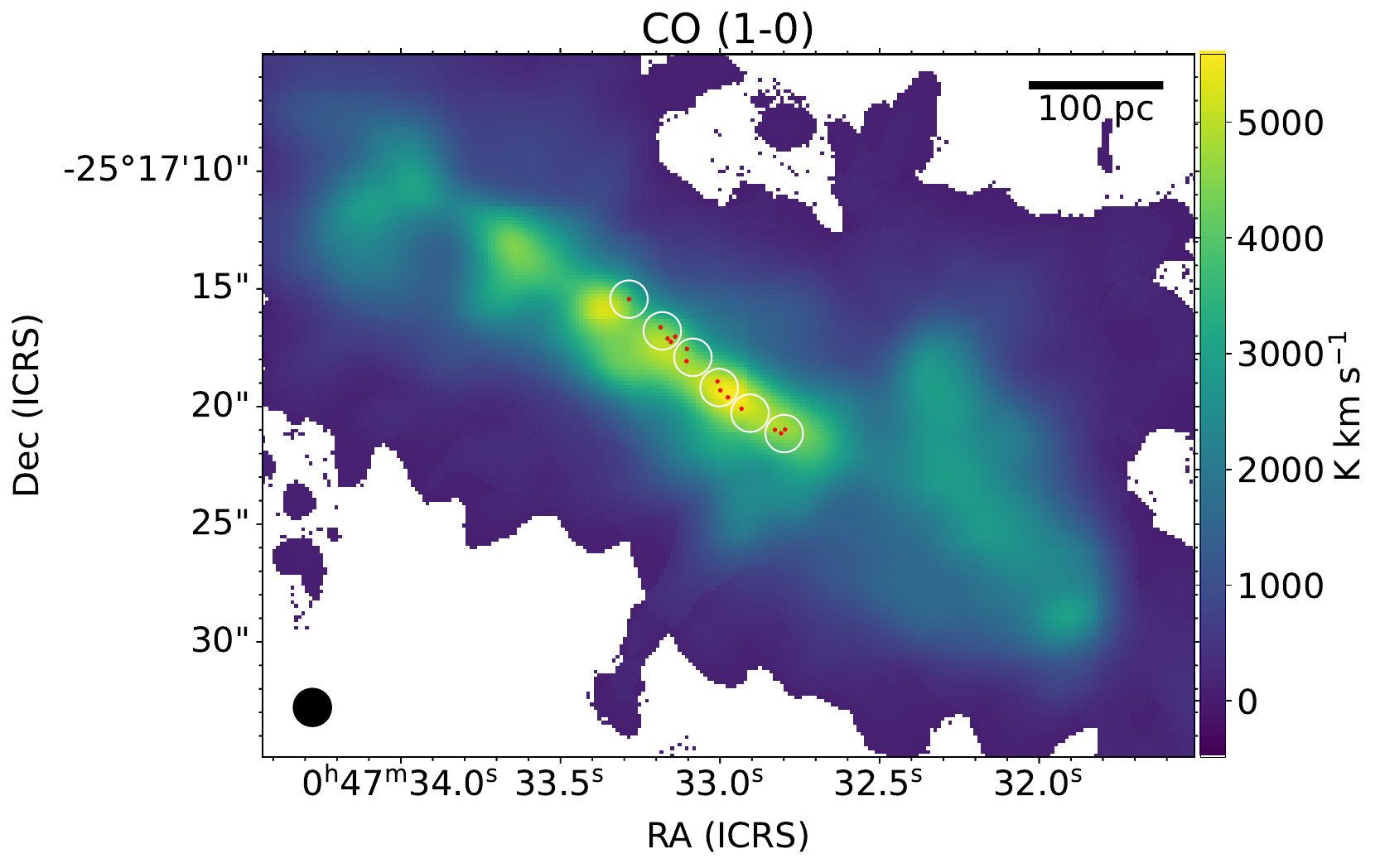} \\
    \centering\small (a) 
  \end{tabular}%
  \quad
\begin{tabular}[b]{@{}p{0.48\textwidth}@{}}
    \centering\includegraphics[width=1.0\linewidth]{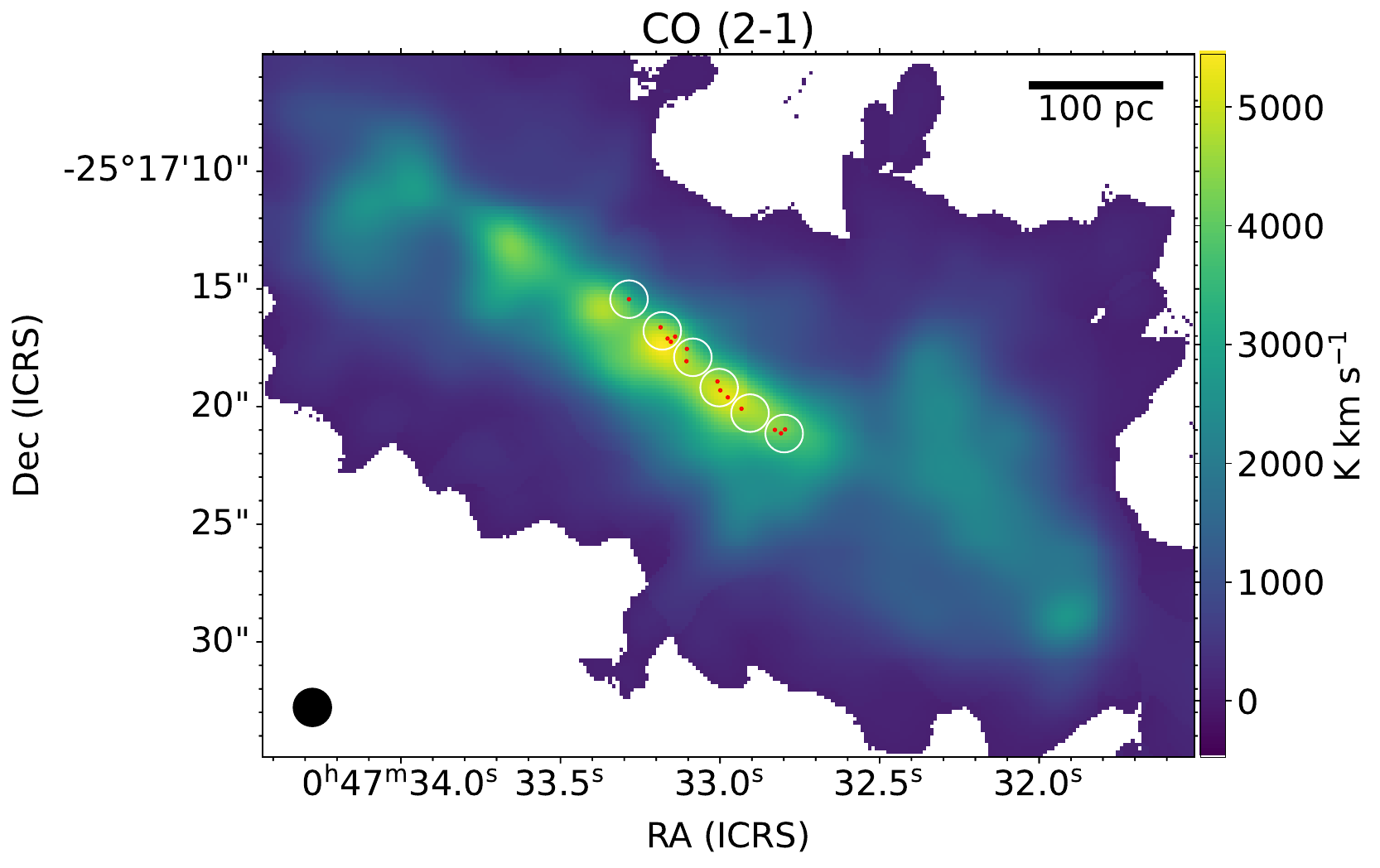} \\
    \centering\small (b)  
  \end{tabular}
  \quad
\begin{tabular}[b]{@{}p{0.48\textwidth}@{}}
    \centering\includegraphics[width=1.0\linewidth]{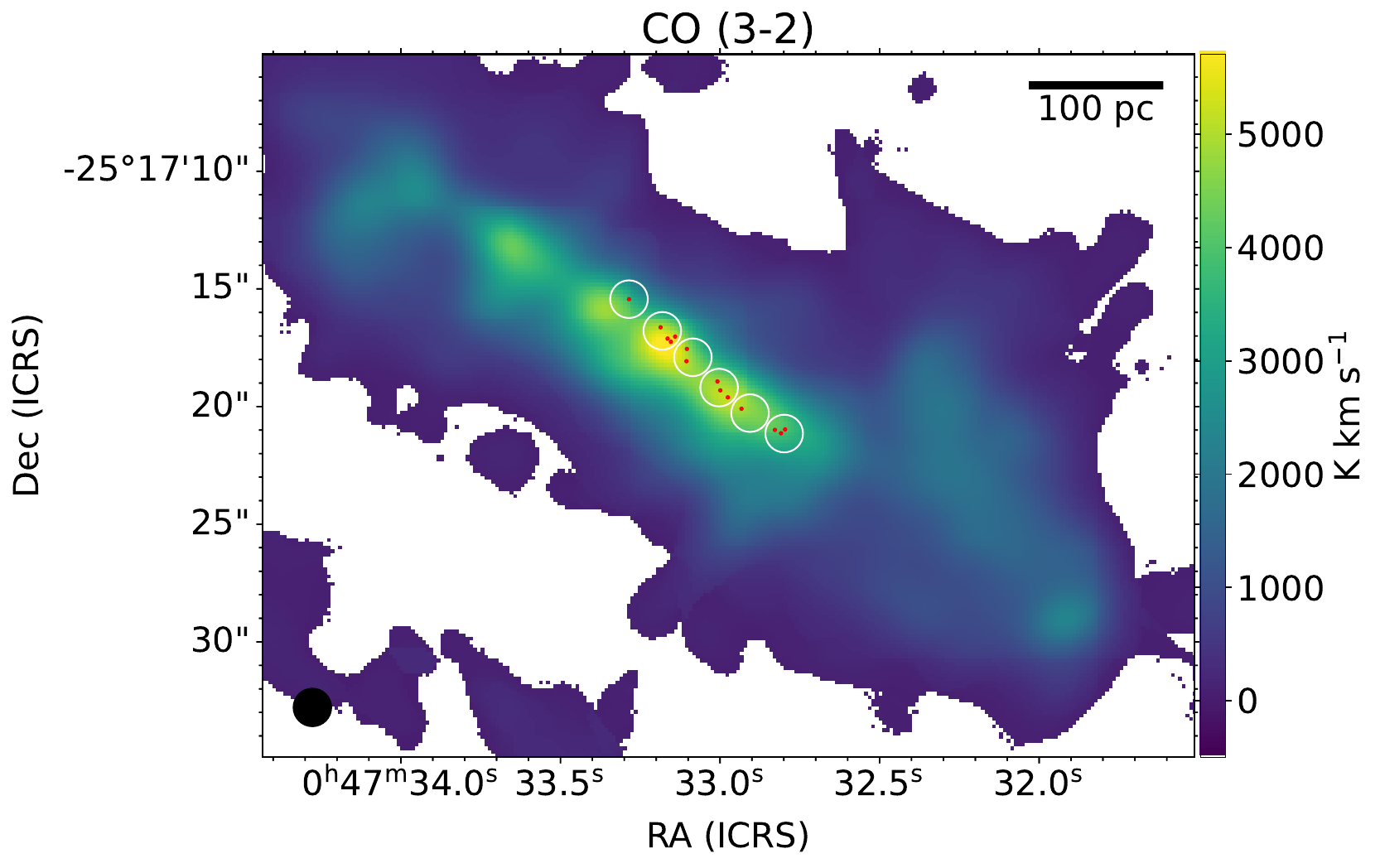} \\
    \centering\small (c)  
  \end{tabular}
  \caption{Velocity-integrated line intensities in [K\,\kms] of CO (1-0), (2-1) and (3-2). Each of the maps shown have been generated using a signal-to-noise cutoff of 3. The studied SSC regions as listed in Tab.~\ref{tab:SSC_locations} are labeled in white texts on the map. The original SSC locations with appropriate beam sizes from \cite{2018Leroy} are shown by the red regions. The ALCHEMI $1''.6 \times 1''.6$ beam is displayed in the lower-left corner of the map. 
  }
  \label{fig:mom0_CO_10_21_32}
\end{figure}

\begin{figure}
  \centering
  \begin{tabular}[b]{@{}p{0.48\textwidth}@{}}
    \centering\includegraphics[width=1.0\linewidth]{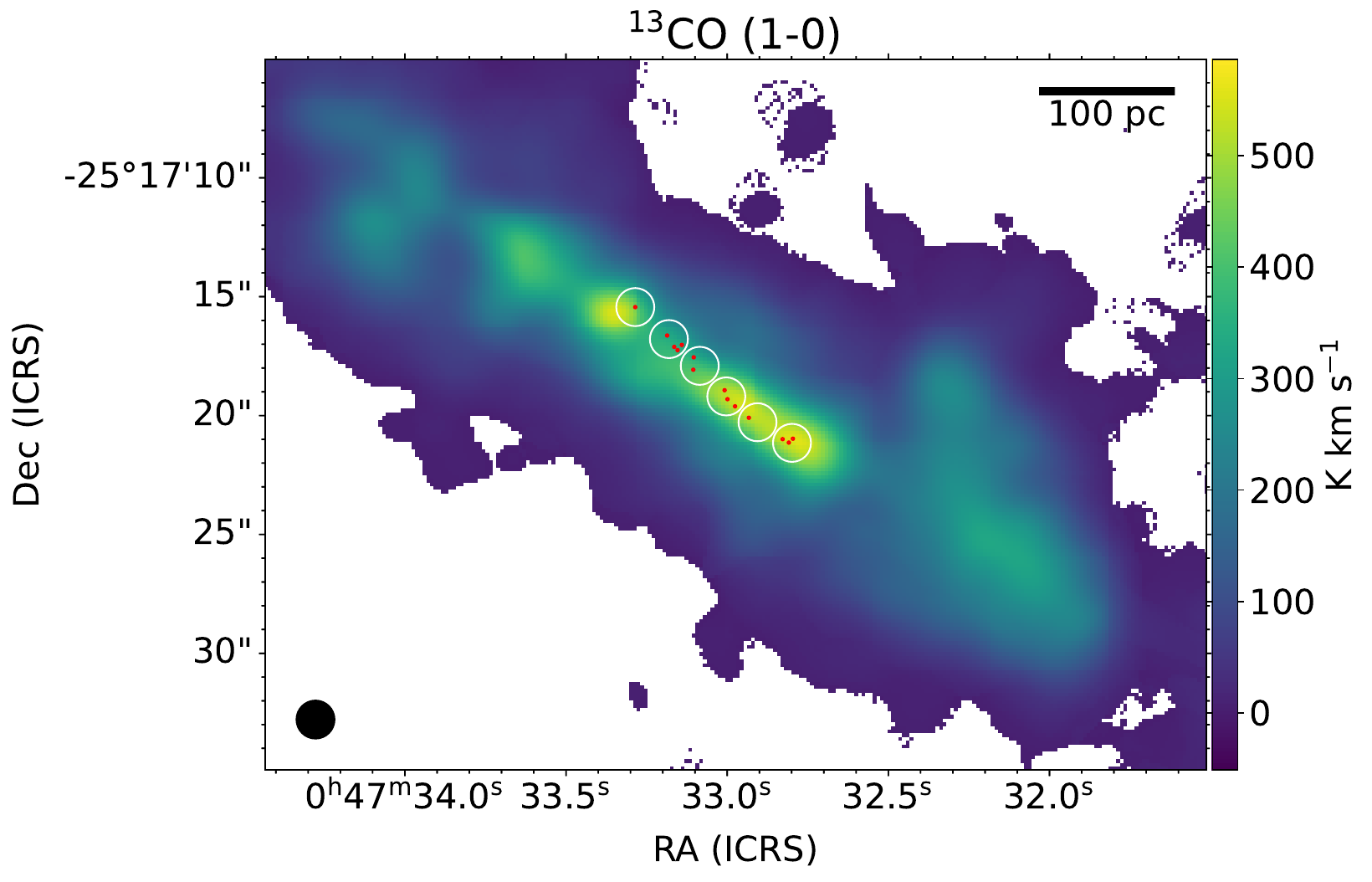} \\
    \centering\small (a) 
  \end{tabular}%
  \quad
\begin{tabular}[b]{@{}p{0.48\textwidth}@{}}
    \centering\includegraphics[width=1.0\linewidth]{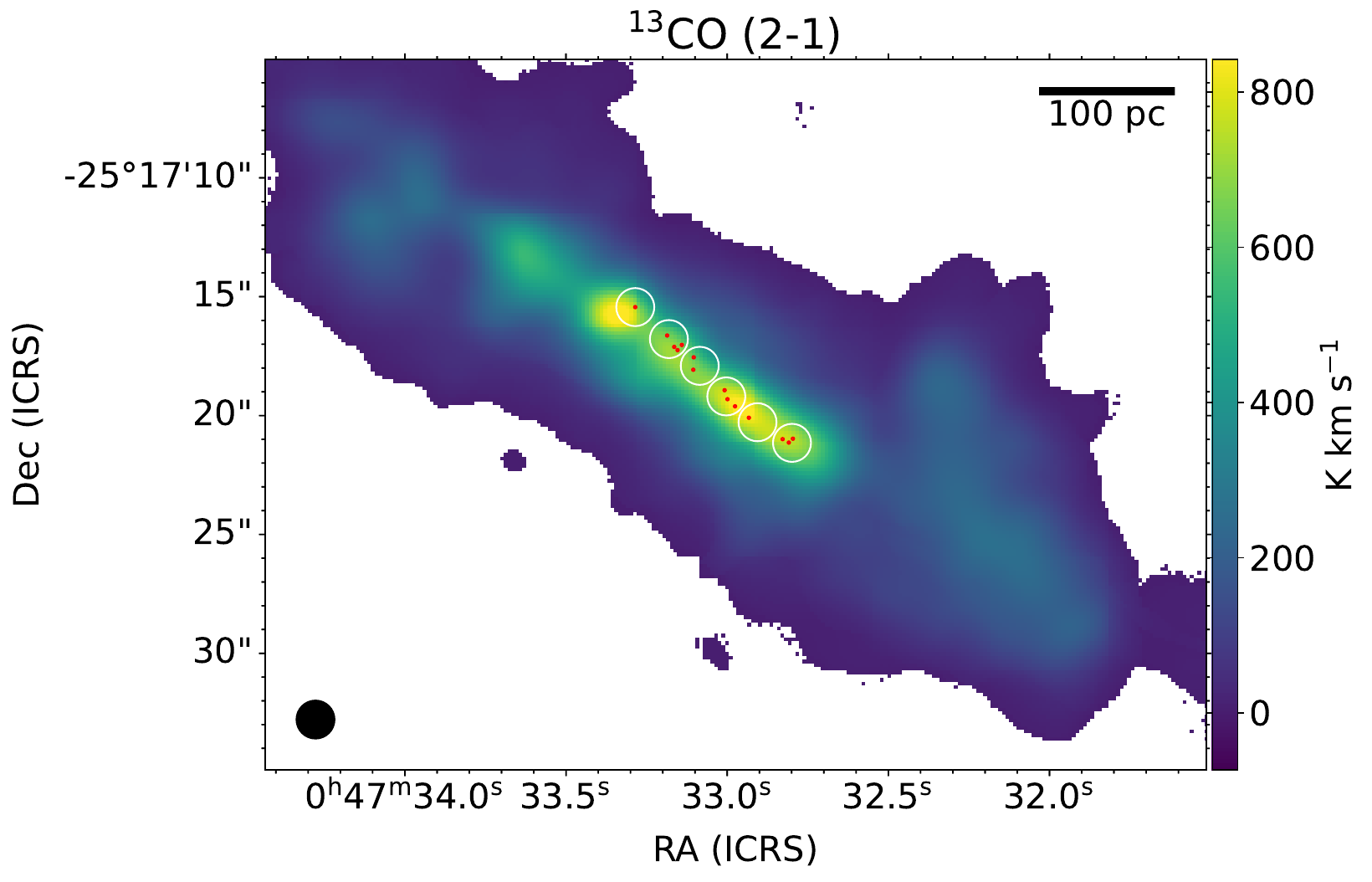} \\
    \centering\small (b)  
  \end{tabular}
  \quad
\begin{tabular}[b]{@{}p{0.48\textwidth}@{}}
    \centering\includegraphics[width=1.0\linewidth]{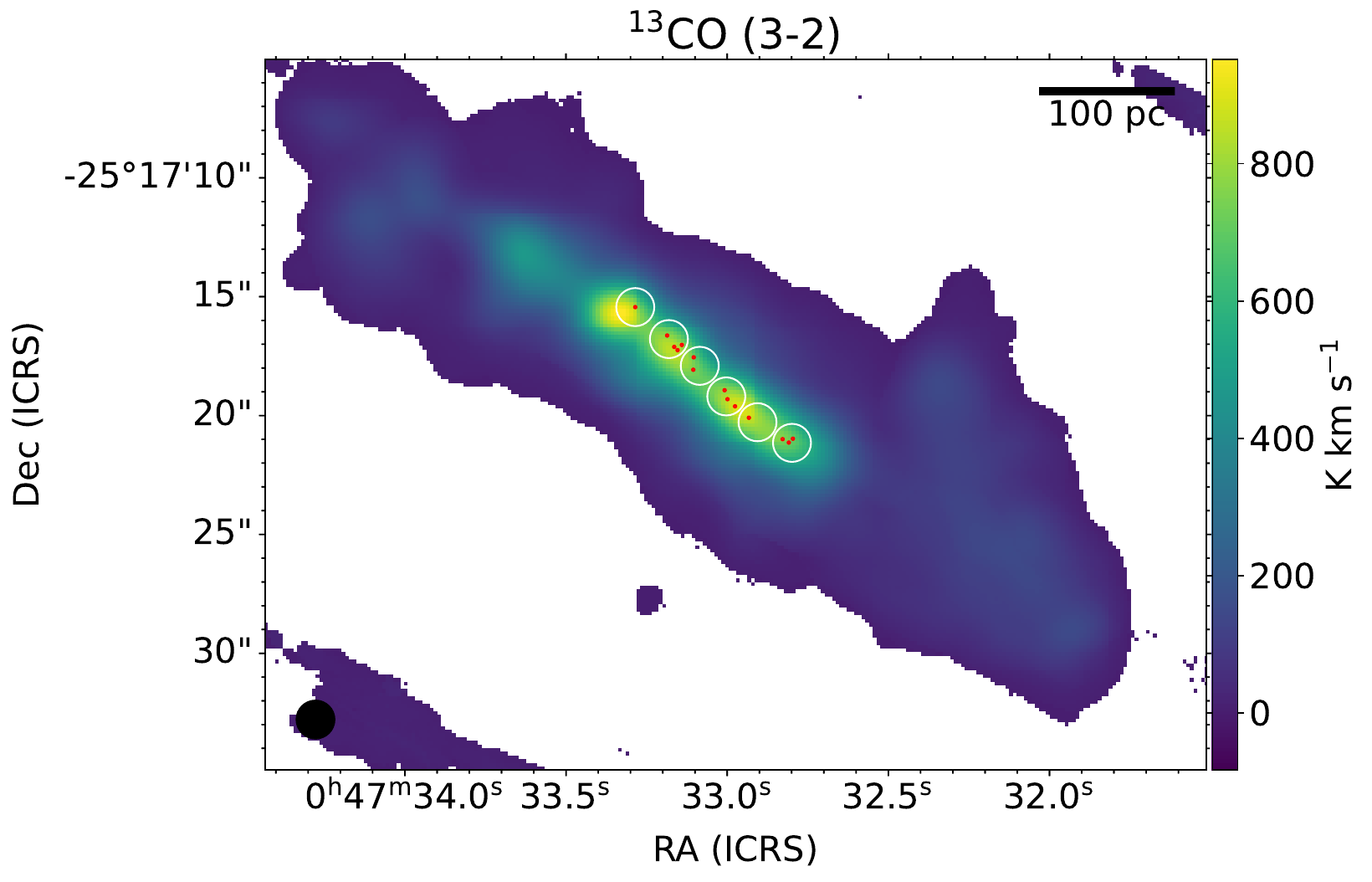} \\
    \centering\small (c)  
  \end{tabular}
  \caption{Velocity-integrated line intensities in [K\,\kms] of \thirteen CO (1-0), (2-1) and (3-2). Each of the maps shown have been generated using a signal-to-noise cutoff of 3. The studied SSC regions as listed in Tab.~\ref{tab:SSC_locations} are labeled in white texts on the map. The original SSC locations with appropriate beam sizes from \cite{2018Leroy} are shown by the red regions. The ALCHEMI $1''.6 \times 1''.6$ beam is displayed in the lower-left corner of the map. 
  }
  \label{fig:mom0_13CO_10_21_32}
\end{figure}

\begin{figure}
  \centering
  \begin{tabular}[b]{@{}p{0.48\textwidth}@{}}
    \centering\includegraphics[width=1.0\linewidth]{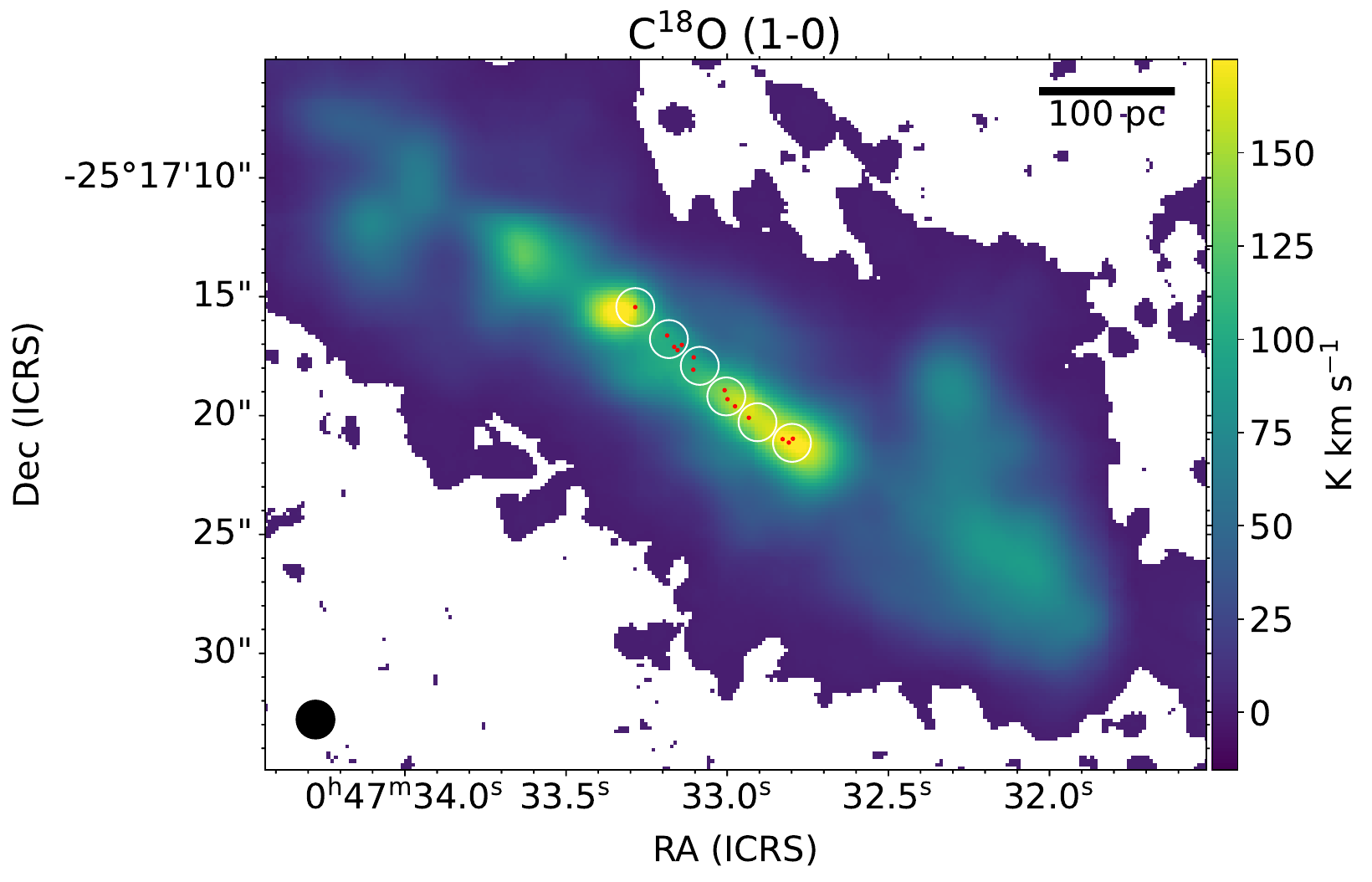} \\
    \centering\small (a) 
  \end{tabular}%
  \quad
\begin{tabular}[b]{@{}p{0.48\textwidth}@{}}
    \centering\includegraphics[width=1.0\linewidth]{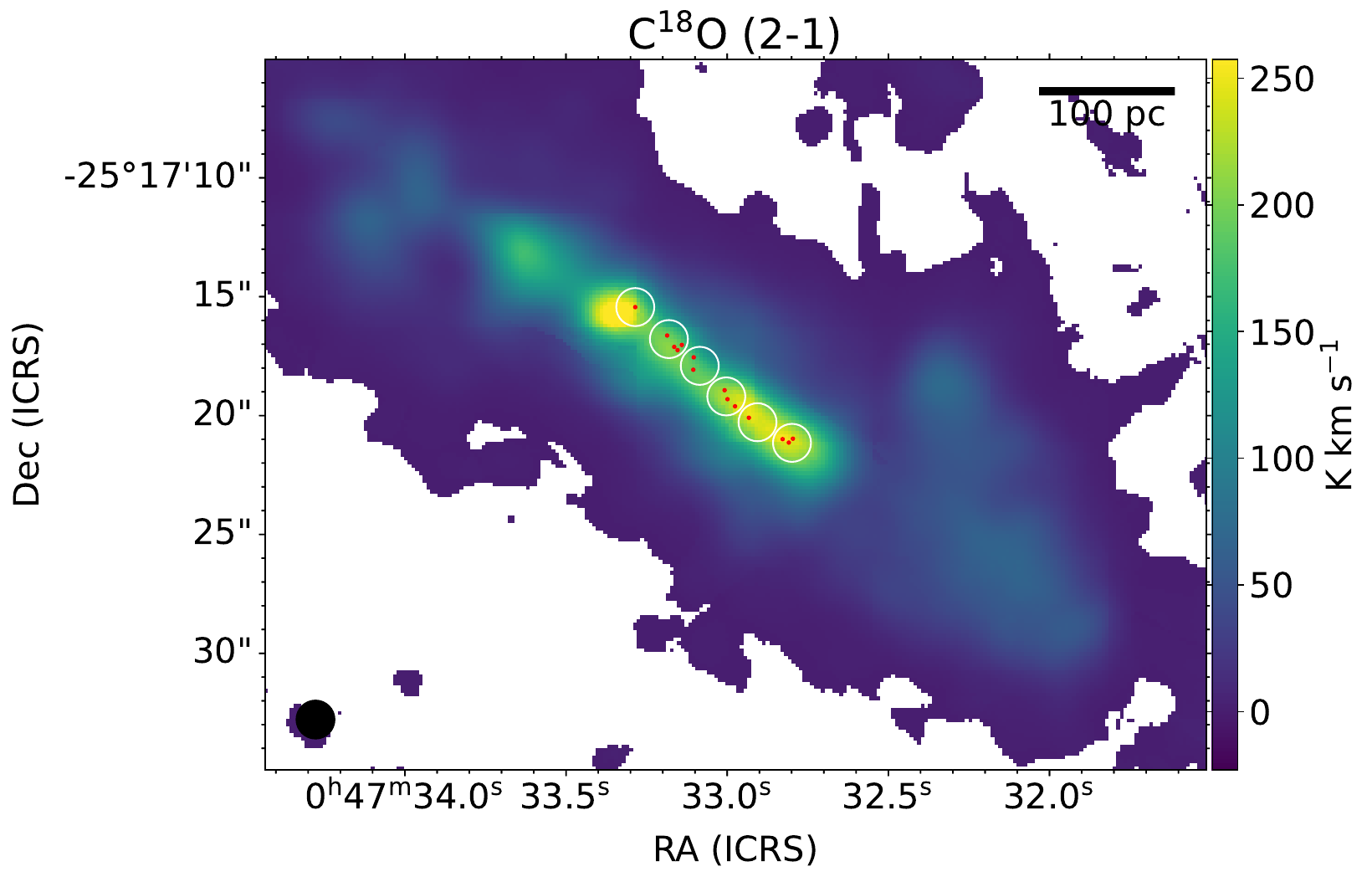} \\
    \centering\small (b)  
  \end{tabular}
  \quad
\begin{tabular}[b]{@{}p{0.48\textwidth}@{}}
    \centering\includegraphics[width=1.0\linewidth]{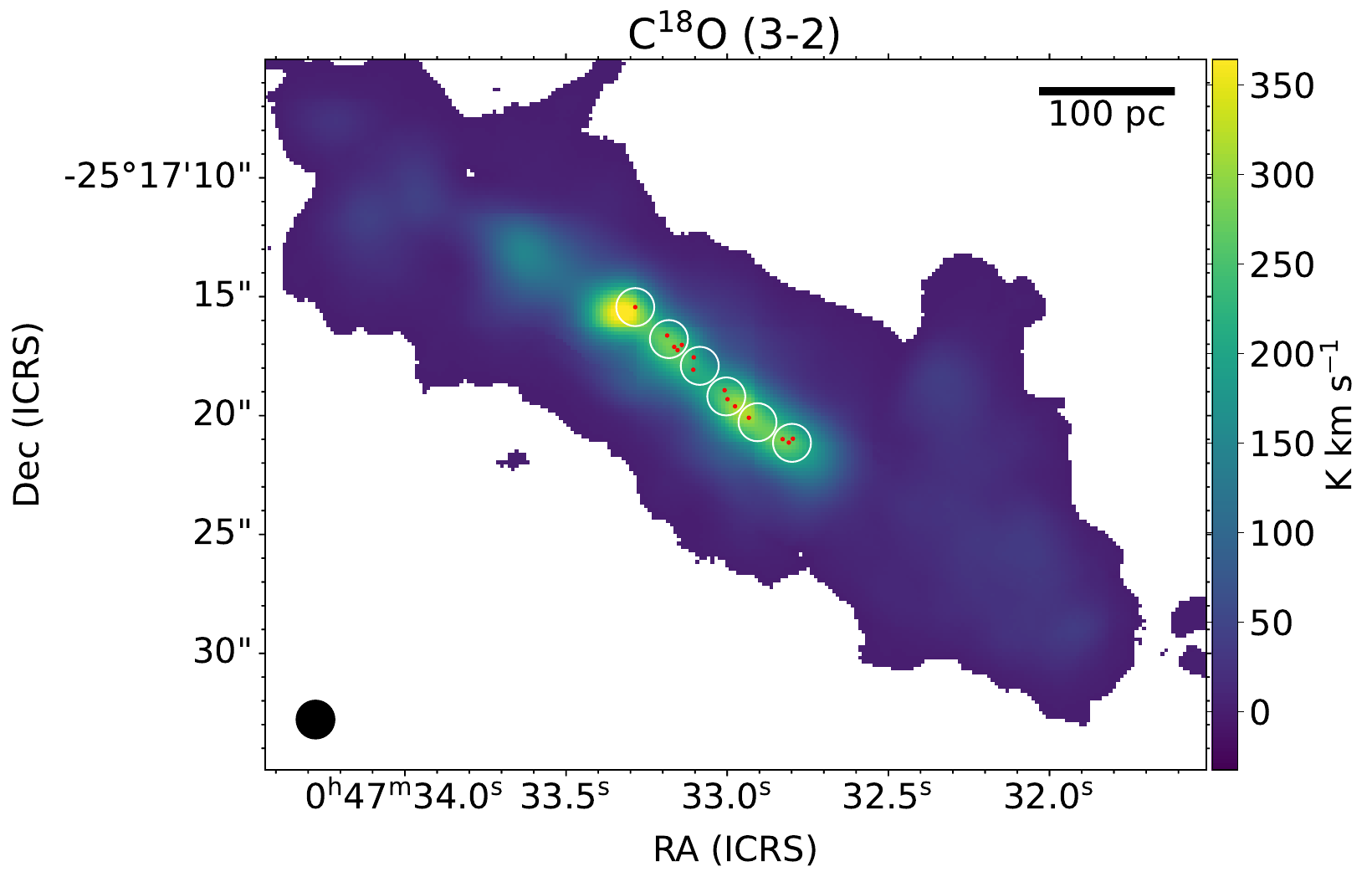} \\
    \centering\small (c)  
  \end{tabular}
  \caption{Velocity-integrated line intensities in [K\,\kms] of C\eighteen O (1-0), (2-1) and (3-2). Each of the maps shown have been generated using a signal-to-noise cutoff of 3. The studied SSC regions as listed in Tab.~\ref{tab:SSC_locations} are labeled in white texts on the map. The original SSC locations with appropriate beam sizes from \cite{2018Leroy} are shown by the red regions. The ALCHEMI $1''.6 \times 1''.6$ beam is displayed in the lower-left corner of the map. 
  }
  \label{fig:mom0_C18O_10_21_32}
\end{figure}

\begin{figure}
  \centering
  \begin{tabular}[b]{@{}p{0.48\textwidth}@{}}
    \centering\includegraphics[width=1.0\linewidth]{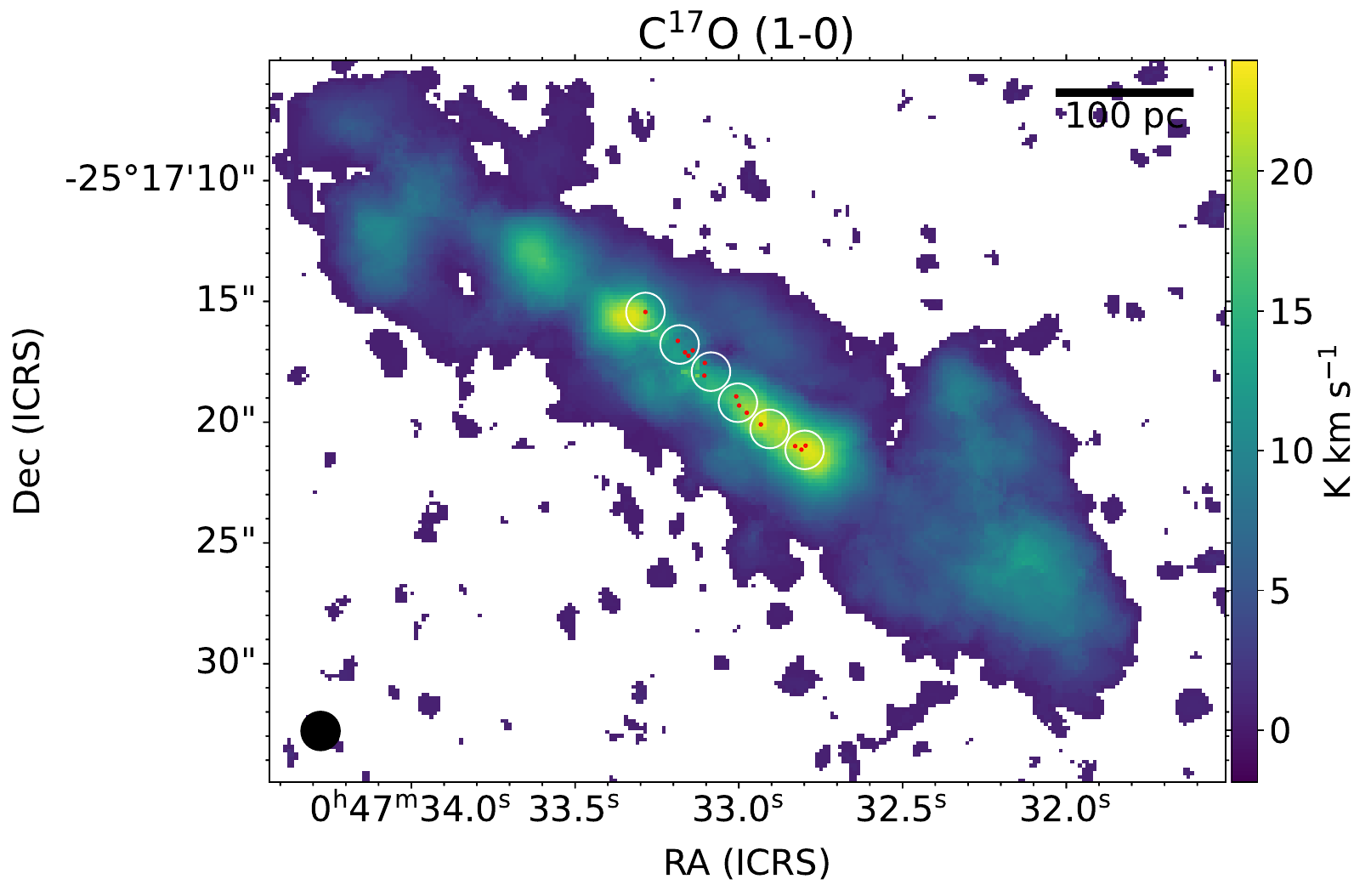} \\
    \centering\small (a) 
  \end{tabular}%
  \quad
\begin{tabular}[b]{@{}p{0.48\textwidth}@{}}
    \centering\includegraphics[width=1.0\linewidth]{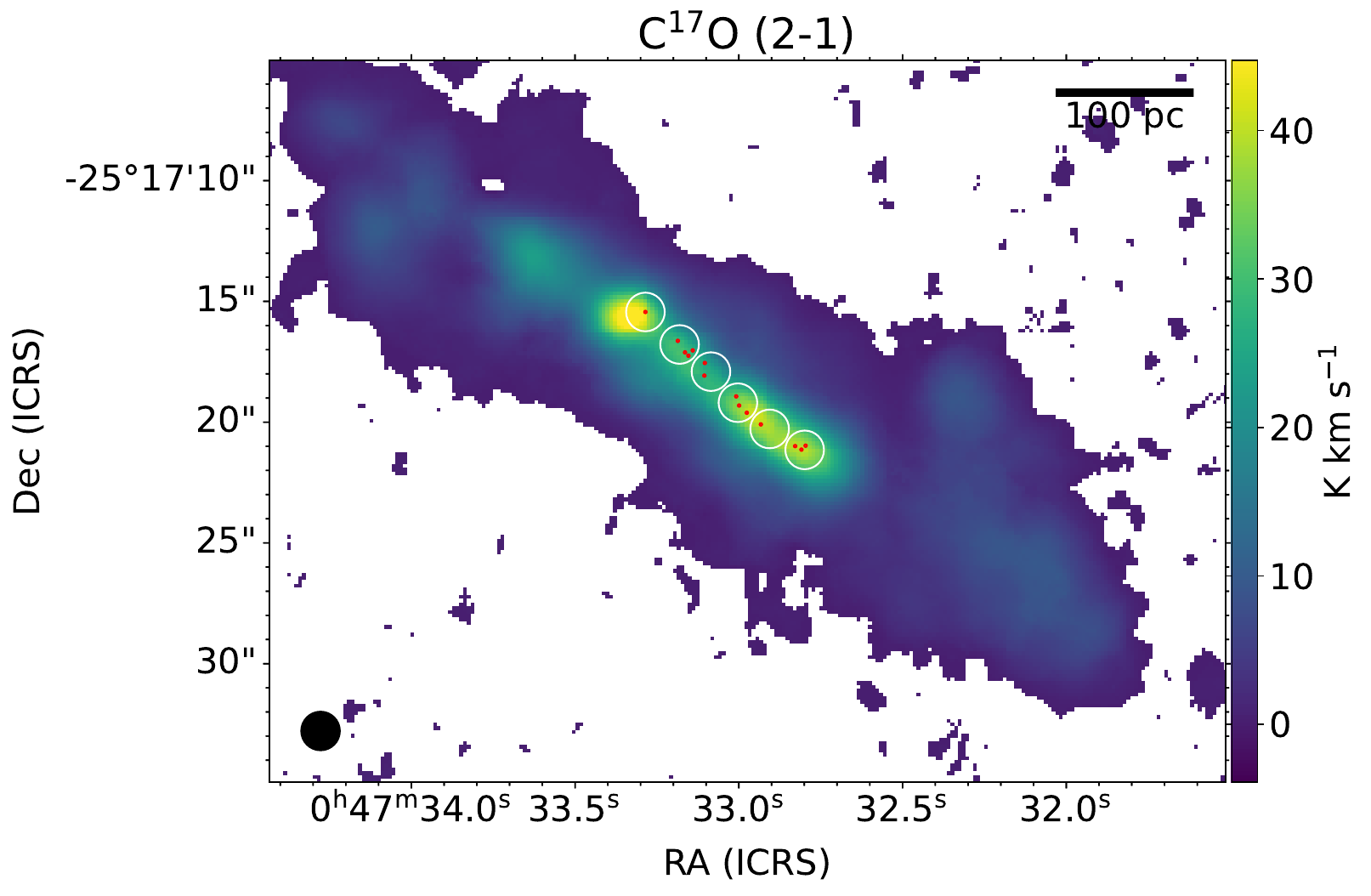} \\
    \centering\small (b)  
  \end{tabular}
  \quad
\begin{tabular}[b]{@{}p{0.48\textwidth}@{}}
    \centering\includegraphics[width=1.0\linewidth]{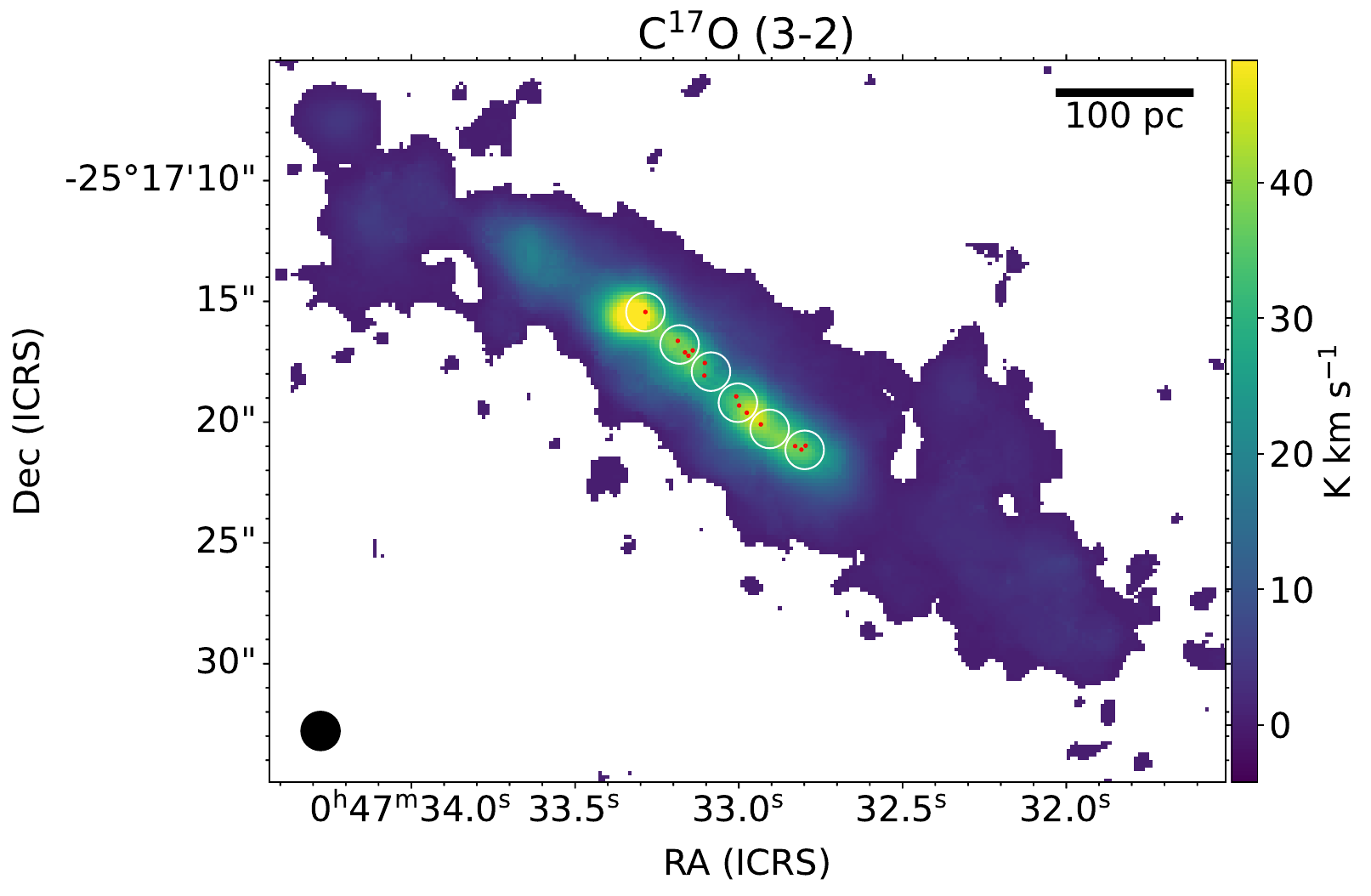} \\
    \centering\small (c)  
  \end{tabular}
  \caption{Velocity-integrated line intensities in [K\,\kms] of C\seventeen O (1-0), (2-1) and (3-2). Each of the maps shown have been generated using a signal-to-noise cutoff of 3. The studied SSC regions as listed in Tab.~\ref{tab:SSC_locations} are labeled in white texts on the map. The original SSC locations with appropriate beam sizes from \cite{2018Leroy} are shown by the red regions. The ALCHEMI $1''.6 \times 1''.6$ beam is displayed in the lower-left corner of the map. 
  }
  \label{fig:mom0_C17O_10_21_32}
\end{figure}

\begin{figure}
  \centering
  \begin{tabular}[b]{@{}p{0.48\textwidth}@{}}
    \centering\includegraphics[width=1.0\linewidth]{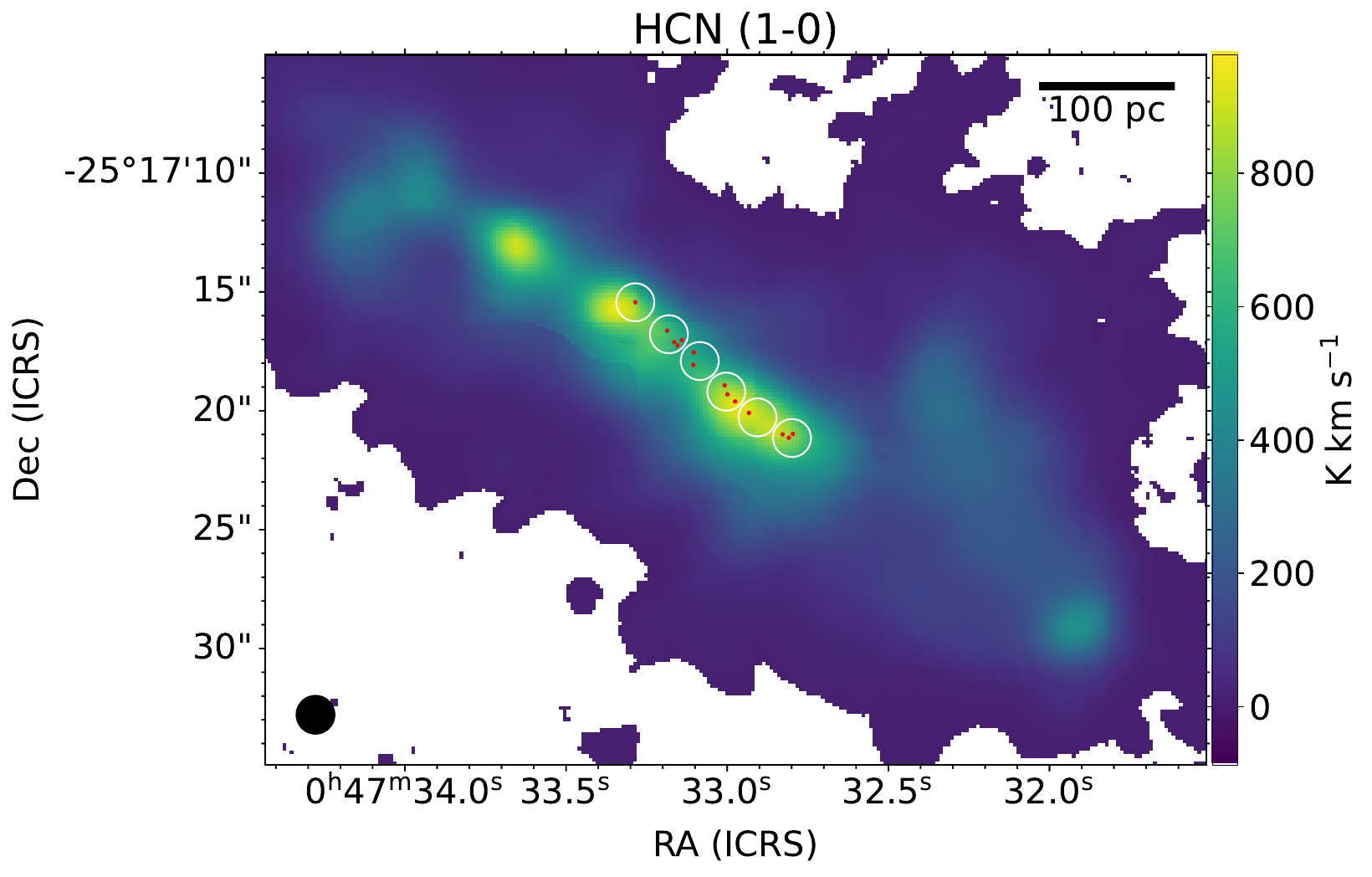} \\
    \centering\small (a) 
  \end{tabular}%
  \quad
\begin{tabular}[b]{@{}p{0.48\textwidth}@{}}
    \centering\includegraphics[width=1.0\linewidth]{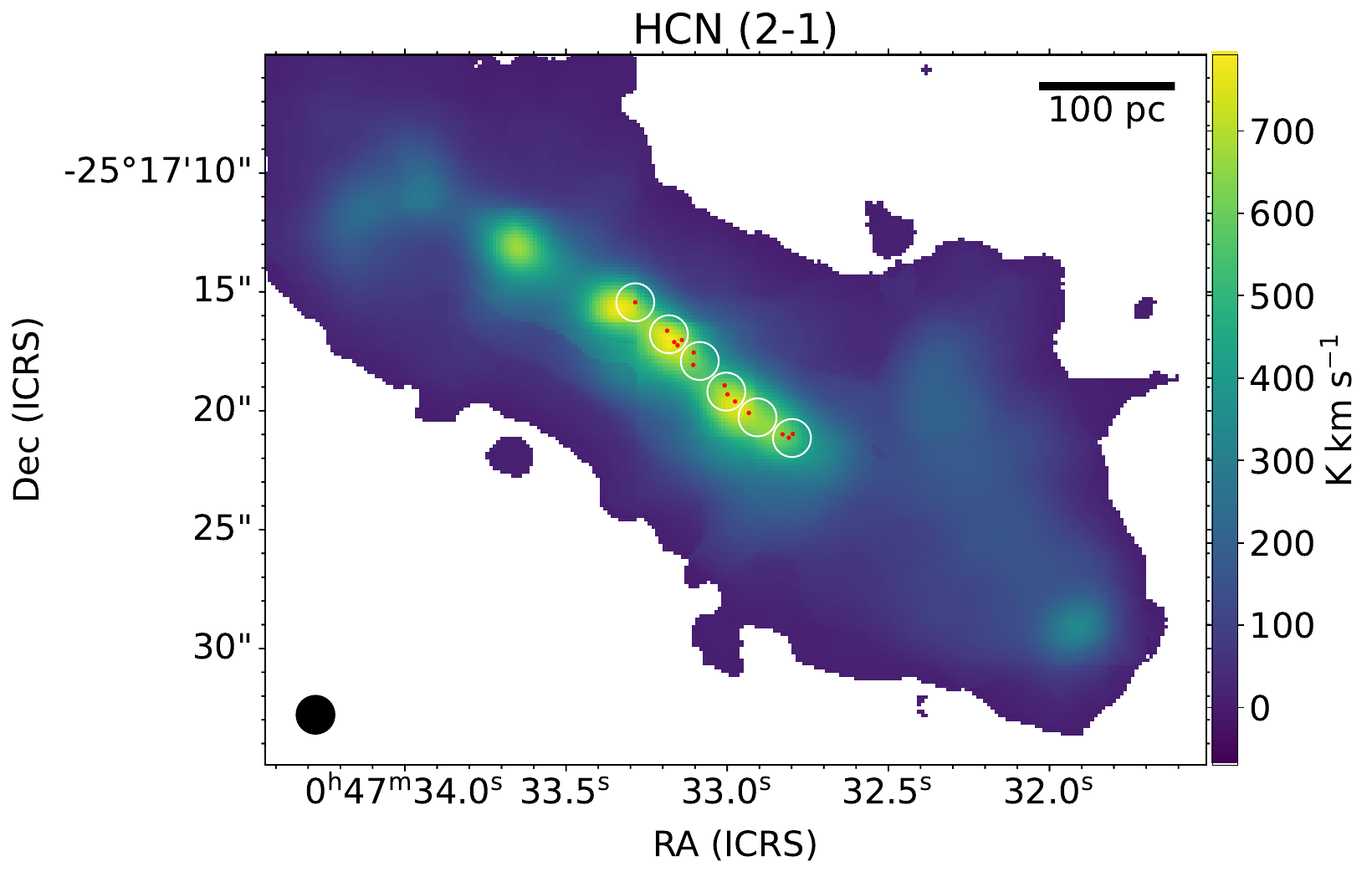} \\
    \centering\small (b)  
  \end{tabular}
  \quad
\begin{tabular}[b]{@{}p{0.48\textwidth}@{}}
    \centering\includegraphics[width=1.0\linewidth]{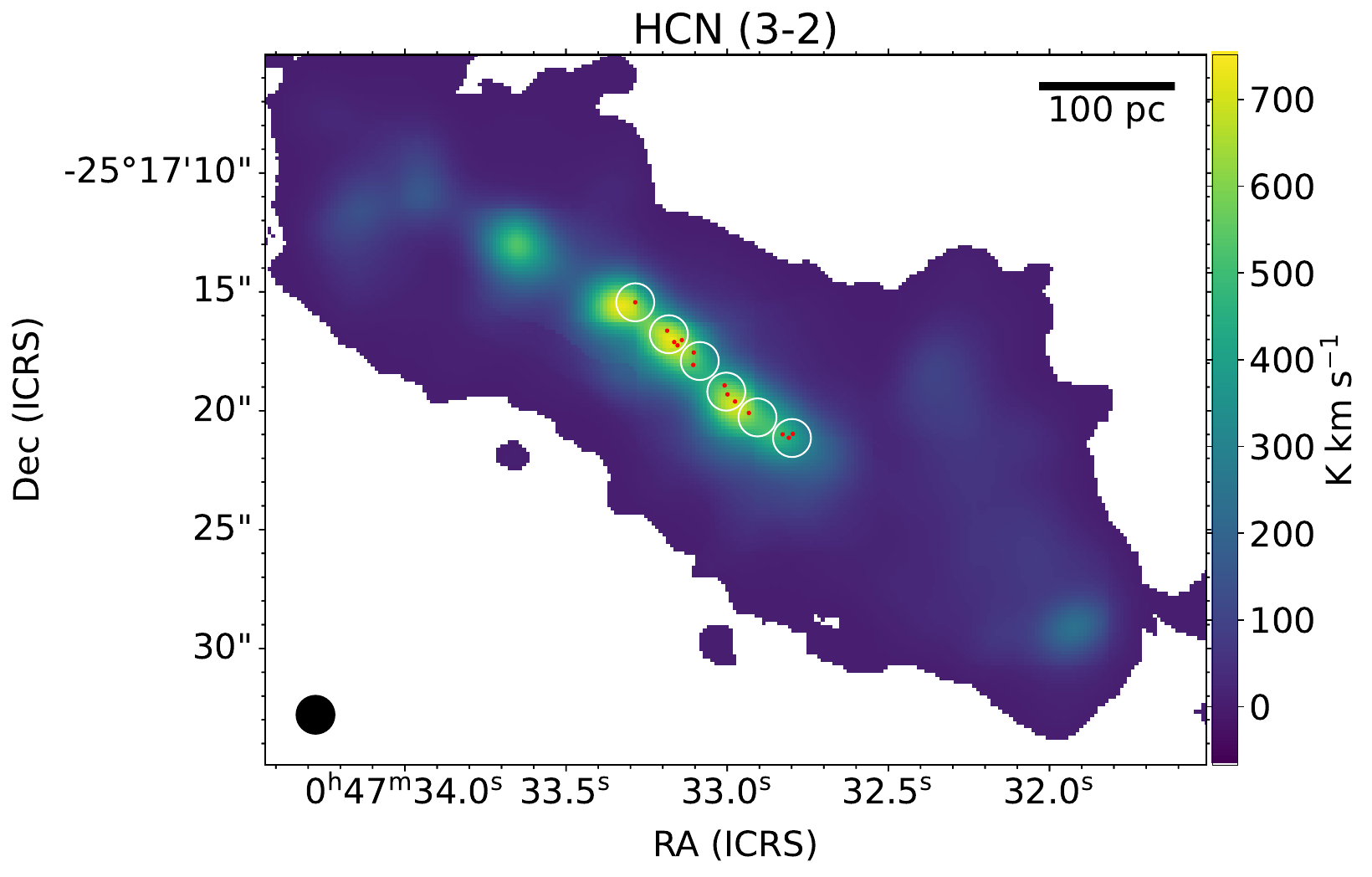} \\
    \centering\small (c)  
  \end{tabular}
  \caption{Velocity-integrated line intensities in [K\,\kms] of HCN (1-0), (2-1) and (3-2). Each of the maps shown have been generated using a signal-to-noise cutoff of 3. The studied SSC regions as listed in Tab.~\ref{tab:SSC_locations} are labeled in white texts on the map. The original SSC locations with appropriate beam sizes from \cite{2018Leroy} are shown by the red regions. The ALCHEMI $1''.6 \times 1''.6$ beam is displayed in the lower-left corner of the map. 
  }
  \label{fig:mom0_HCN_10_21_32}
\end{figure}

\begin{figure}
  \centering
  \begin{tabular}[b]{@{}p{0.48\textwidth}@{}}
    \centering\includegraphics[width=1.0\linewidth]{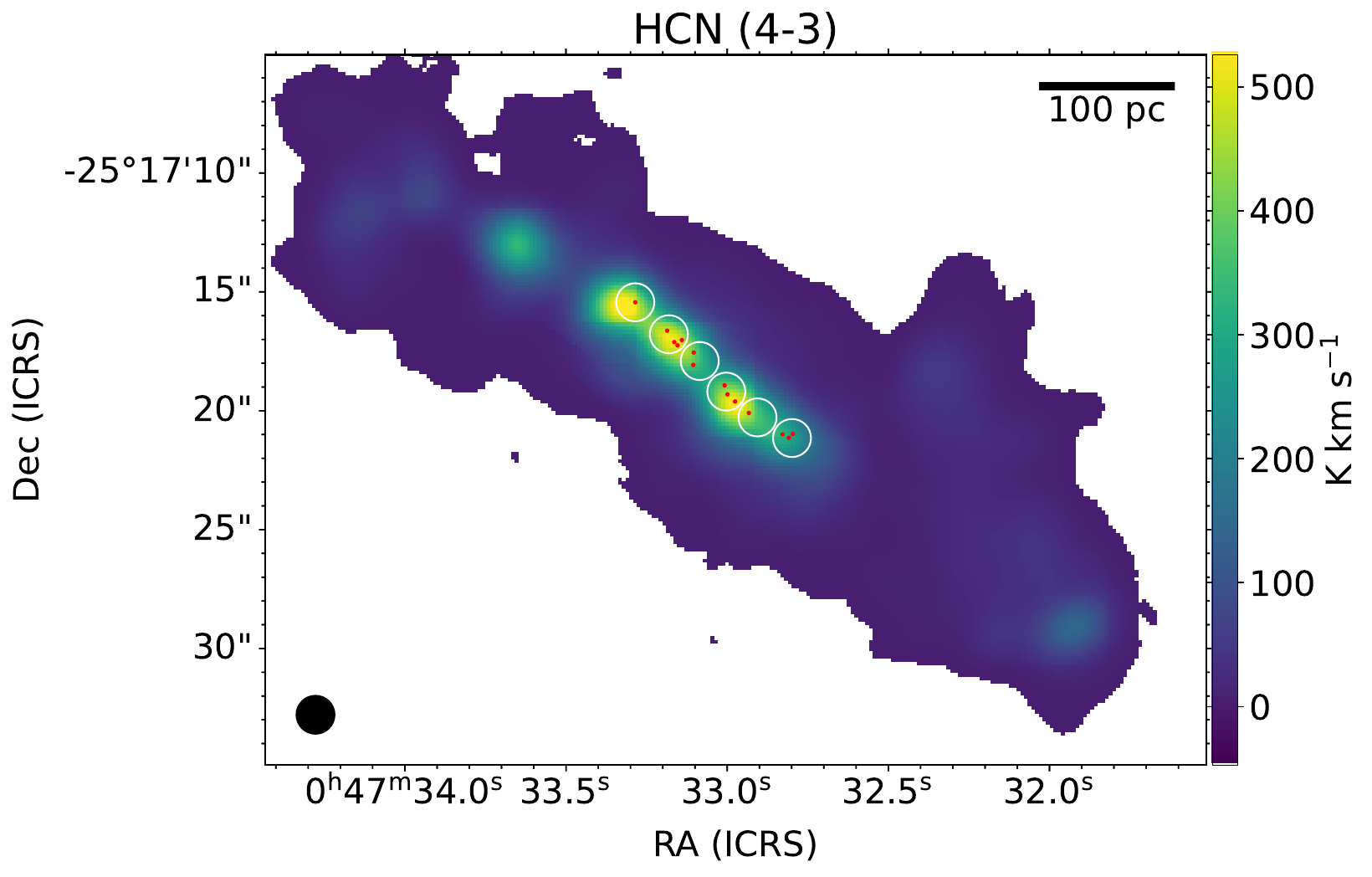} \\
    \centering\small (a) 
  \end{tabular}%
  \quad
\begin{tabular}[b]{@{}p{0.48\textwidth}@{}}
    \centering\includegraphics[width=1.0\linewidth]{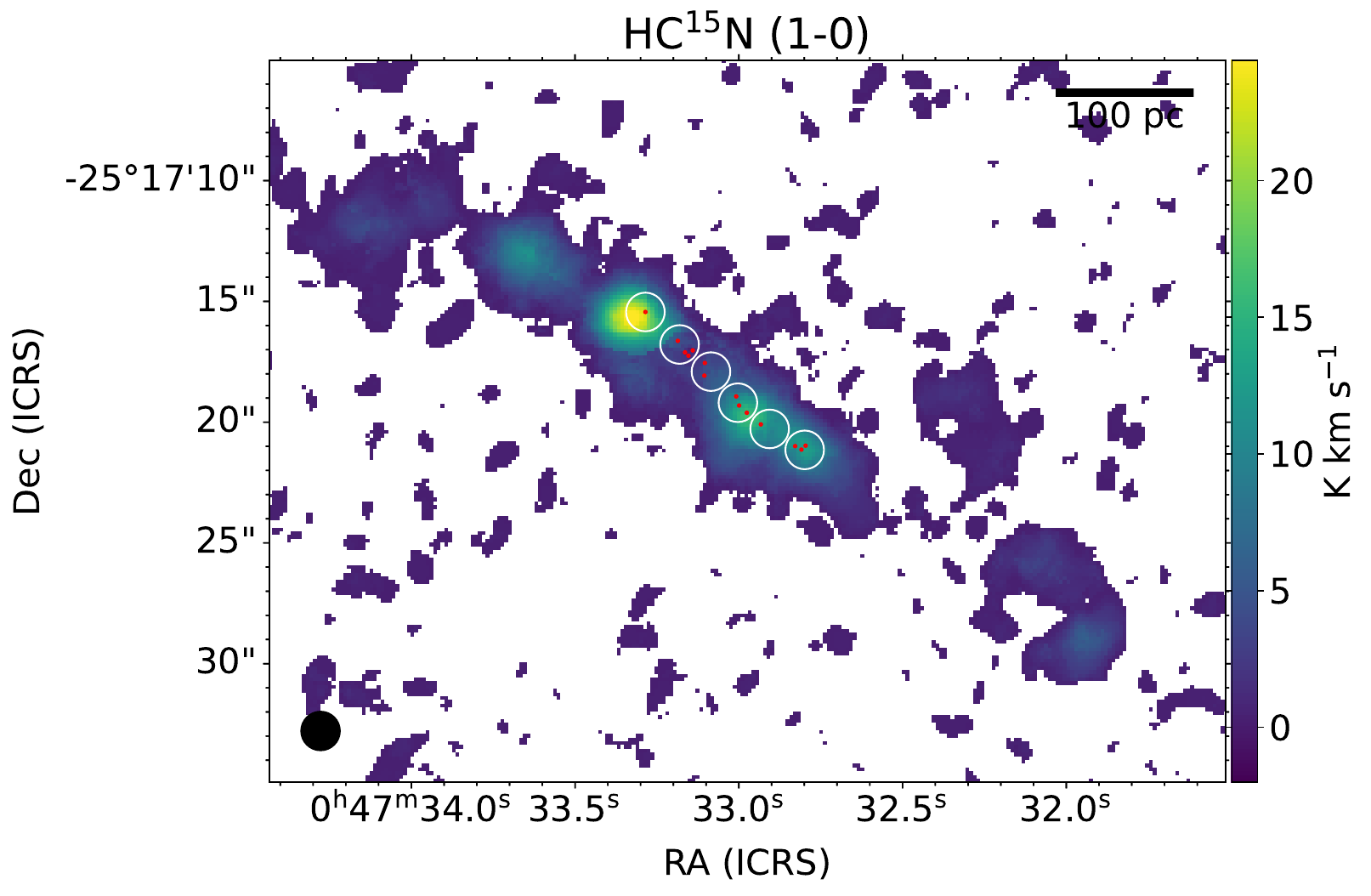} \\
    \centering\small (b)  
  \end{tabular}
  \quad
\begin{tabular}[b]{@{}p{0.48\textwidth}@{}}
    \centering\includegraphics[width=1.0\linewidth]{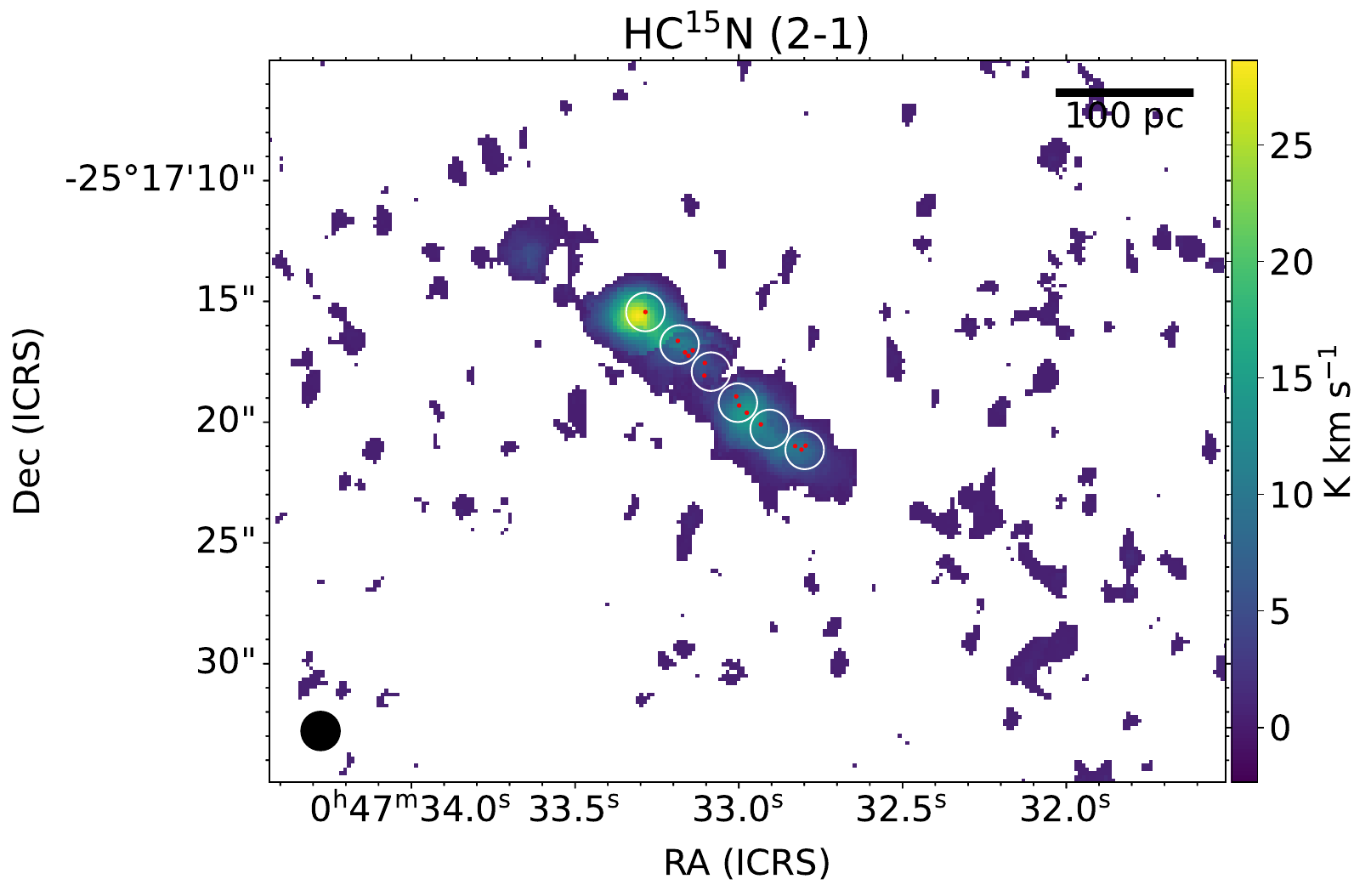} \\
    \centering\small (c) 
  \end{tabular}
  \caption{Velocity-integrated line intensities in [K\,\kms] of HCN (4-3), HC\fifteen N(1-0) and (2-1). Each of the maps shown have been generated using a signal-to-noise cutoff of 3. The studied SSC regions as listed in Tab.~\ref{tab:SSC_locations} are labeled in white texts on the map. The original SSC locations with appropriate beam sizes from \cite{2018Leroy} are shown by the red regions. The ALCHEMI $1''.6 \times 1''.6$ beam is displayed in the lower-left corner of the map. 
  }
  \label{fig:mom0_HCN_HC15N}
\end{figure}

\begin{figure}
  \centering
  \begin{tabular}[b]{@{}p{0.48\textwidth}@{}}
    \centering\includegraphics[width=1.0\linewidth]{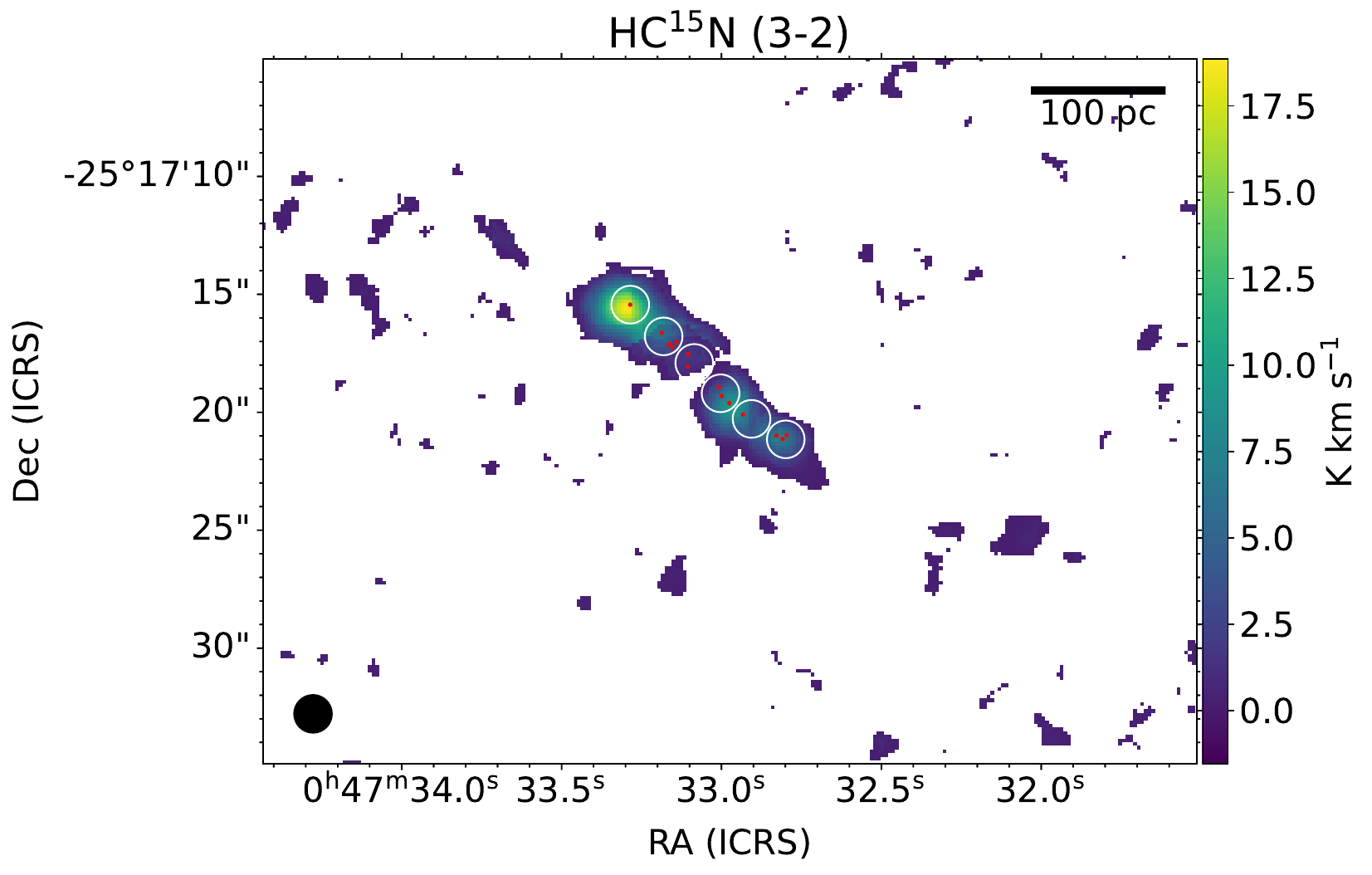} \\
    \centering\small (a) 
  \end{tabular}%
  \quad
\begin{tabular}[b]{@{}p{0.48\textwidth}@{}}
    \centering\includegraphics[width=1.0\linewidth]{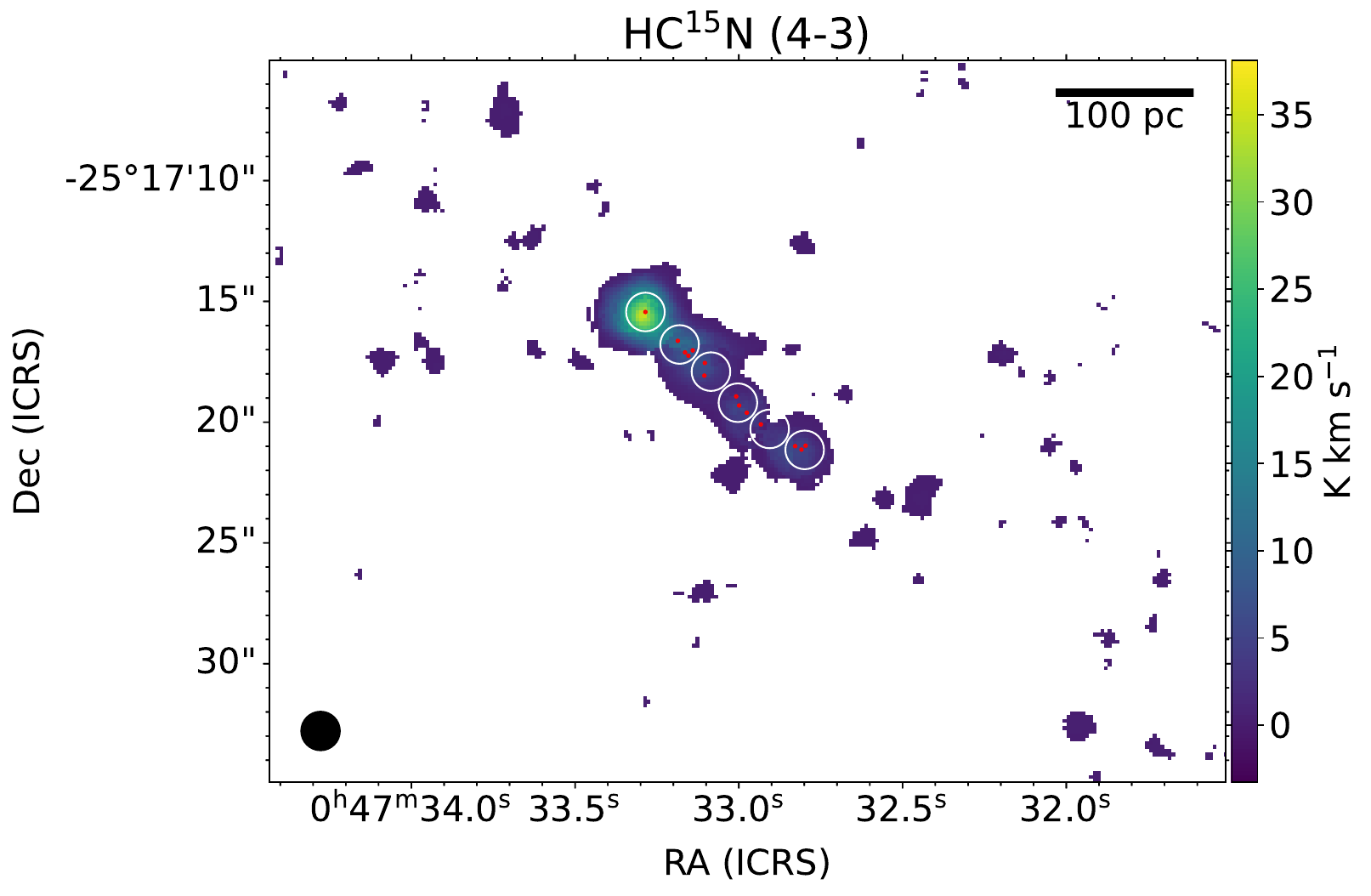} \\
    \centering\small (b)  
  \end{tabular}
  \quad
\begin{tabular}[b]{@{}p{0.48\textwidth}@{}}
    \centering\includegraphics[width=1.0\linewidth]{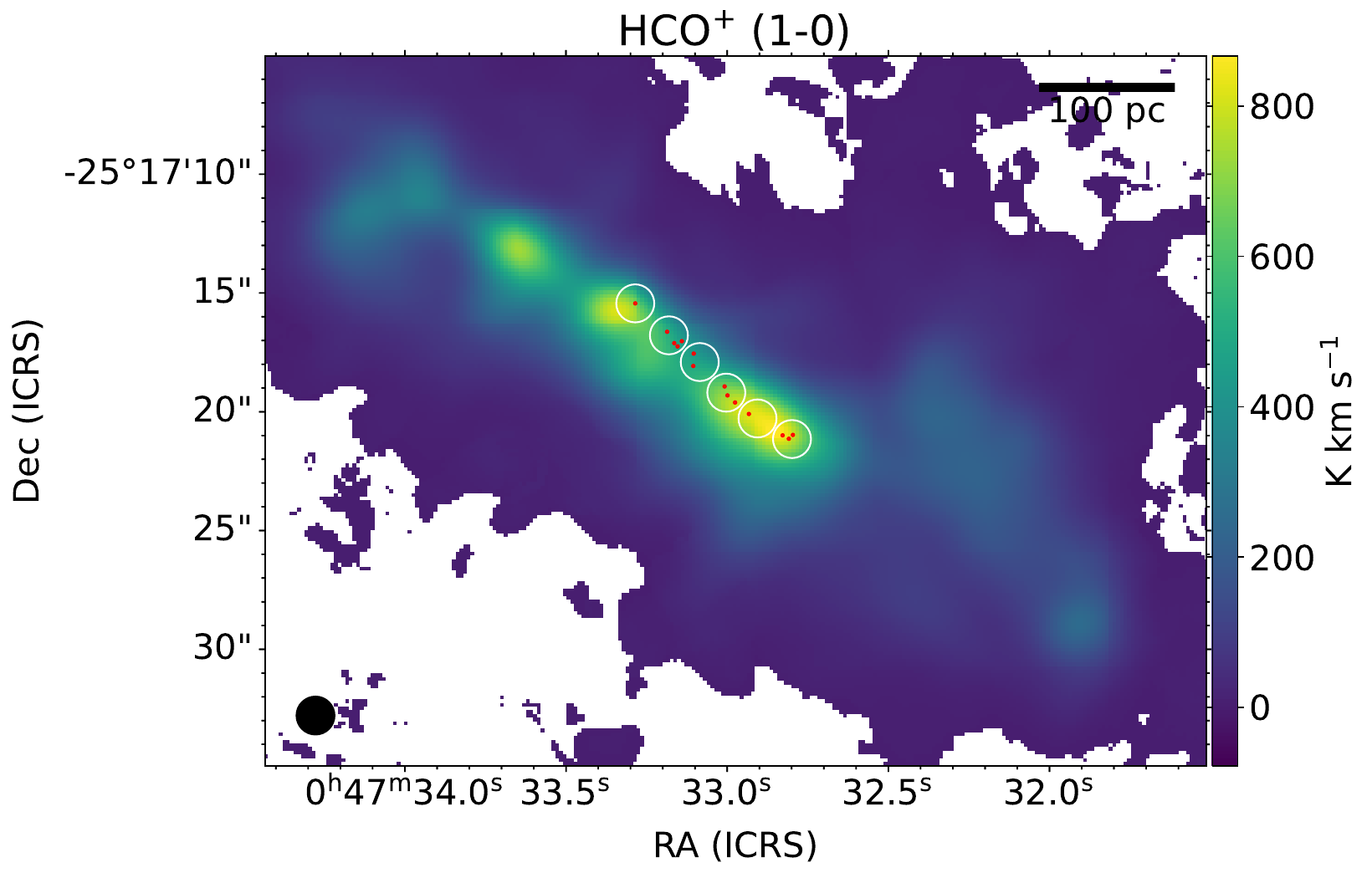} \\
    \centering\small (c)  
  \end{tabular}
  \caption{Velocity-integrated line intensities in [K\,\kms] of HC\fifteen N (3-2), (4-3) and HCO\plus\ (1-0). Each of the maps shown have been generated using a signal-to-noise cutoff of 3. The studied SSC regions as listed in Tab.~\ref{tab:SSC_locations} are labeled in white texts on the map. The original SSC locations with appropriate beam sizes from \cite{2018Leroy} are shown by the red regions. The ALCHEMI $1''.6 \times 1''.6$ beam is displayed in the lower-left corner of the map. 
  }
  \label{fig:mom0_HC15N_HCOP}
\end{figure}

\begin{figure}
  \centering
  \begin{tabular}[b]{@{}p{0.48\textwidth}@{}}
    \centering\includegraphics[width=1.0\linewidth]{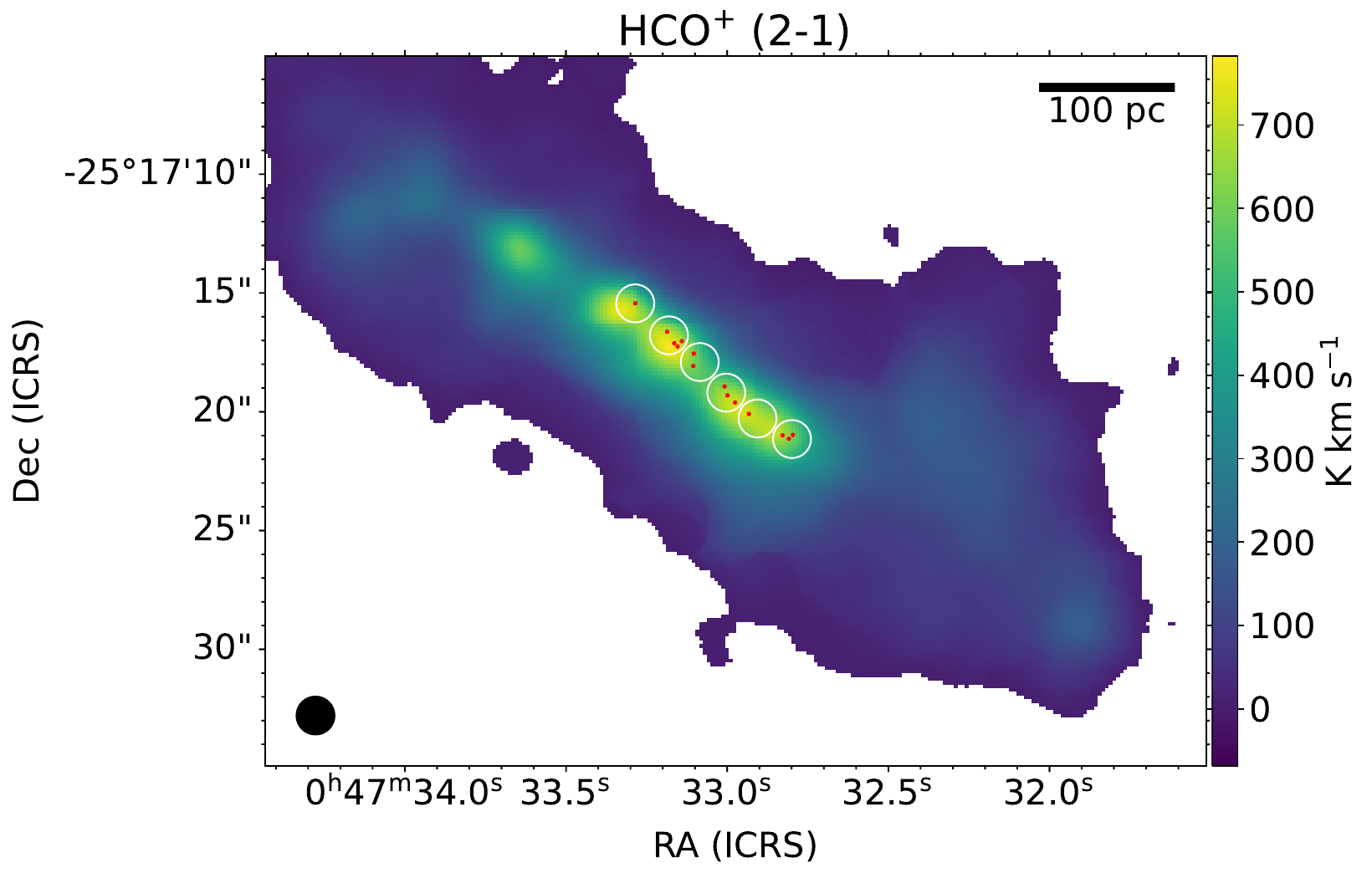} \\
    \centering\small (a) 
  \end{tabular}%
  \quad
\begin{tabular}[b]{@{}p{0.48\textwidth}@{}}
    \centering\includegraphics[width=1.0\linewidth]{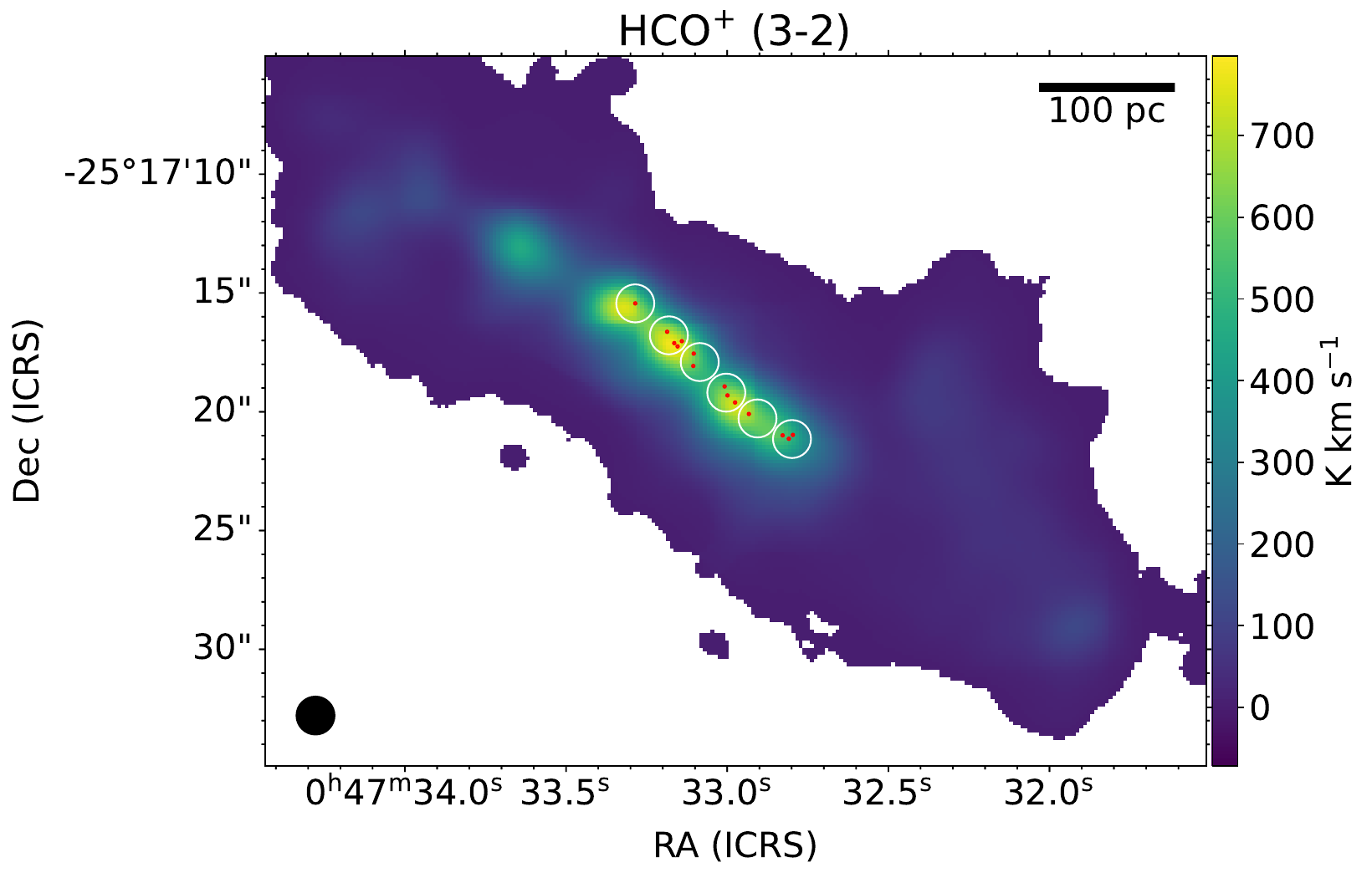} \\
    \centering\small (b)  
  \end{tabular}
  \quad
\begin{tabular}[b]{@{}p{0.48\textwidth}@{}}
    \centering\includegraphics[width=1.0\linewidth]{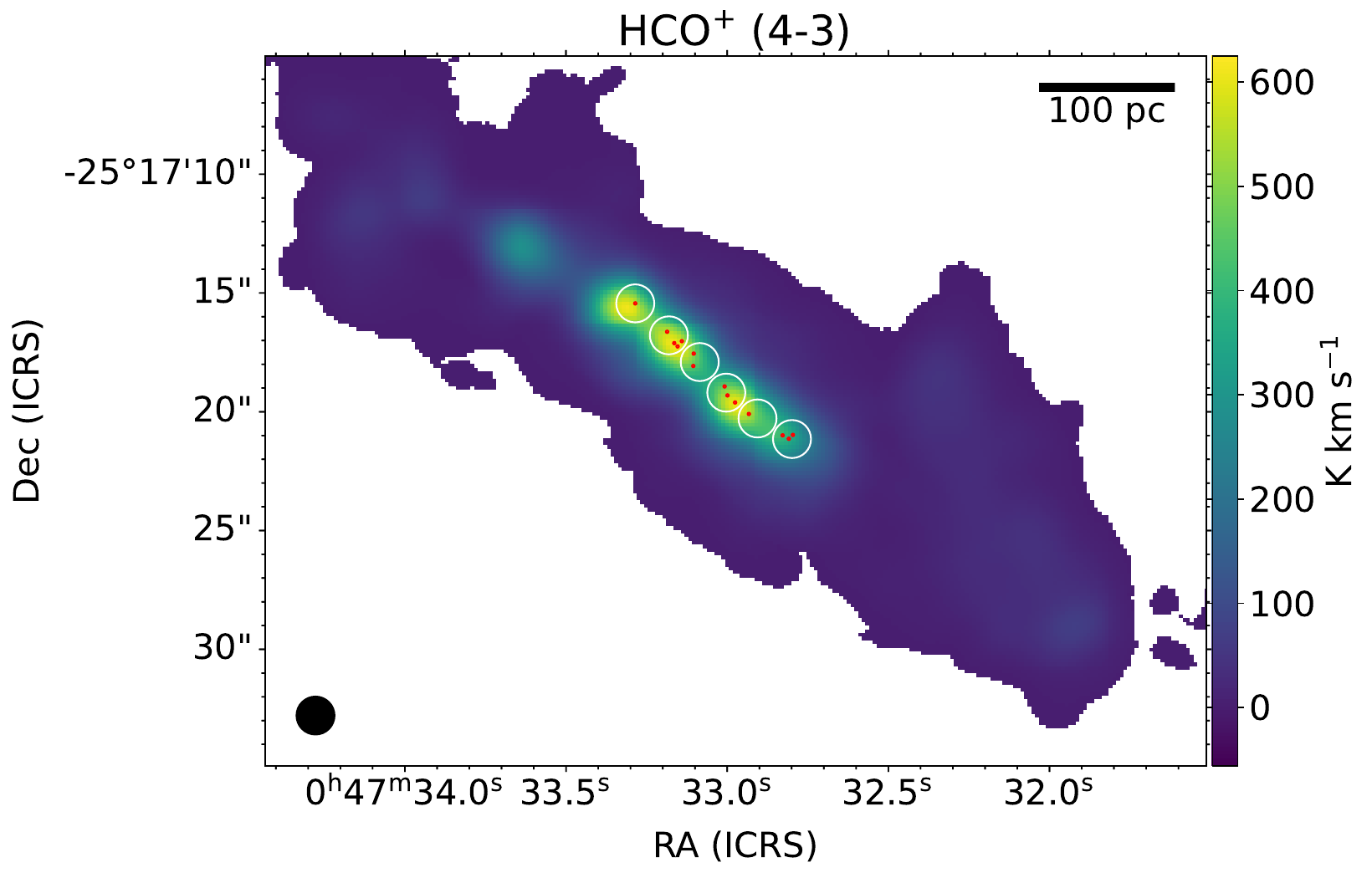} \\
    \centering\small (c)  
  \end{tabular}
  \caption{Velocity-integrated line intensities in [K\,\kms] of HCO\plus\ (1-0), (2-1) and (3-2). Each of the maps shown have been generated using a signal-to-noise cutoff of 3. The studied SSC regions as listed in Tab.~\ref{tab:SSC_locations} are labeled in white texts on the map. The original SSC locations with appropriate beam sizes from \cite{2018Leroy} are shown by the red regions. The ALCHEMI $1''.6 \times 1''.6$ beam is displayed in the lower-left corner of the map. 
  }
  \label{fig:mom0_HCOP}
\end{figure}

\begin{figure}
  \centering
  \begin{tabular}[b]{@{}p{0.48\textwidth}@{}}
    \centering\includegraphics[width=1.0\linewidth]{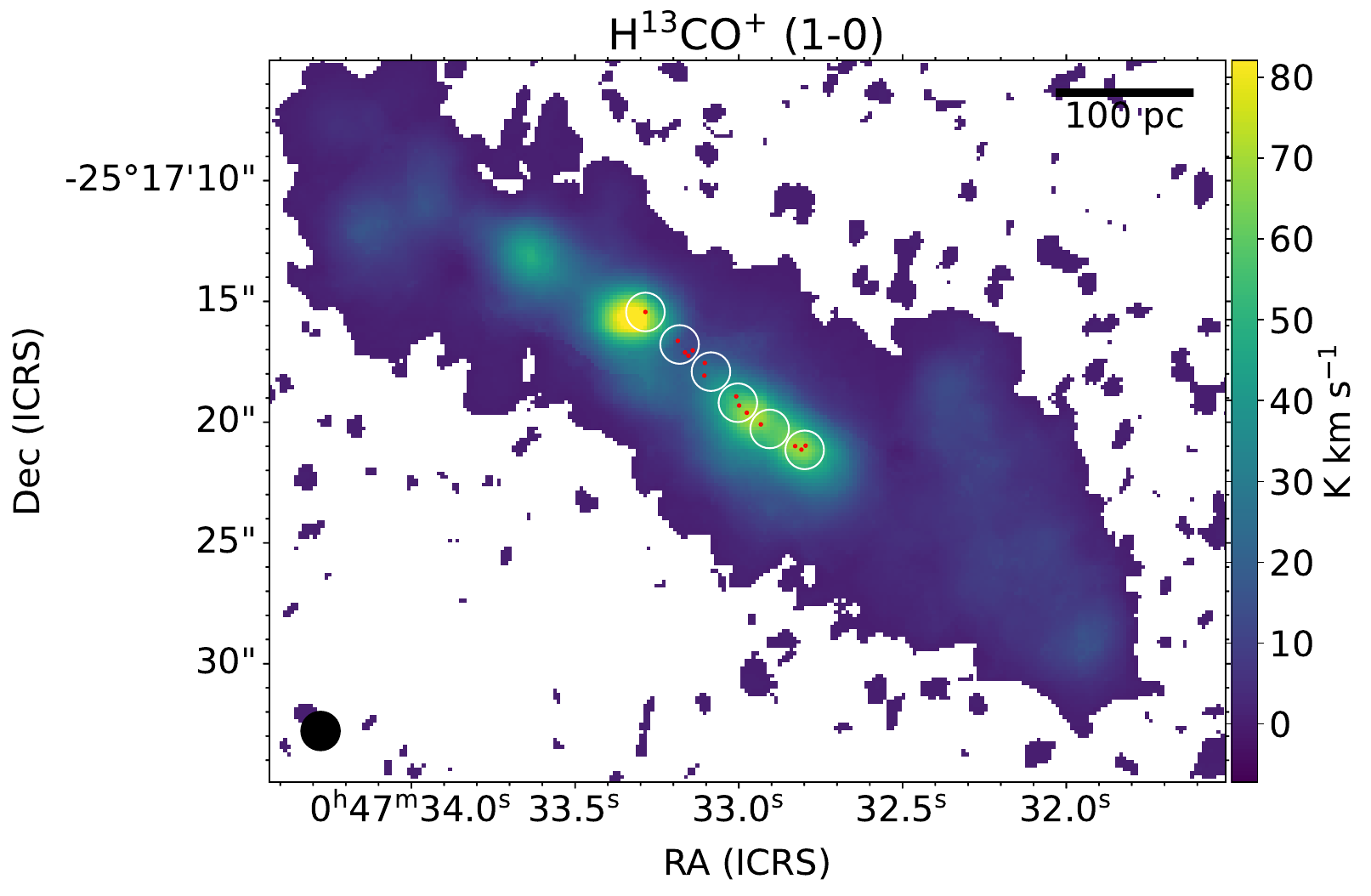} \\
    \centering\small (a) 
  \end{tabular}%
  \quad
\begin{tabular}[b]{@{}p{0.48\textwidth}@{}}
    \centering\includegraphics[width=1.0\linewidth]{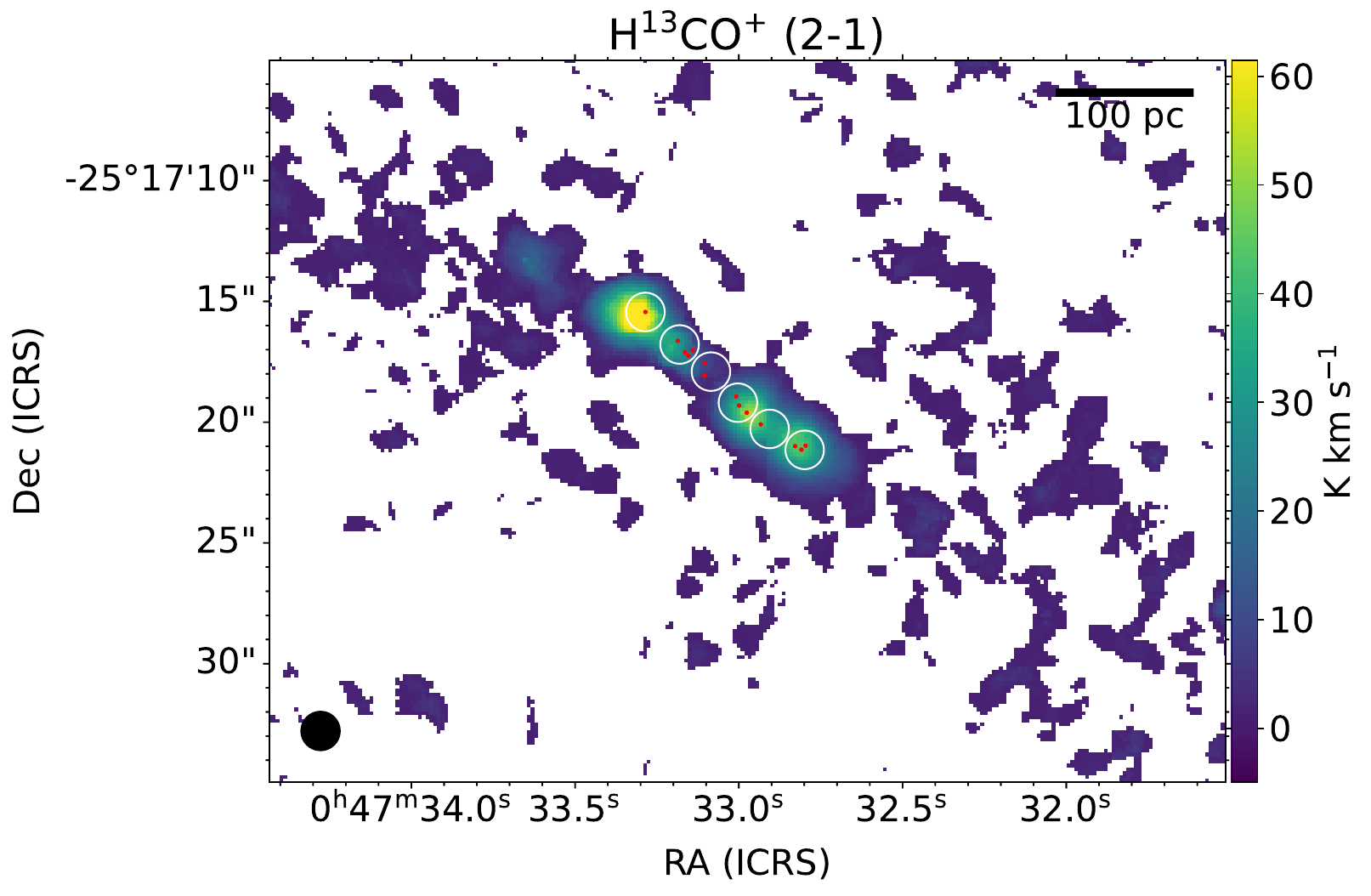} \\
    \centering\small (b)  
  \end{tabular}
  \quad
\begin{tabular}[b]{@{}p{0.48\textwidth}@{}}
    \centering\includegraphics[width=1.0\linewidth]{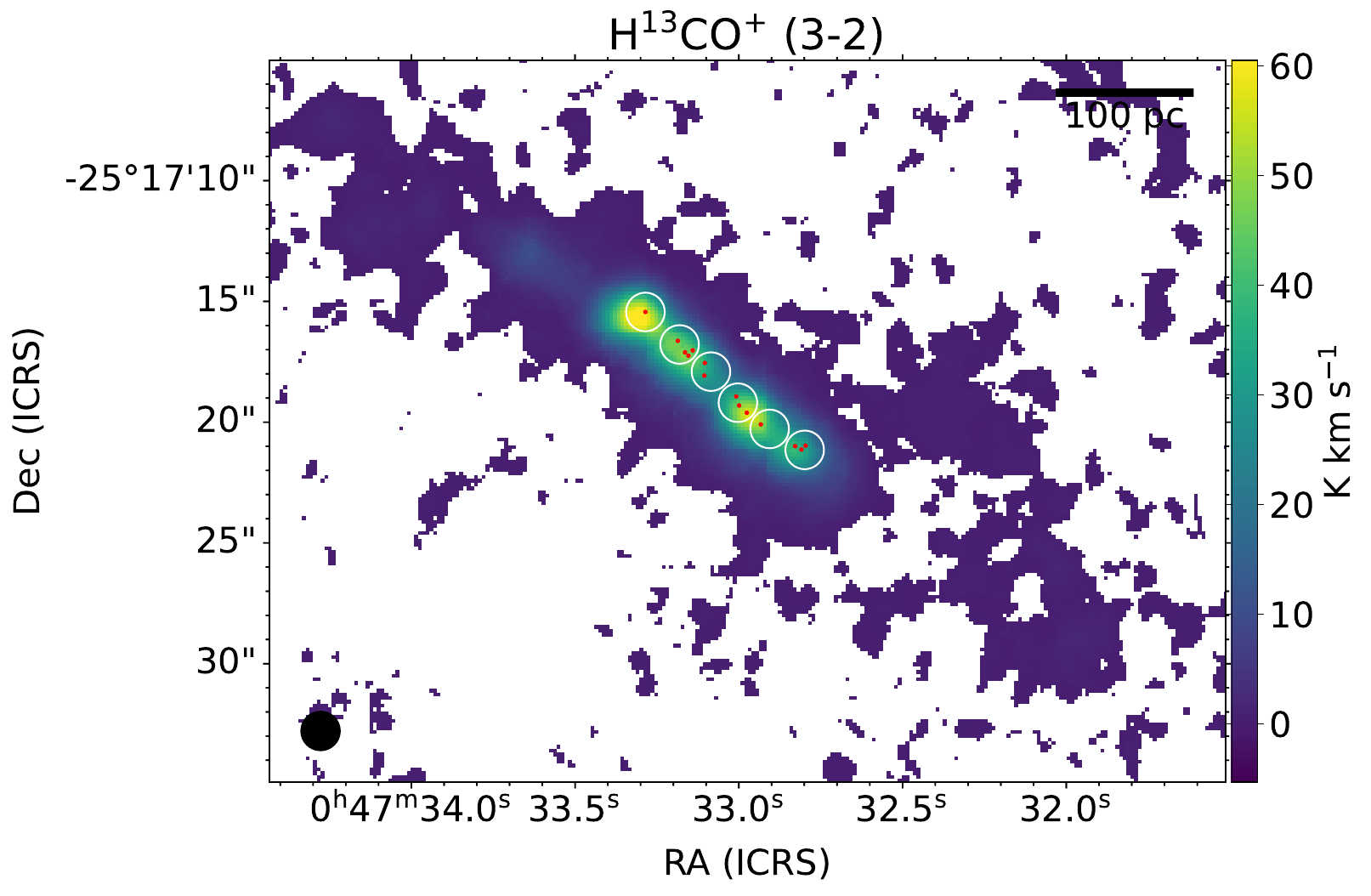} \\
    \centering\small (c)  
  \end{tabular}
  \caption{Velocity-integrated line intensities in [K\,\kms] of H\thirteen CO\plus\ (1-0), (2-1) and (3-2). Each of the maps shown have been generated using a signal-to-noise cutoff of 3. The studied SSC regions as listed in Tab.~\ref{tab:SSC_locations} are labeled in white texts on the map. The original SSC locations with appropriate beam sizes from \cite{2018Leroy} are shown by the red regions. The ALCHEMI $1''.6 \times 1''.6$ beam is displayed in the lower-left corner of the map. 
  }
  \label{fig:mom0_H13COP_10_21_32}
\end{figure}

\begin{figure}
  \centering
  \begin{tabular}[b]{@{}p{0.48\textwidth}@{}}
    \centering\includegraphics[width=1.0\linewidth]{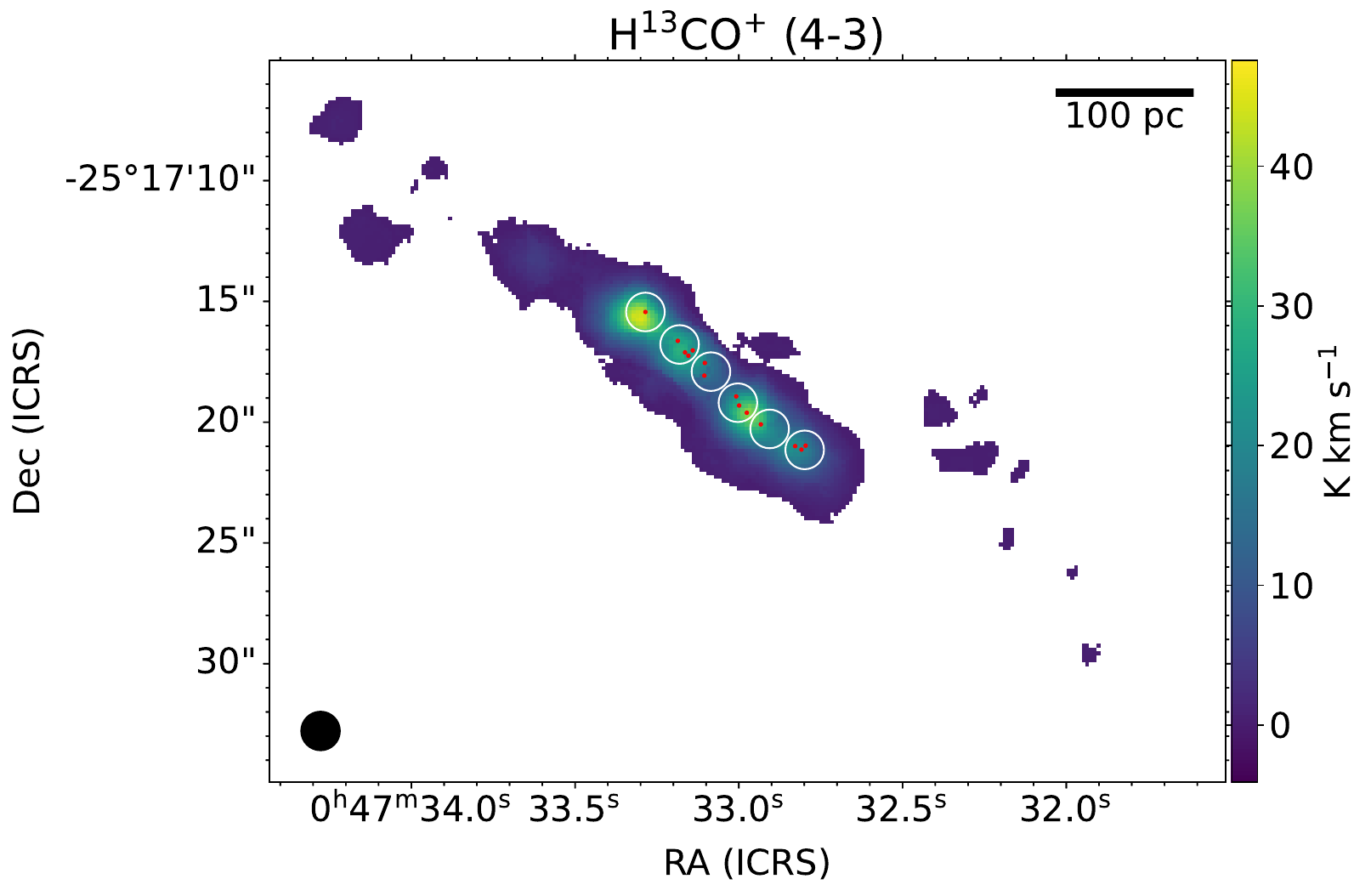} \\
    \centering\small (a) 
  \end{tabular}%
  \quad
\begin{tabular}[b]{@{}p{0.48\textwidth}@{}}
    \centering\includegraphics[width=1.0\linewidth]{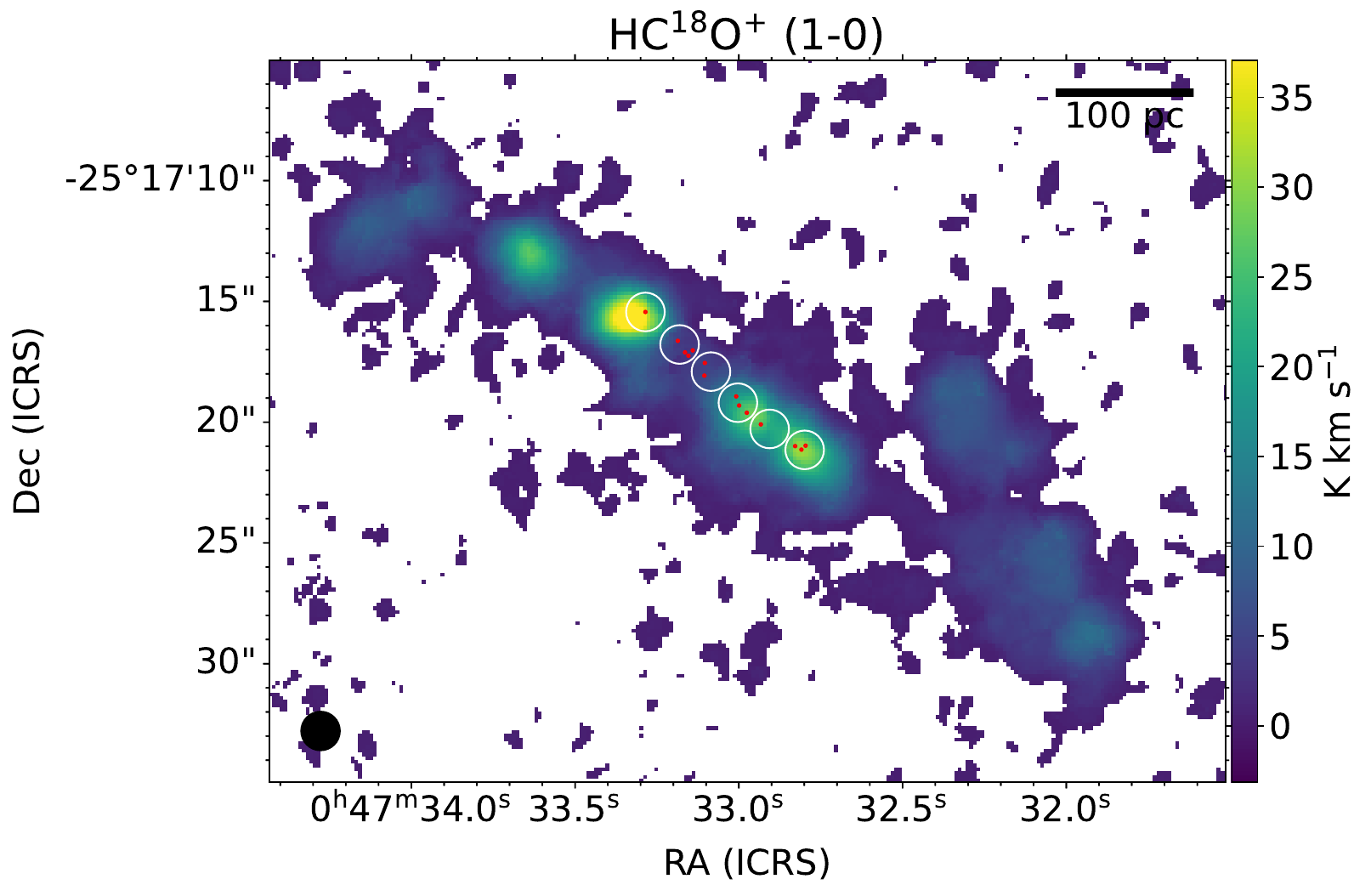} \\
    \centering\small (b)  
  \end{tabular}
  \quad
\begin{tabular}[b]{@{}p{0.48\textwidth}@{}}
    \centering\includegraphics[width=1.0\linewidth]{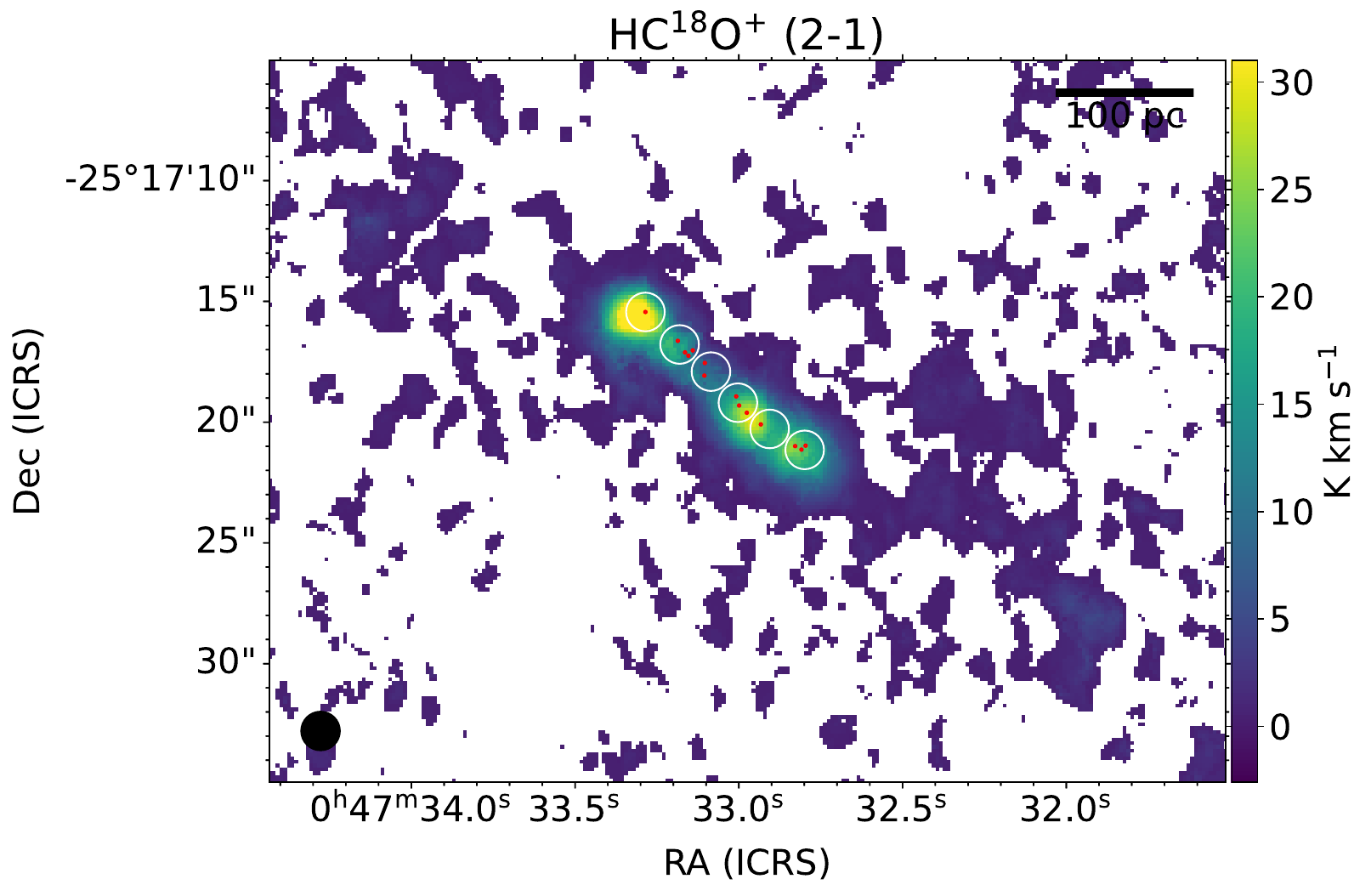} \\
    \centering\small (c)  
  \end{tabular}
  \caption{Velocity-integrated line intensities in [K\,\kms] of H\thirteen CO\plus\ (4-3), HC\eighteen O\plus\ (1-0) and (2-1). Each of the maps shown have been generated using a signal-to-noise cutoff of 3. The studied SSC regions as listed in Tab.~\ref{tab:SSC_locations} are labeled in white texts on the map. The original SSC locations with appropriate beam sizes from \cite{2018Leroy} are shown by the red regions. The ALCHEMI $1''.6 \times 1''.6$ beam is displayed in the lower-left corner of the map. 
  }
  \label{fig:mom0_H13COP_HC18OP}
\end{figure}
\begin{figure}
\begin{tabular}[b]{@{}p{0.48\textwidth}@{}}
    \centering\includegraphics[width=1.0\linewidth]{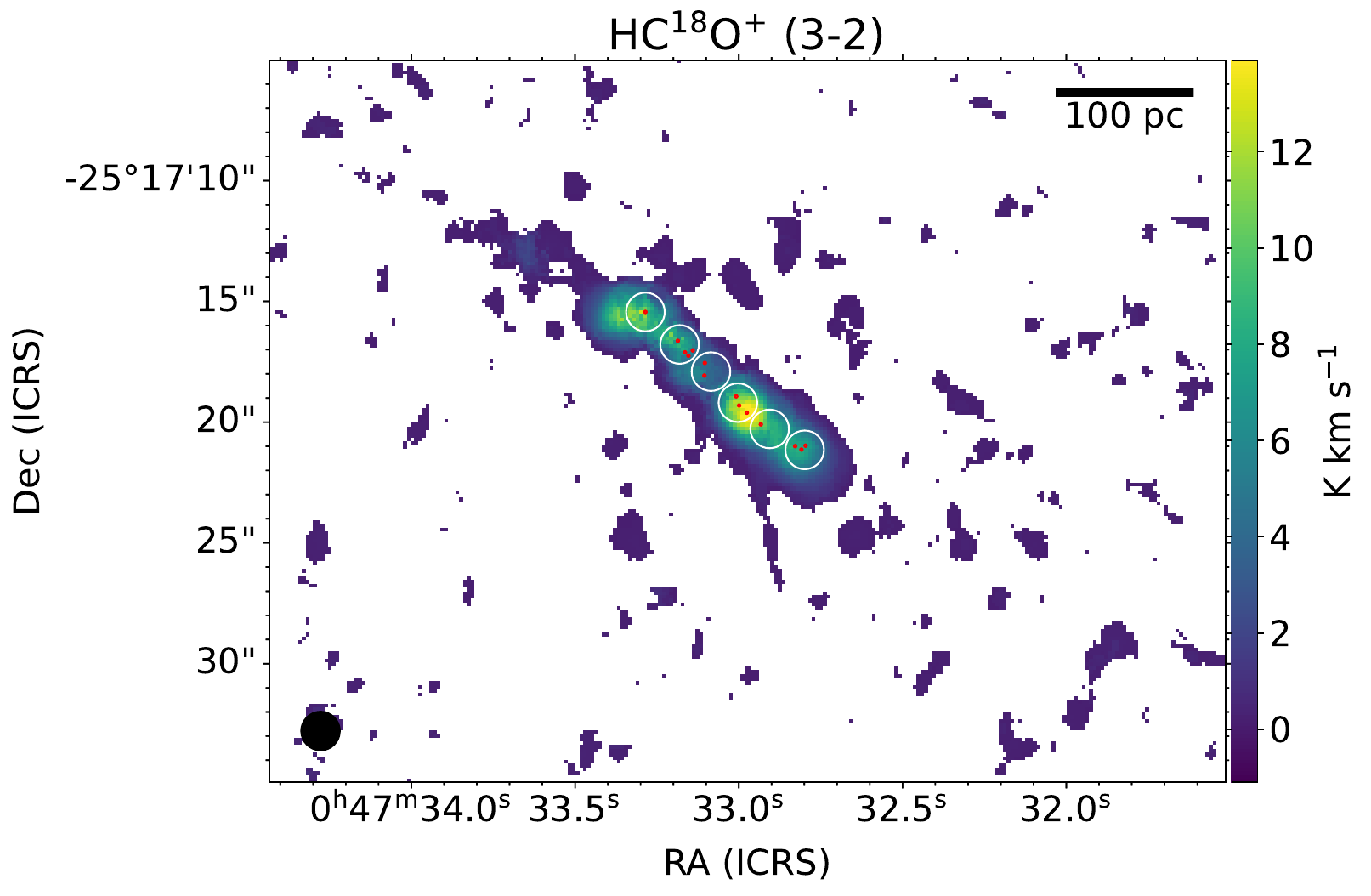} \\ 
  \end{tabular}
  \caption{Velocity-integrated line intensities in [K\,\kms] of HC\eighteen O\plus\ (3-2). Each of the maps shown have been generated using a signal-to-noise cutoff of 3. The studied SSC regions as listed in Tab.~\ref{tab:SSC_locations} are labeled in white texts on the map. The original SSC locations with appropriate beam sizes from \cite{2018Leroy} are shown by the red regions. The ALCHEMI $1''.6 \times 1''.6$ beam is displayed in the lower-left corner of the map. 
  }
  \label{fig:mom0_HC18OP}
\end{figure}

\section{Absorption at the TH2 Position}
\label{app:absorption}
Within this section the effect of the `TH2' region upon SSC-13* and SSC-9* are shown. This effect is shown both upon the affected spectra and column density ratio vs age diagrams.
\subsection{Affected Spectra}

\begin{figure}
  \centering
  \includegraphics[width=0.5\textwidth]{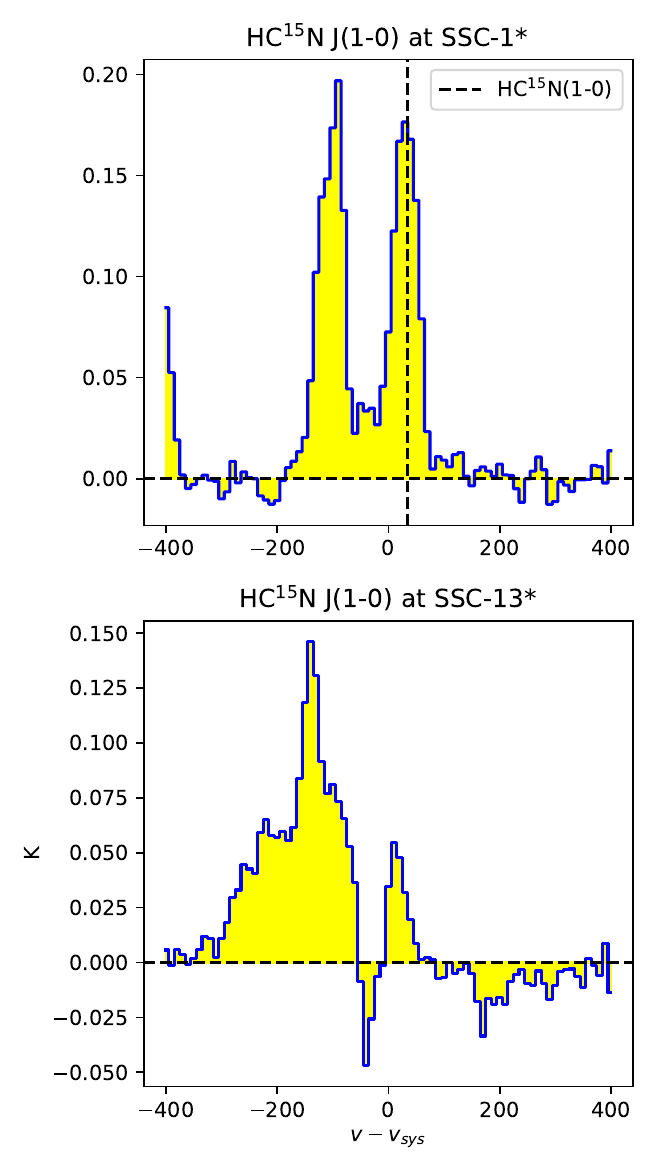}
  \caption{The systemic velocity subtracted velocity spectra for HC\fifteen(1-0) shown for regions SSC-1* (representative for unaffected regions) and SSC-13* (representative for absorption affected regions).}
  \label{fig:Absorption comparison}
\end{figure}

\subsection{Affected Abundance Ratio vs Age Diagrams}
\label{app:more:ratio_age_plots}
\begin{figure}
  \centering
  \includegraphics[width=0.5\textwidth]{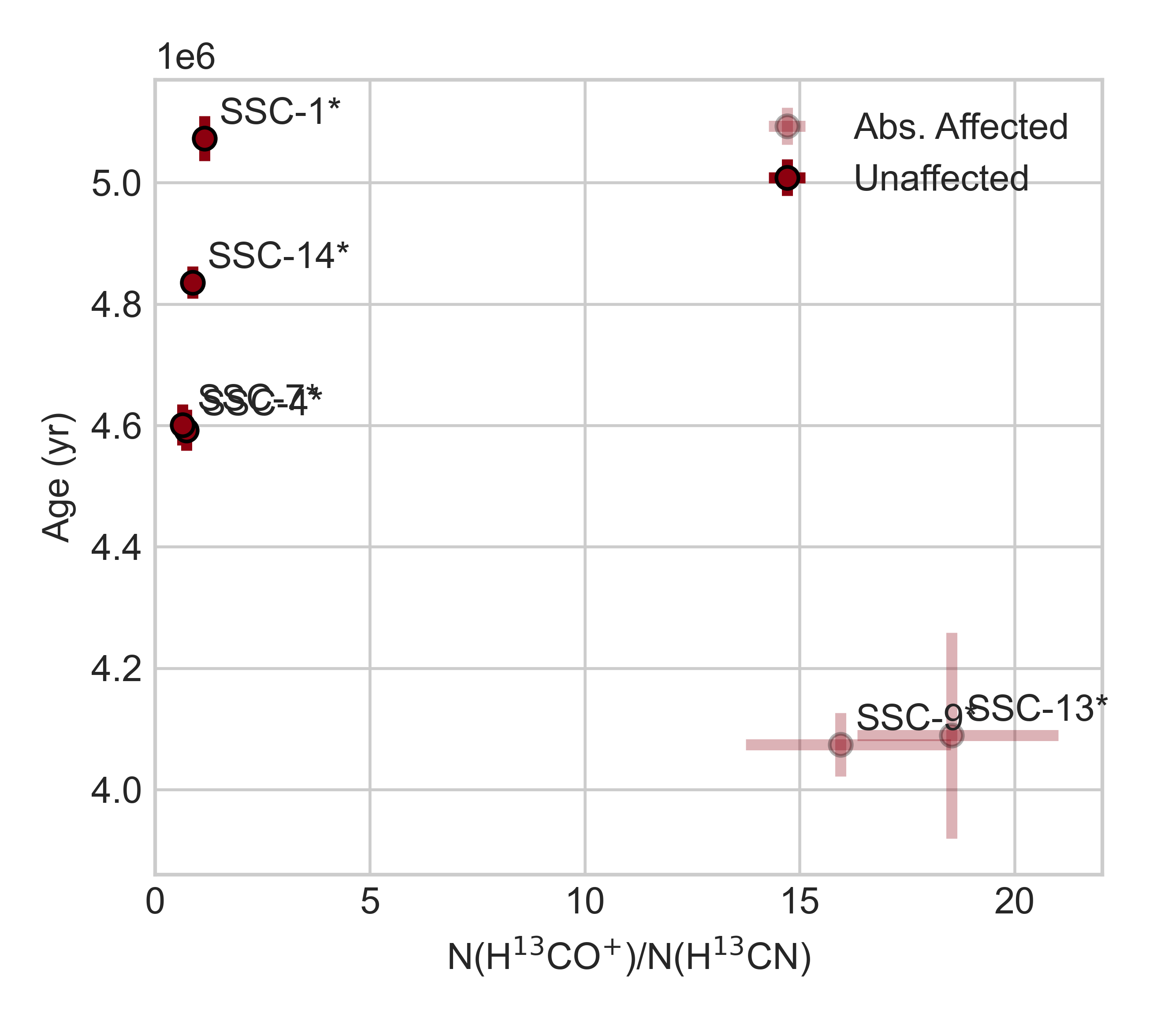}
  \caption{The H\thirteen CO\plus/H\thirteen CN column density ratio derived from \texttt{RADEX}. The regions unaffected by the absorption effects observed near GMC-5 in NGC~253 are shown in full colour, whereas those regions that are effected are shown shaded.}
  \label{fig:Full_H13COP_H13CN}
\end{figure}
\begin{figure}
  \centering
  \includegraphics[width=0.5\textwidth]{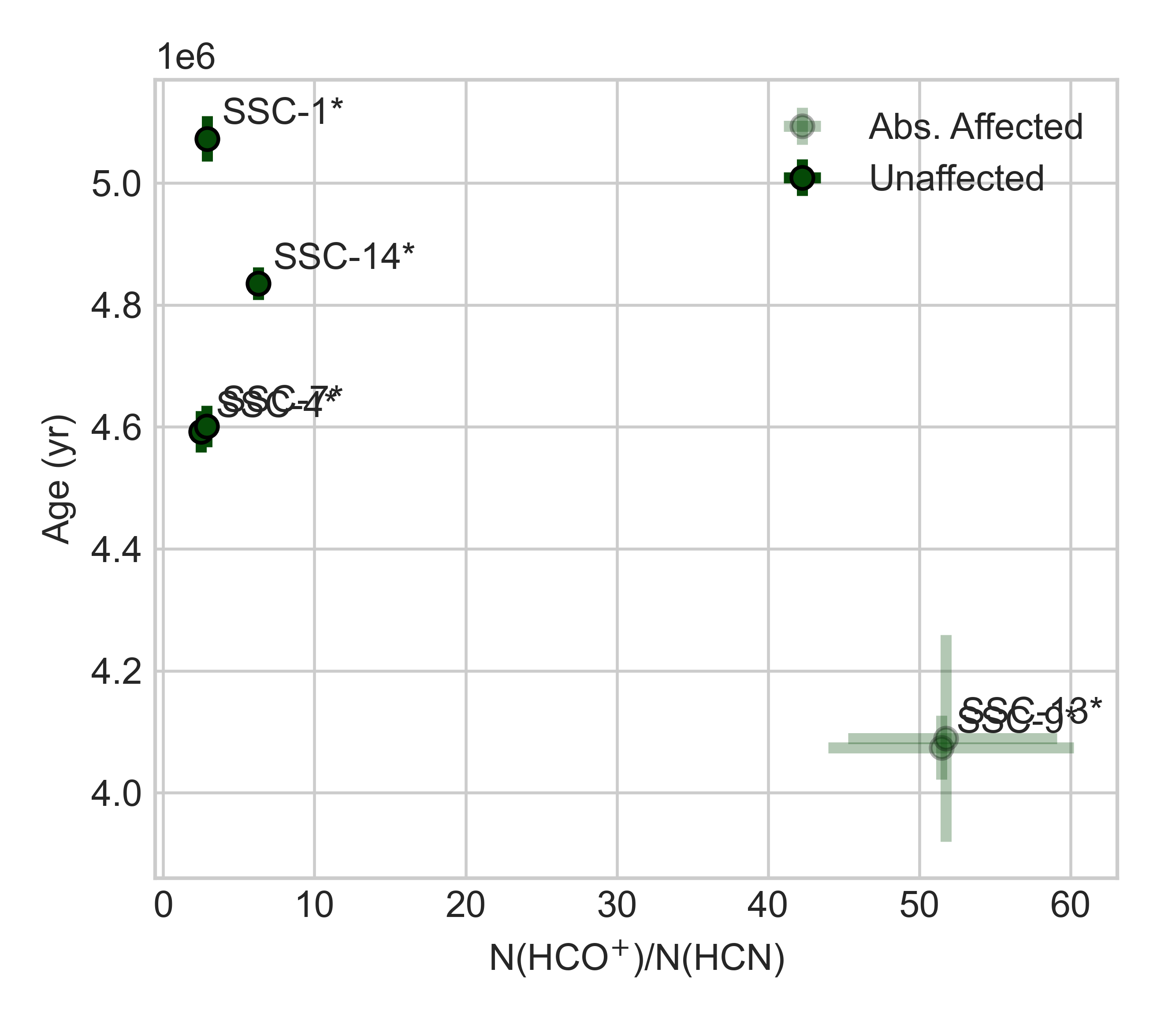}
  \caption{The HCO\plus/HCN column density ratio derived from \texttt{RADEX}. The regions unaffected by the absorption effects observed near GMC-5 in NGC~253 are shown in full colour, whereas those regions that are effected are shown shaded.}
  \label{fig:Full_HCOP_HCN}
\end{figure}
\newpage
\section{Additional Posterior Distributions}
\label{app:Corners}
This Section contains the remaining Posterior distributions generated from the nested sampling of the \texttt{RADEX} modelling.
\subsection{SSC-1*}
\begin{figure*}
  \centering
  \includegraphics[width=\textwidth]{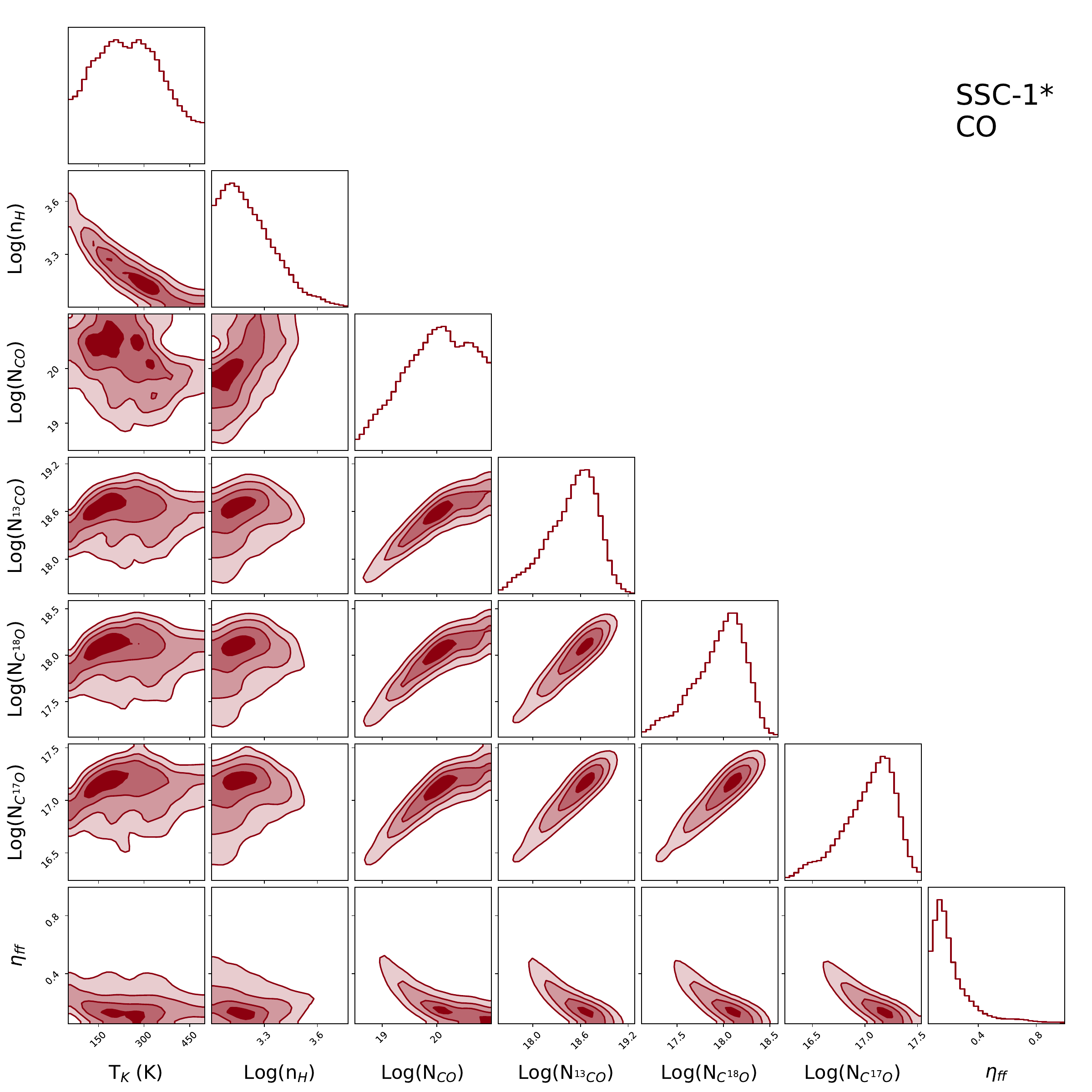}
  \caption{The posterior distributions for SSC-1* of the kinetic temperature, neutral H\textsubscript{2} number density, beam filling factor and column densities of CO, \thirteen CO, C\eighteen O, and C\seventeen O, as predicted by \texttt{RADEX}. }
  \label{fig:RADEX_corner_SSC1_CO}
\end{figure*}

\begin{figure*}
  \centering
  \includegraphics[width=\textwidth]{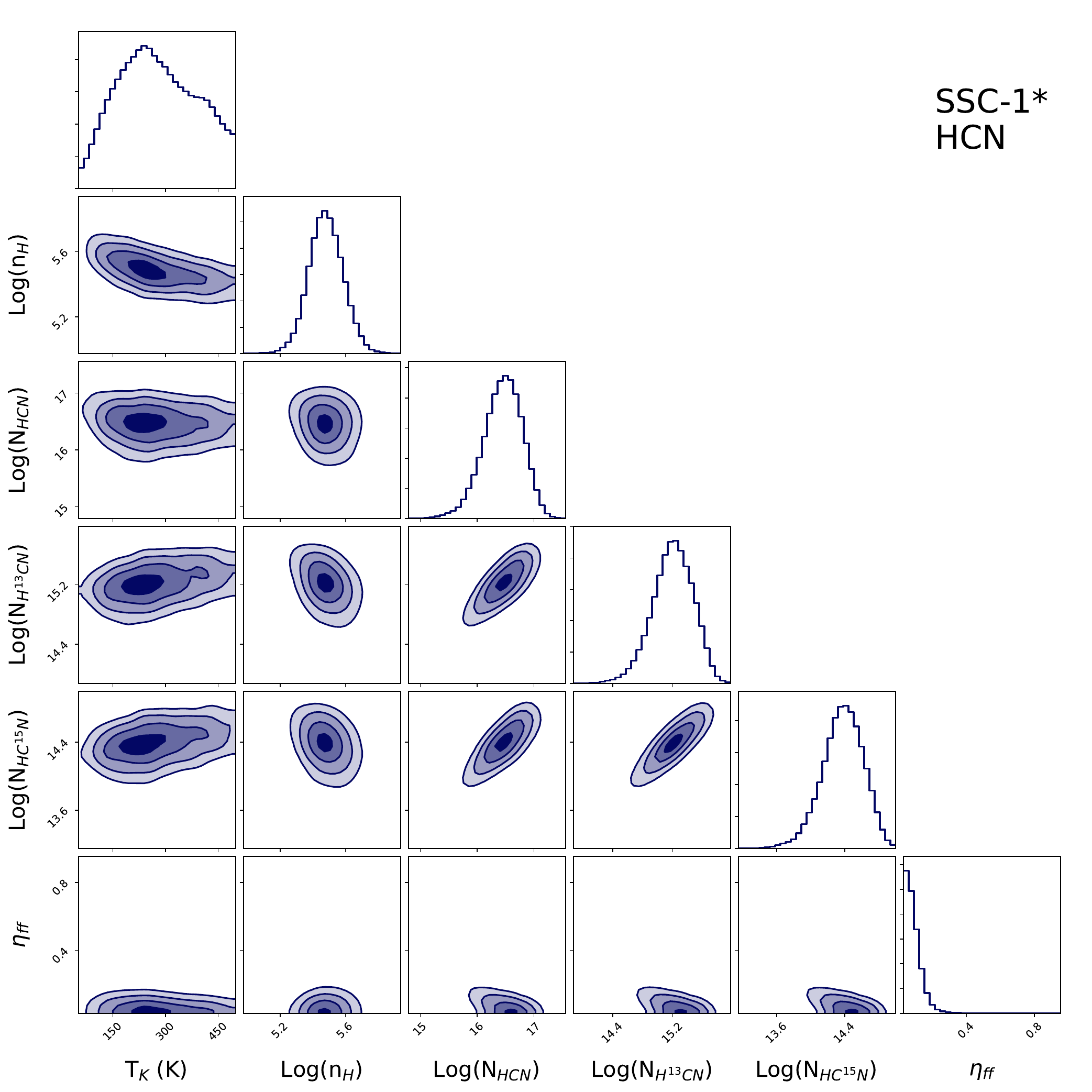}
  \caption{The posterior distributions for SSC-1* of the kinetic temperature, neutral H\textsubscript{2} number density, beam filling factor and column densities of HCN, H\thirteen CN and HC \fifteen N, as predicted by \texttt{RADEX}. }
  \label{fig:RADEX_corner_SSC1_HCN}
\end{figure*}

\begin{figure*}
  \centering
  \includegraphics[width=\textwidth]{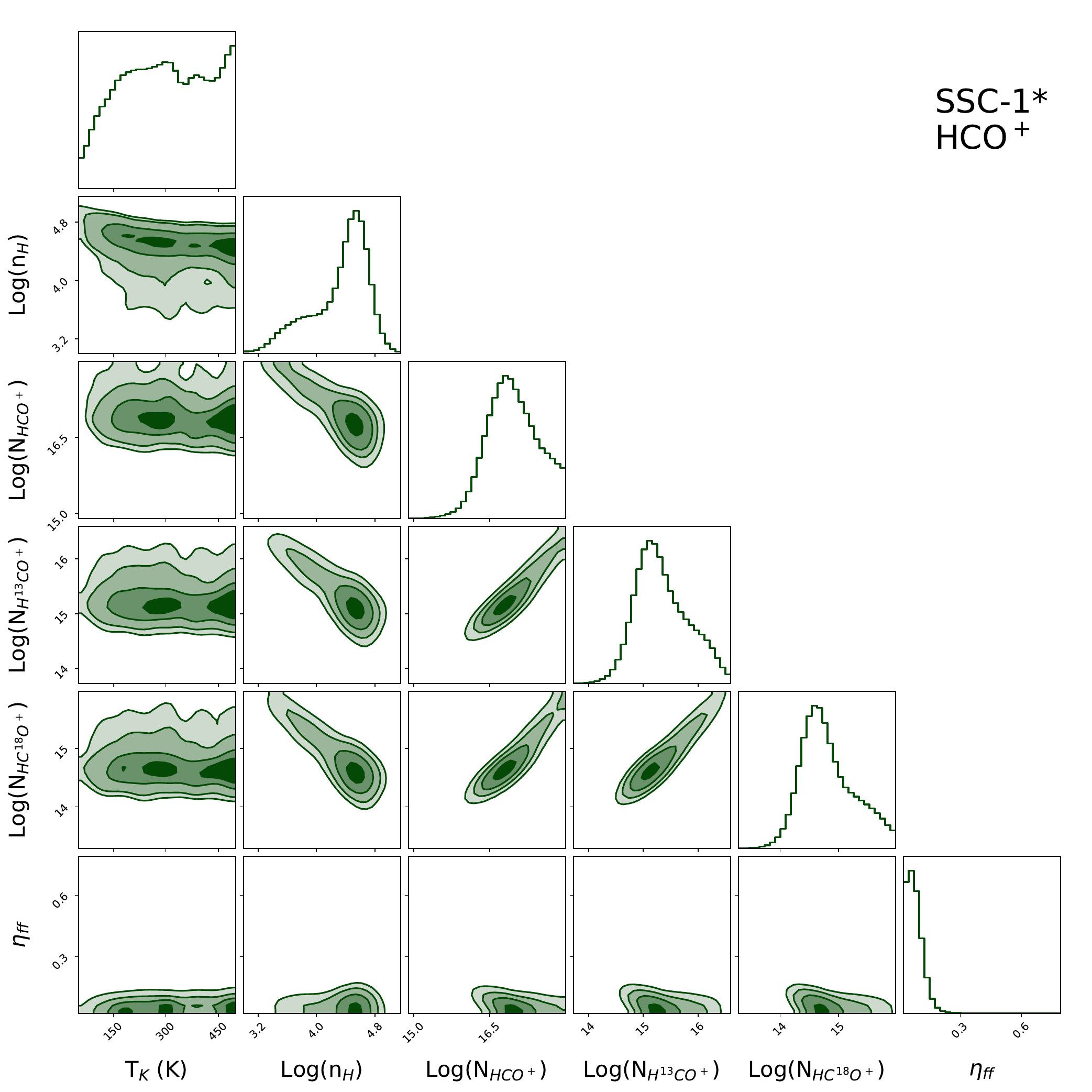}
  \caption{The posterior distributions for SSC-1* of the kinetic temperature, neutral H\textsubscript{2} number density, beam filling factor and column densities of HCO\plus, H\thirteen CO\plus\ and HC\eighteen O\plus, as predicted by \texttt{RADEX}. }
  \label{fig:RADEX_corner_SSC1_HCOP}
\end{figure*}

\subsection{SSC-7*}
\begin{figure*}
  \centering
  \includegraphics[width=\textwidth]{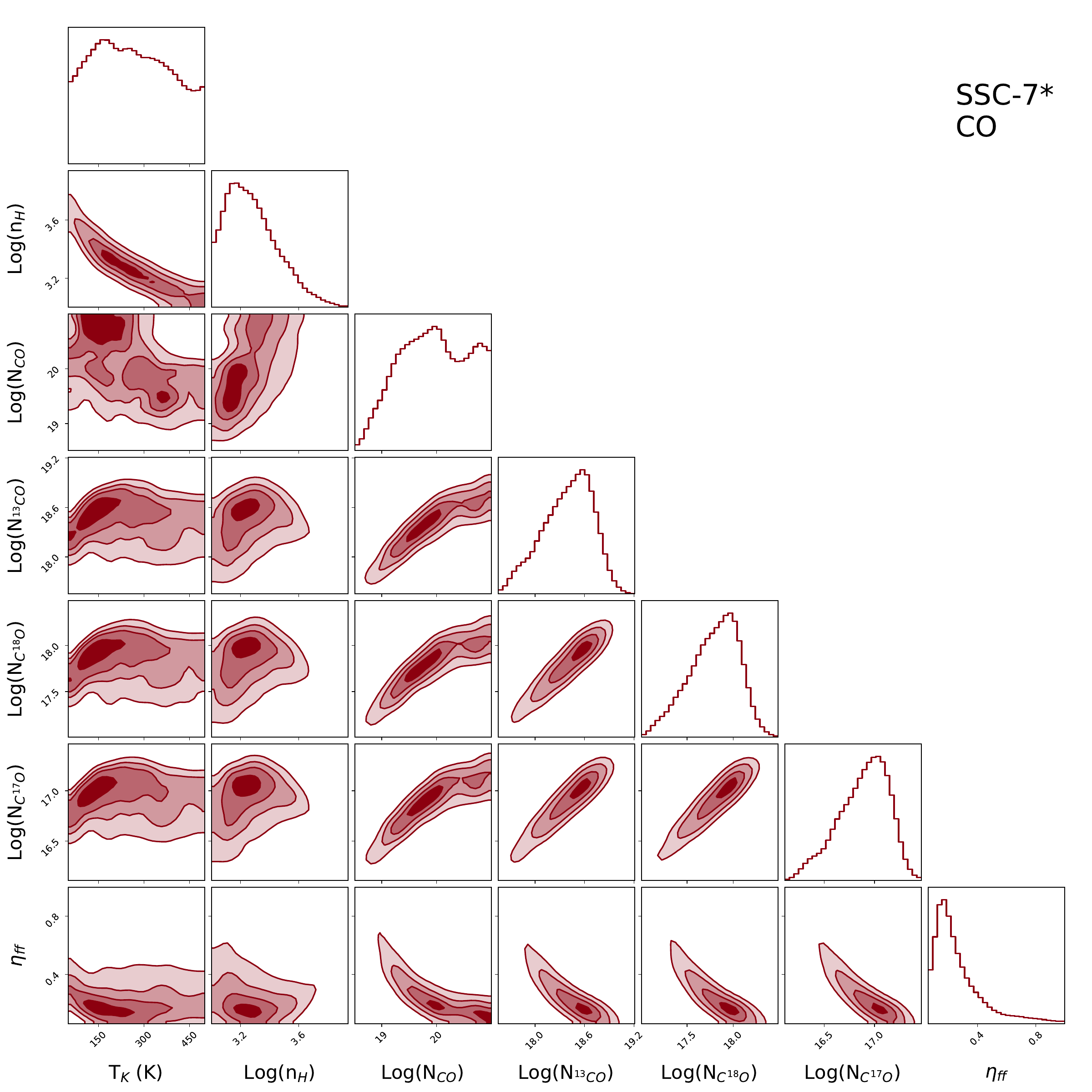}
  \caption{The posterior distributions for SSC-7* of the kinetic temperature, neutral H\textsubscript{2} number density, beam filling factor and column densities of CO, \thirteen CO, C\eighteen O, and C\seventeen O, as predicted by \texttt{RADEX}. }
  \label{fig:RADEX_corner_SSC7_CO}
\end{figure*}

\begin{figure*}
  \centering
  \includegraphics[width=\textwidth]{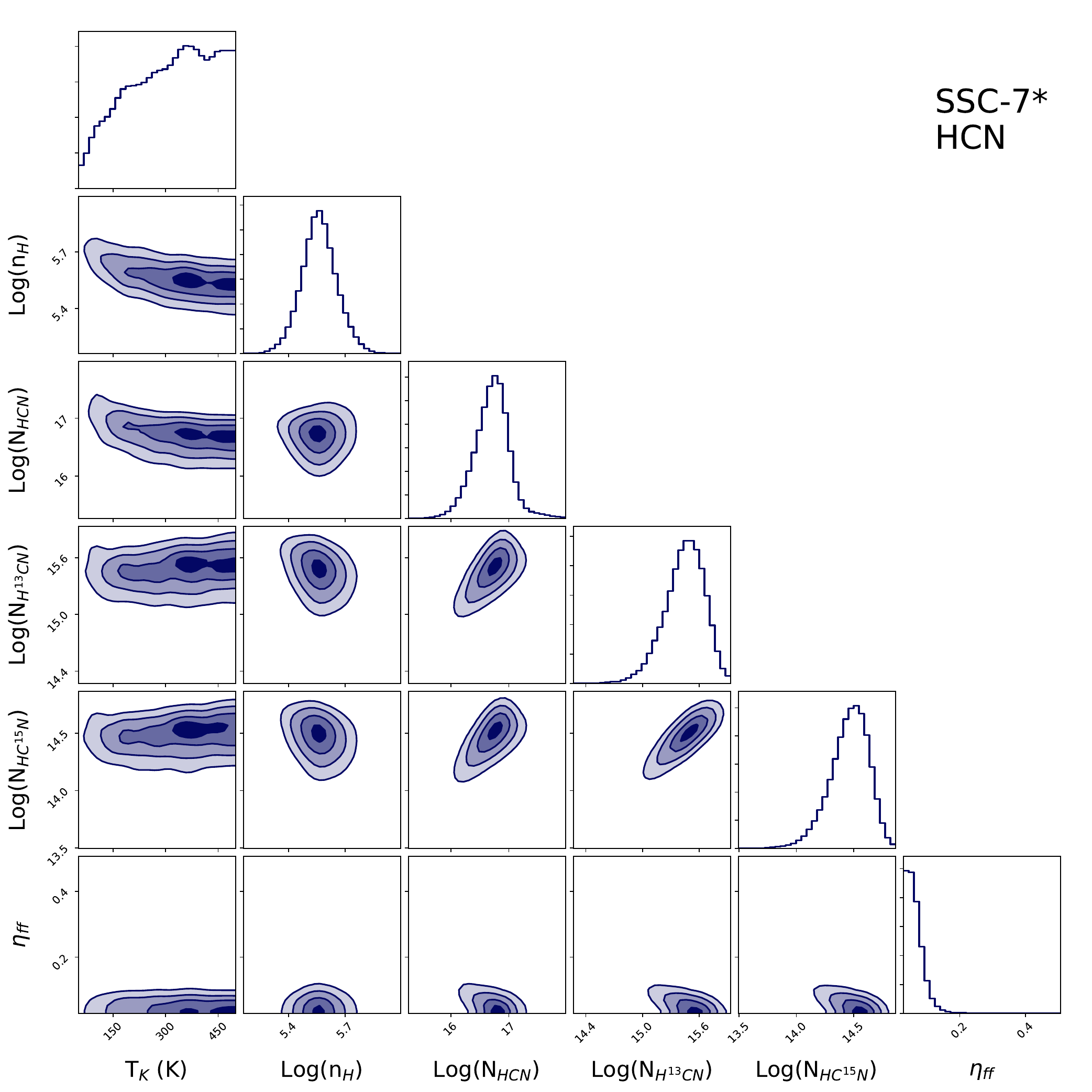}
  \caption{The posterior distributions for SSC-7* of the kinetic temperature, neutral H\textsubscript{2} number density, beam filling factor and column densities of HCN, H\thirteen CN and HC \fifteen N, as predicted by \texttt{RADEX}. }
  \label{fig:RADEX_corner_SSC7_HCN}
\end{figure*}

\begin{figure*}
  \centering
  \includegraphics[width=\textwidth]{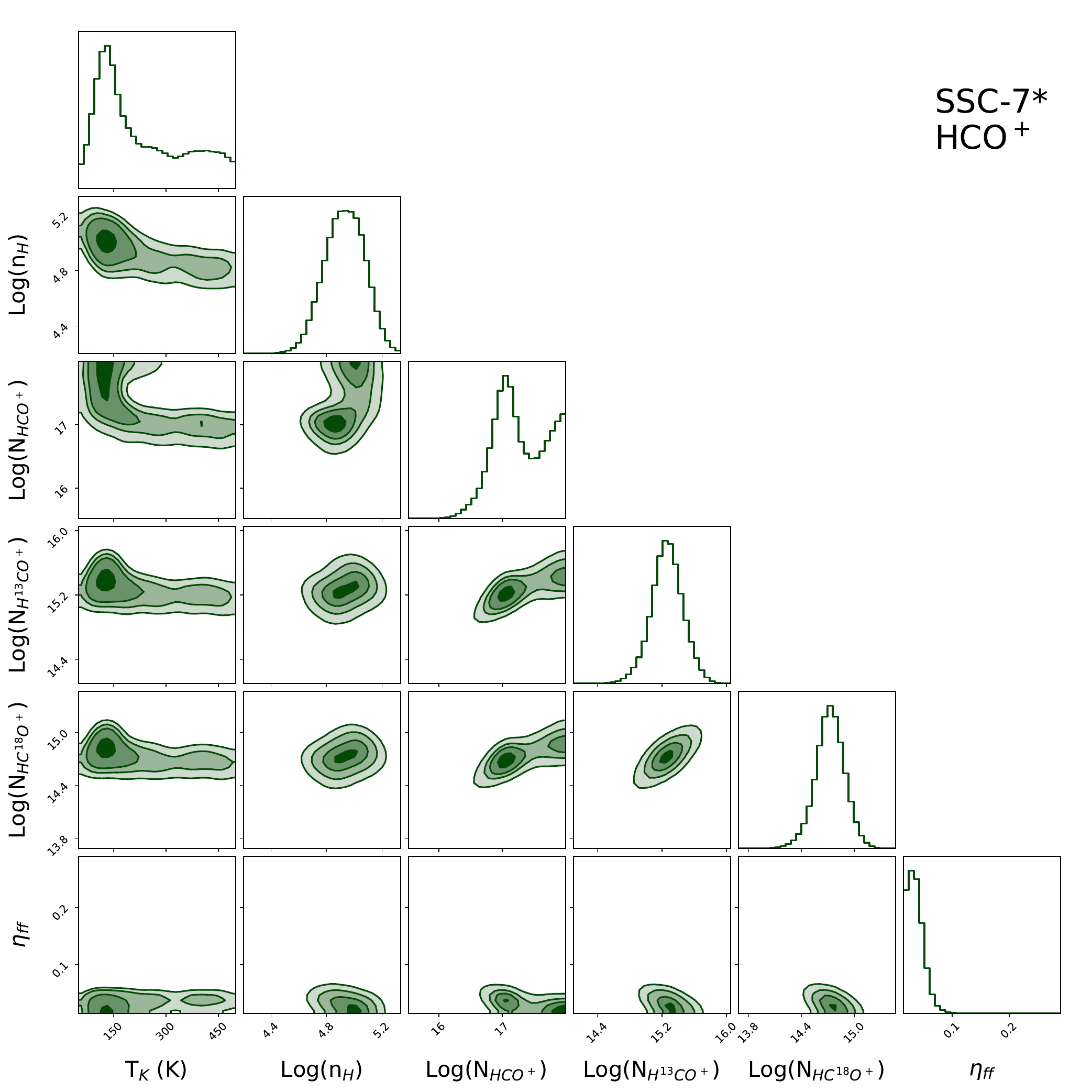}
  \caption{The posterior distributions for SSC-7* of the kinetic temperature, neutral H\textsubscript{2} number density, beam filling factor and column densities of HCO\plus, H\thirteen CO\plus\ and HC\eighteen O\plus, as predicted by \texttt{RADEX}. }
  \label{fig:RADEX_corner_SSC7_HCOP}
\end{figure*}

\subsection{SSC-9*}
\begin{figure*}
  \centering
  \includegraphics[width=\textwidth]{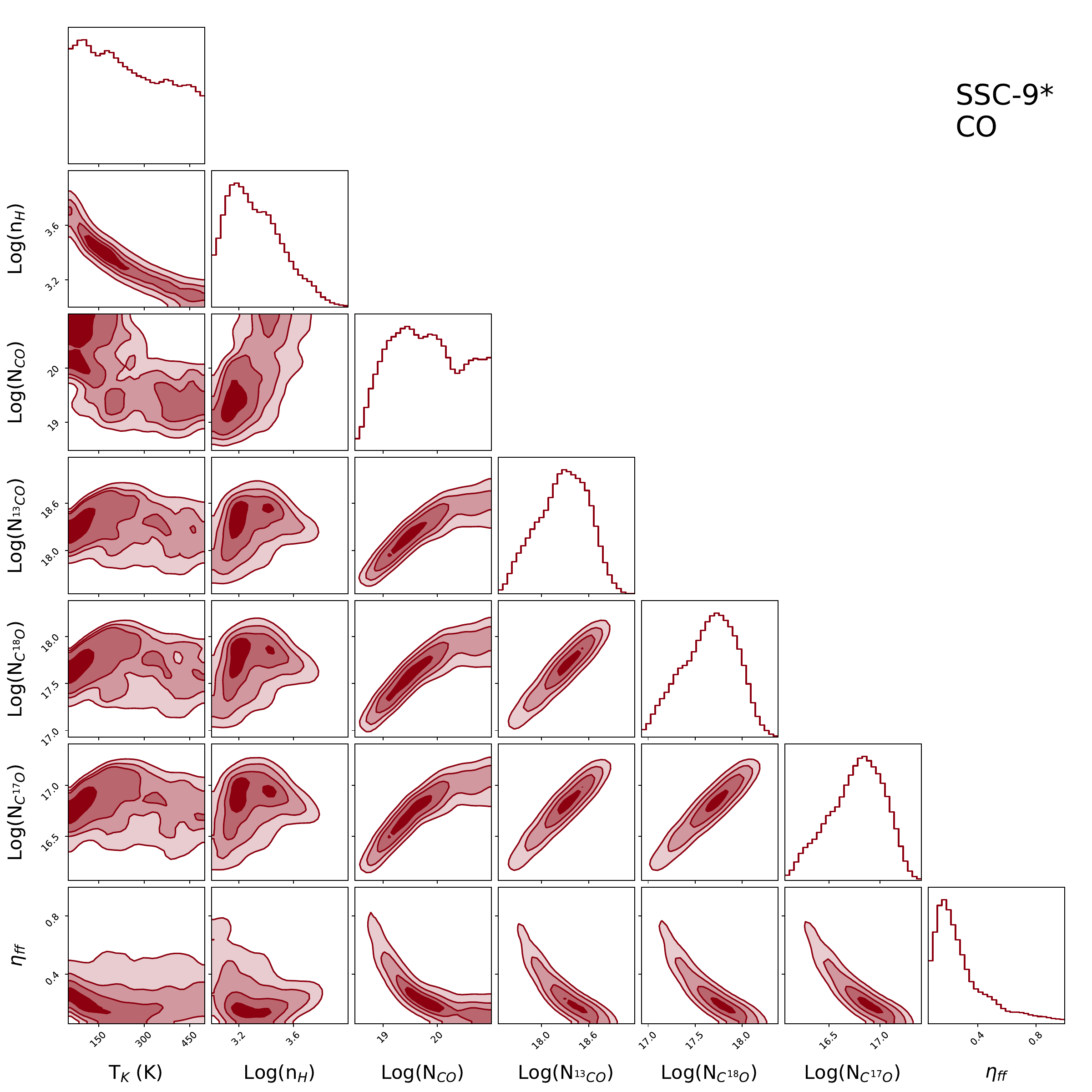}
  \caption{The posterior distributions for SSC-9* of the kinetic temperature, neutral H\textsubscript{2} number density, beam filling factor and column densities of CO, \thirteen CO, C\eighteen O, and C\seventeen O, as predicted by \texttt{RADEX}. }
  \label{fig:RADEX_corner_SSC9_CO}
\end{figure*}

\begin{figure*}
  \centering
  \includegraphics[width=\textwidth]{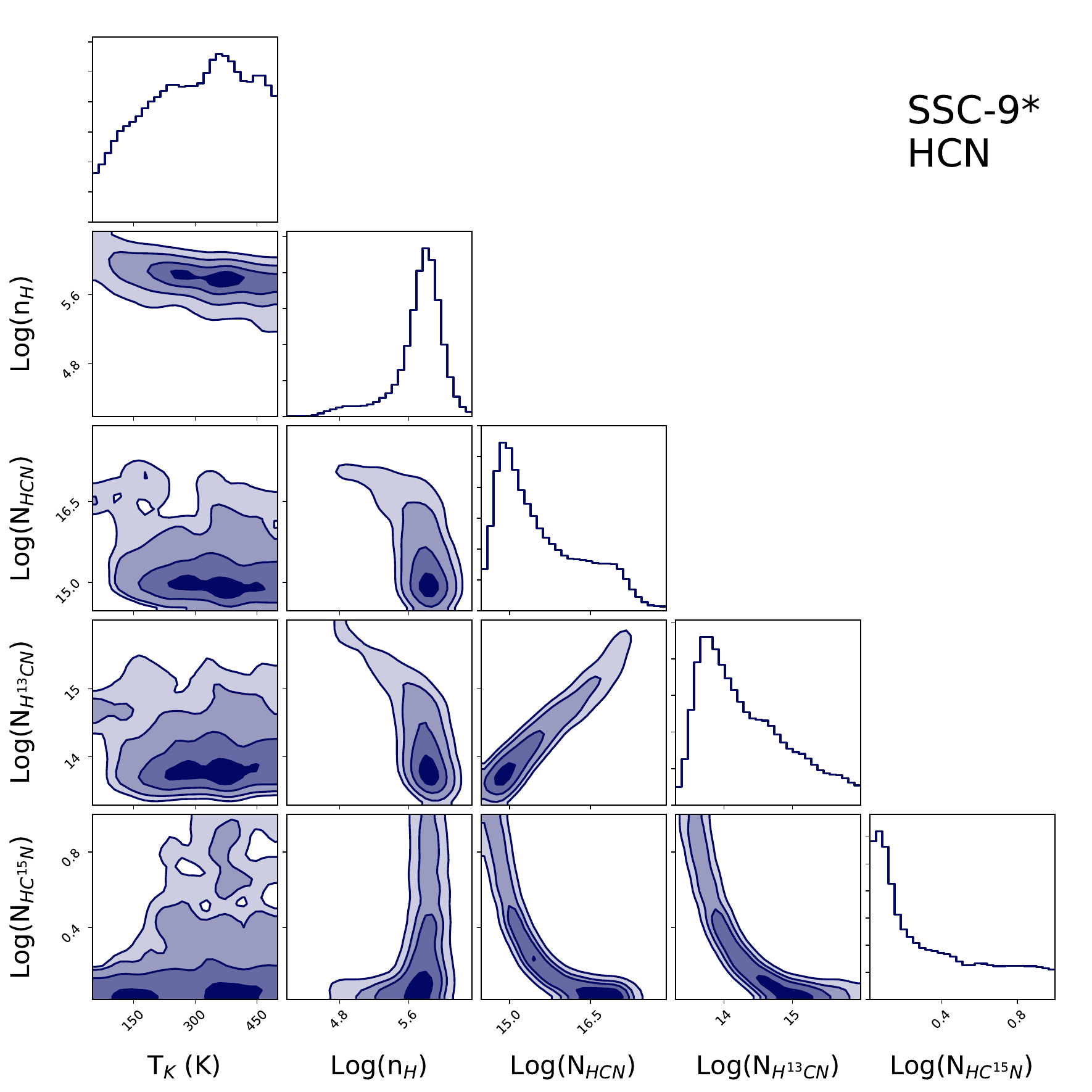}
  \caption{The posterior distributions for SSC-9* of the kinetic temperature, neutral H\textsubscript{2} number density, beam filling factor and column densities of HCN, H\thirteen CN and HC \fifteen N, as predicted by \texttt{RADEX}. }
  \label{fig:RADEX_corner_SSC9_HCN}
\end{figure*}

\begin{figure*}
  \centering
  \includegraphics[width=\textwidth]{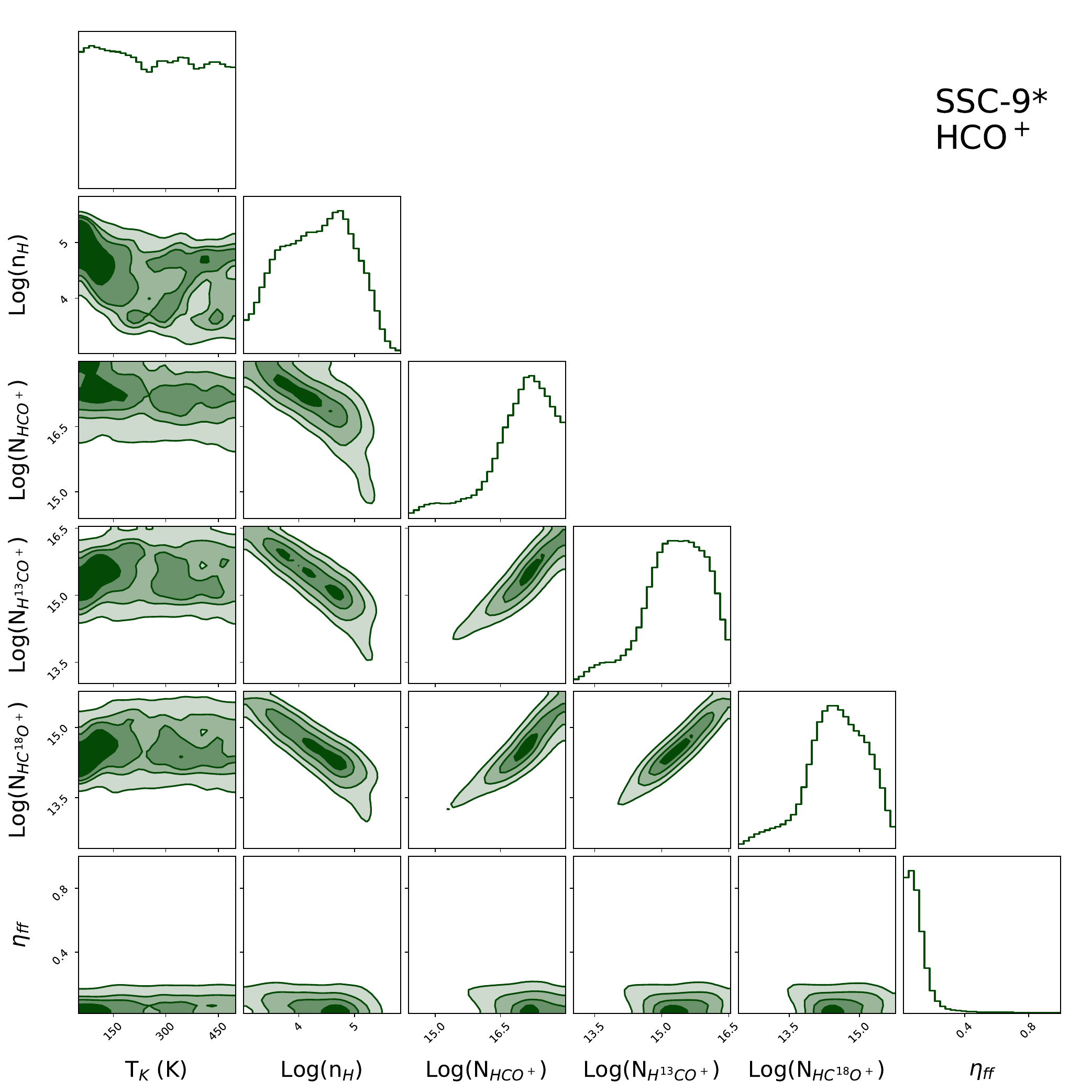}
  \caption{The posterior distributions for SSC-9* of the kinetic temperature, neutral H\textsubscript{2} number density, beam filling factor and column densities of HCO\plus, H\thirteen CO\plus\ and HC\eighteen O\plus, as predicted by \texttt{RADEX}. }
  \label{fig:RADEX_corner_SSC9_HCOP}
\end{figure*}

\subsection{SSC-13*}
\begin{figure*}
  \centering
  \includegraphics[width=\textwidth]{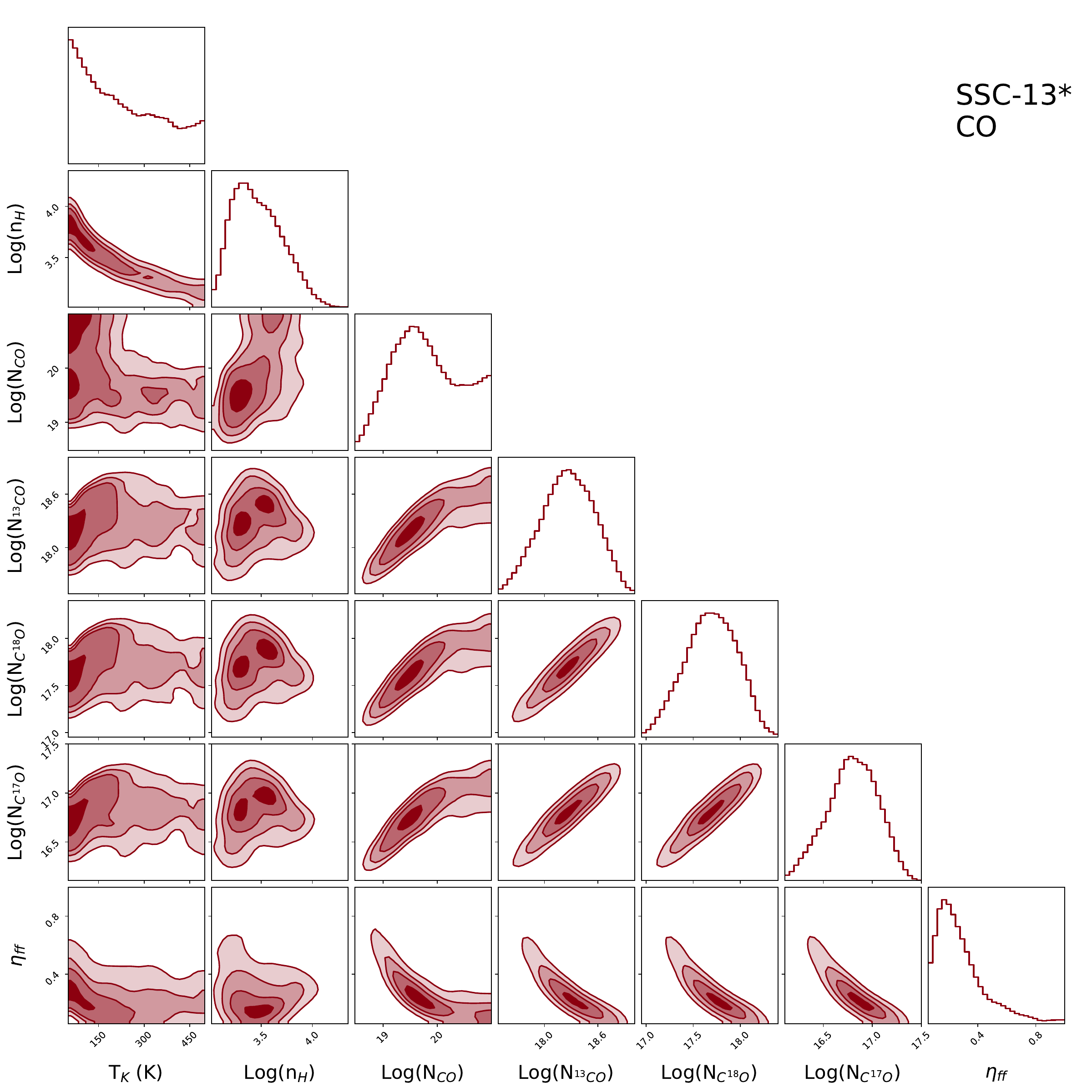}
  \caption{The posterior distributions for SSC-13* of the kinetic temperature, neutral H\textsubscript{2} number density, beam filling factor and column densities of CO, \thirteen CO, C\eighteen O, and C\seventeen O, as predicted by \texttt{RADEX}. }
  \label{fig:RADEX_corner_SSC13_CO}
\end{figure*}

\begin{figure*}
  \centering
  \includegraphics[width=\textwidth]{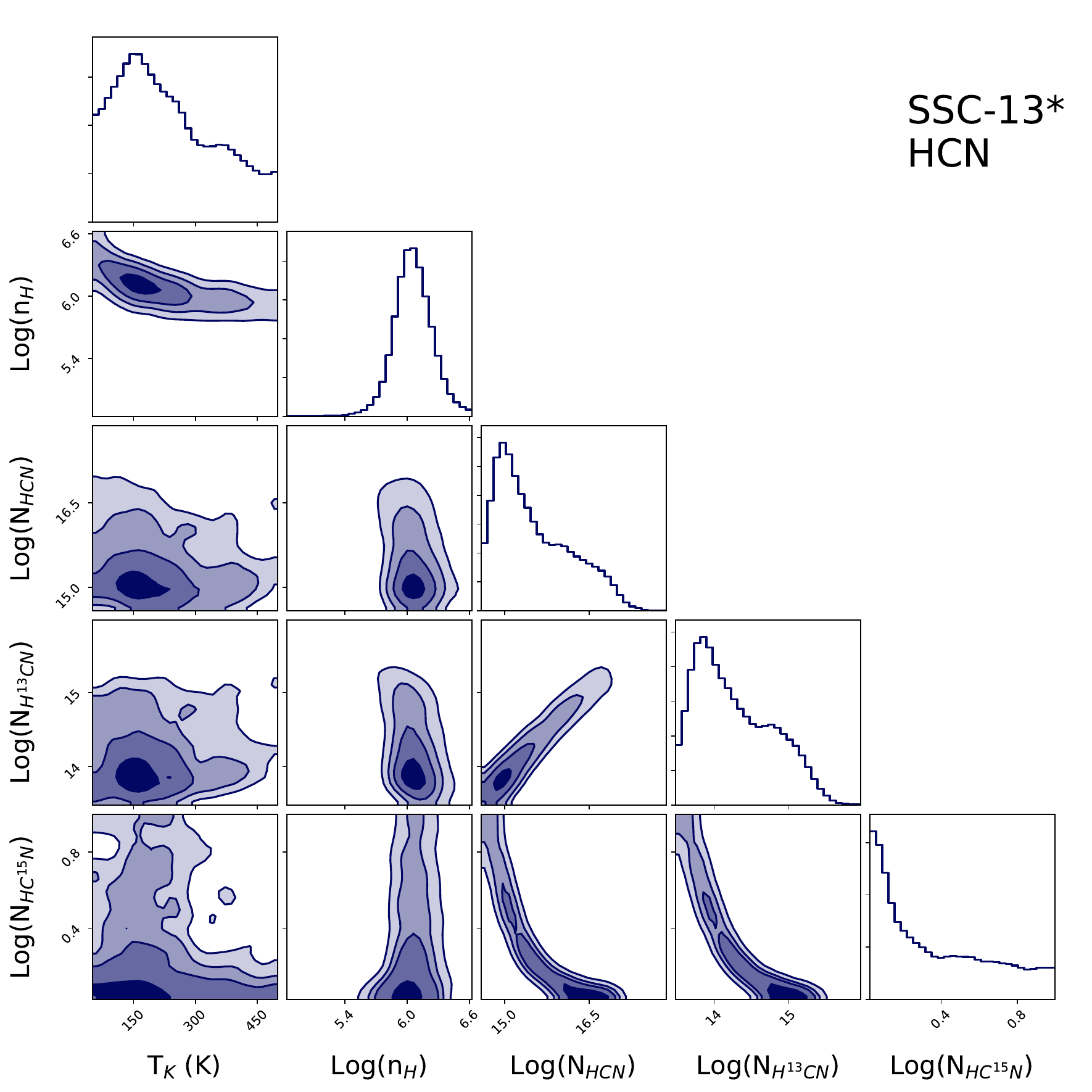}
  \caption{The posterior distributions for SSC-13* of the kinetic temperature, neutral H\textsubscript{2} number density, beam filling factor and column densities of HCN, H\thirteen CN and HC \fifteen N, as predicted by \texttt{RADEX}. }
  \label{fig:RADEX_corner_SSC13_HCN}
\end{figure*}

\begin{figure*}
  \centering
  \includegraphics[width=\textwidth]{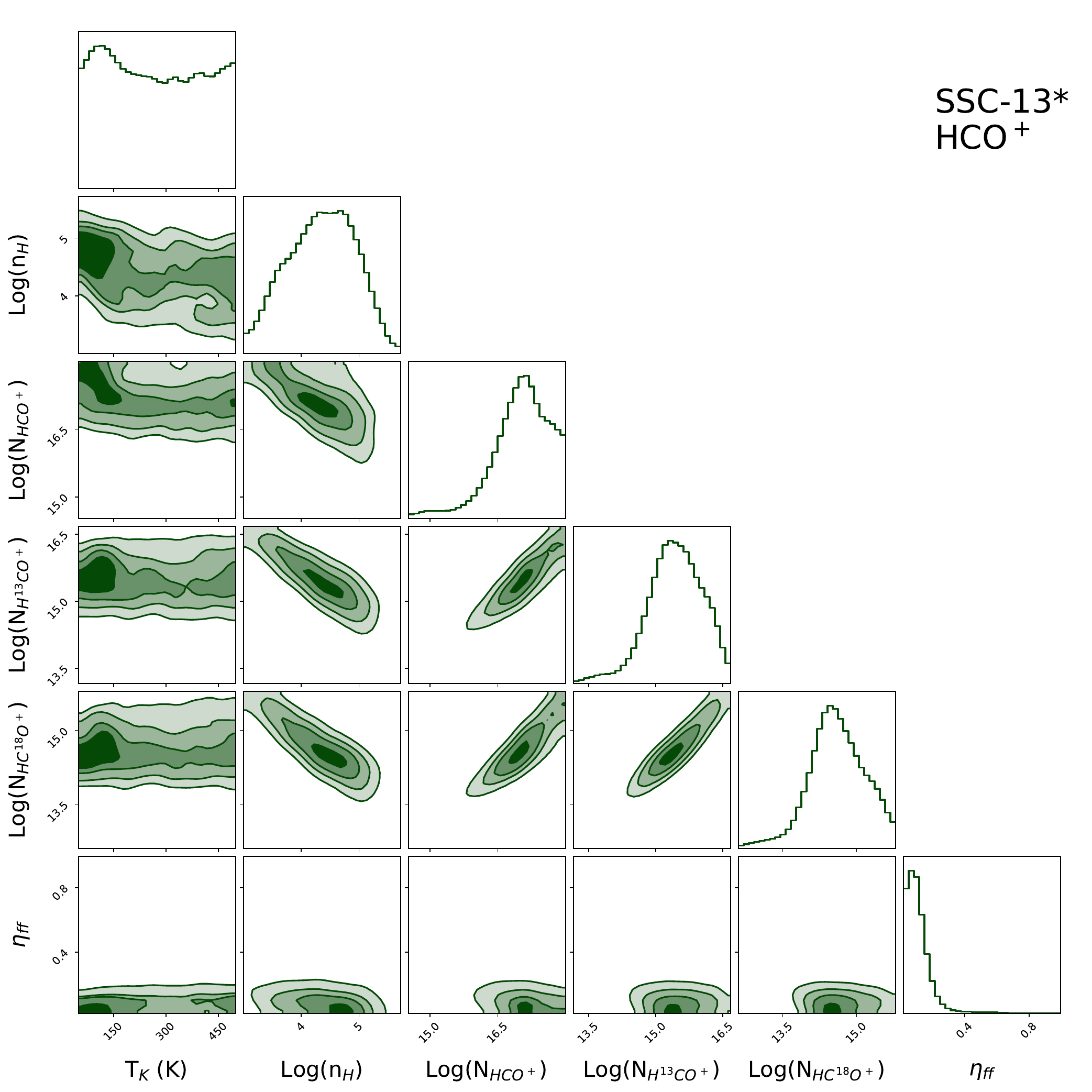}
  \caption{The posterior distributions for SSC-13* of the kinetic temperature, neutral H\textsubscript{2} number density, beam filling factor and column densities of HCO\plus, H\thirteen CO\plus\ and HC\eighteen O\plus, as predicted by \texttt{RADEX}. }
  \label{fig:RADEX_corner_SSC13_HCOP}
\end{figure*}

\subsection{SSC-14*}
\begin{figure*}
  \centering
  \includegraphics[width=\textwidth]{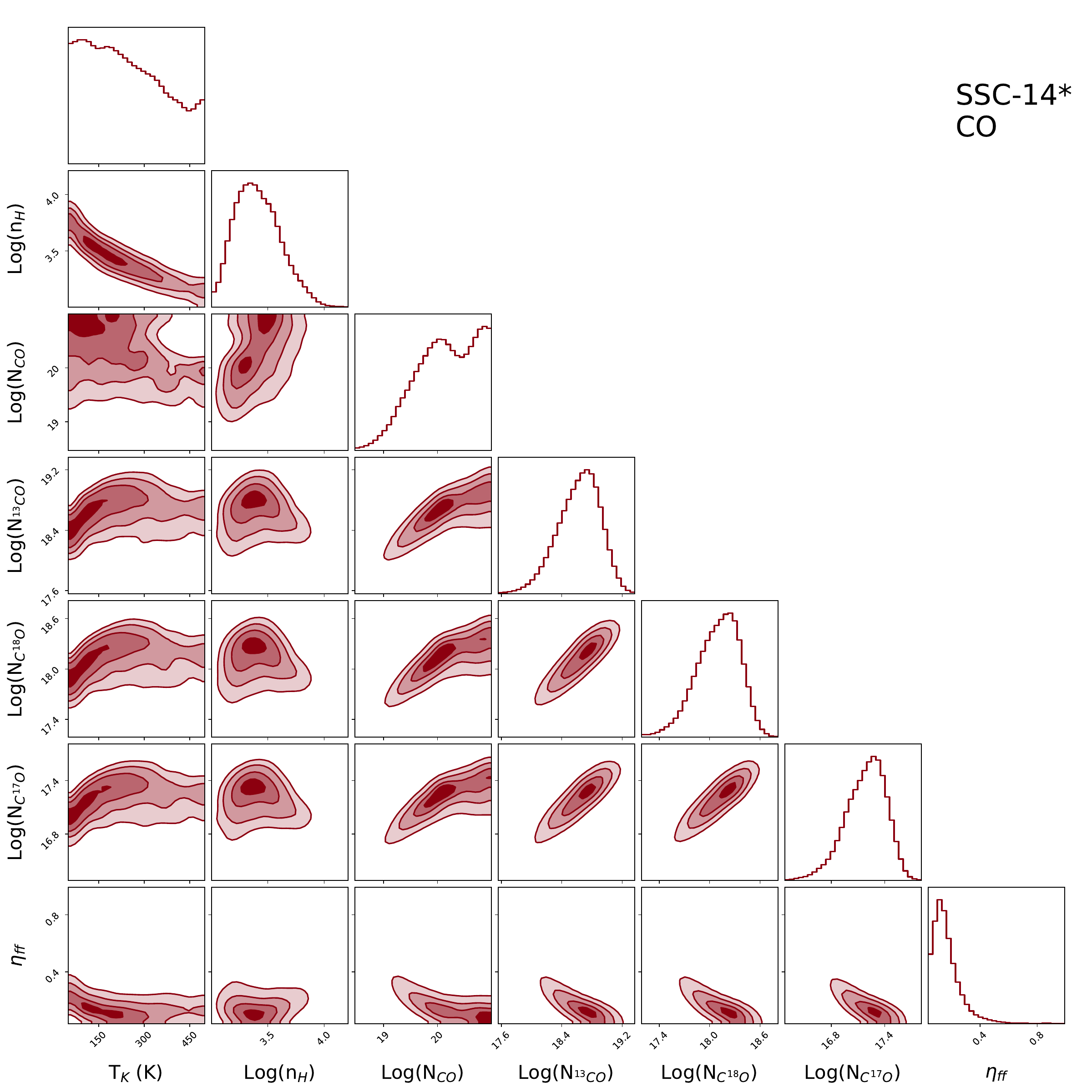}
  \caption{The posterior distributions for SSC-14* of the kinetic temperature, neutral H\textsubscript{2} number density, beam filling factor and column densities of CO, \thirteen CO, C\eighteen O, and C\seventeen O, as predicted by \texttt{RADEX}. }
  \label{fig:RADEX_corner_SSC14_CO}
\end{figure*}

\begin{figure*}
  \centering
  \includegraphics[width=\textwidth]{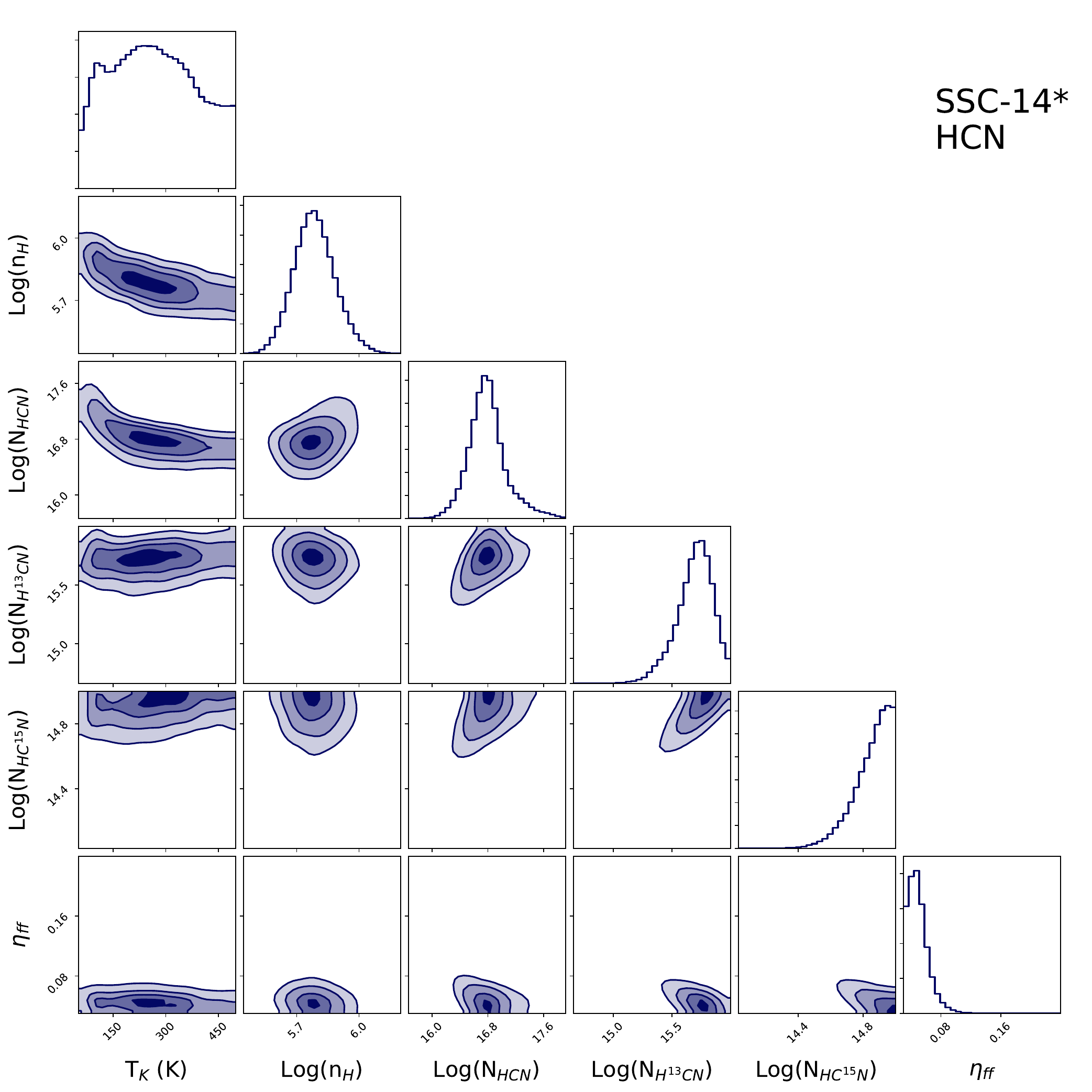}
  \caption{The posterior distributions for SSC-14* of the kinetic temperature, neutral H\textsubscript{2} number density, beam filling factor and column densities of HCN, H\thirteen CN and HC \fifteen N, as predicted by \texttt{RADEX}. }
  \label{fig:RADEX_corner_SSC14_HCN}
\end{figure*}

\begin{figure*}
  \centering
  \includegraphics[width=\textwidth]{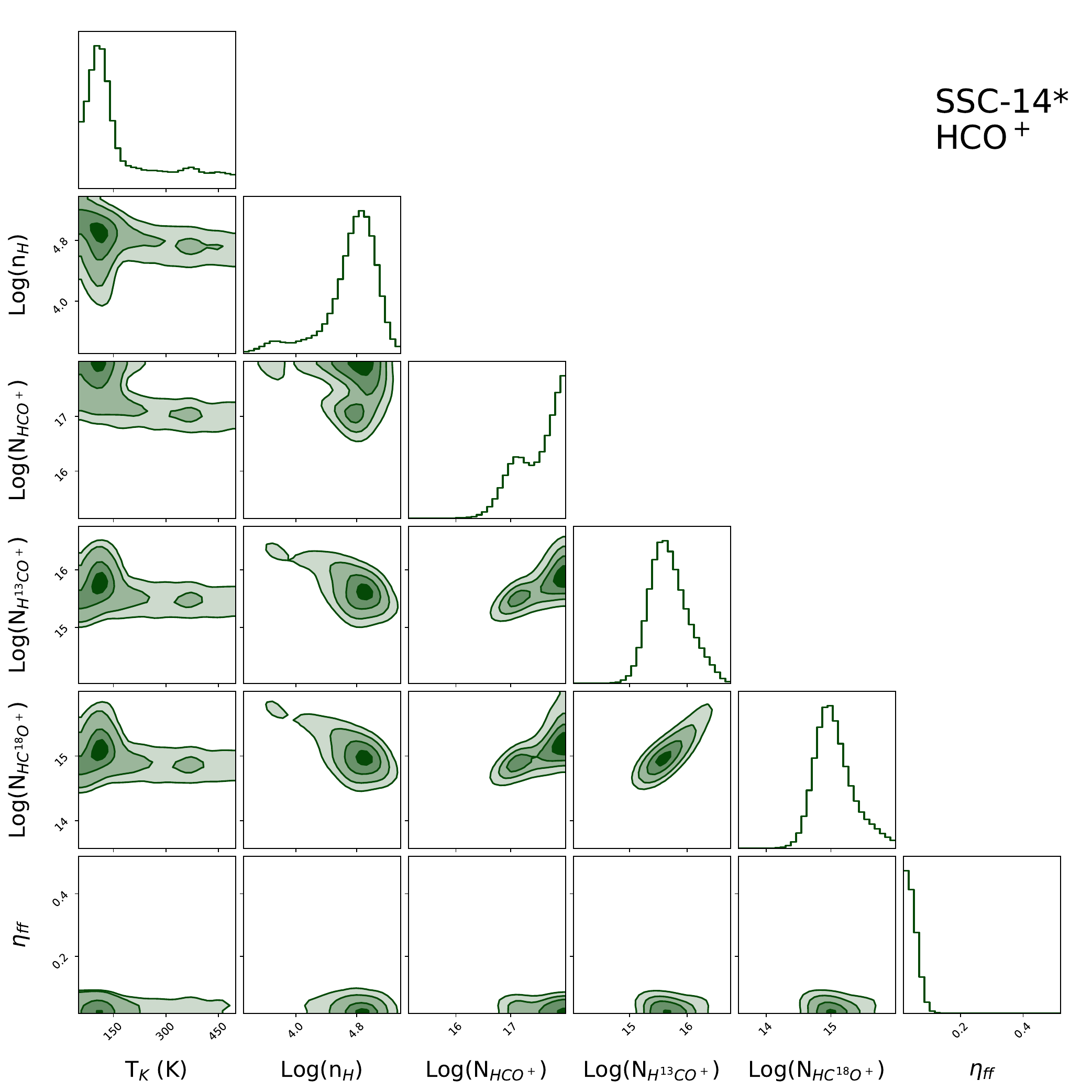}
  \caption{The posterior distributions for SSC-14* of the kinetic temperature, neutral H\textsubscript{2} number density, beam filling factor and column densities of HCO\plus, H\thirteen CO\plus\ and HC\eighteen O\plus, as predicted by \texttt{RADEX}. }
  \label{fig:RADEX_corner_SSC14_HCOP}
\end{figure*}


\section{Optical Depths}
\label{app:tau}

This section contains the predicted optical depths of the best fitting \texttt{RADEX} model for each region for each transition.
\begin{table*}[]

\caption{The transition optical depth, $\tau_{ul}$, predicted for each region from the best fitting \texttt{RADEX} model.}
\label{tab:Taus}
\centering
\begin{tabular}{ccccccc}
\hline
        Transition              & SSC-1*                        & SSC-4*                        & SSC-7*                         & SSC*-9                        & SSC-13*                       & SSC-14*                         \\ \hline \hline
                      & & & CO & & & \\ \hline \hline
(1-0)               & 0.31  & 0.00   & 0.00   & 0.16  & 0.22   & 0.2    \\
(2-1)               & 4.87  & 3.93  & 3.94  & 3.33  & 2.86   & 4.45   \\
(3-2)               & 10.11 & 8.27  & 8.32  & 7.04  & 6.14   & 9.41         \\ \hline \hline
& & & \thirteen CO & & & \\ \hline \hline
(1-0)        & 0.29  & 0.28  & 0.27  & 0.22  & 0.18   & 0.33   \\
(2-1)        & 1.13  & 0.83  & 0.79  & 0.61  & 0.44   & 0.83   \\
(3-2)        & 2.02  & 1.60   & 1.53  & 1.18  & 1.02   & 1.92   \\ \hline \hline
& & & C\eighteen O & & & \\ \hline \hline
(1-0)        & 0.17  & 0.13  & 0.12  & 0.09  & 0.08   & 0.18   \\
(2-1)        & 0.57  & 0.38  & 0.33  & 0.24  & 0.18   & 0.42   \\
(3-2)        & 0.88  & 0.65  & 0.58  & 0.42  & 0.40    & 0.95   \\ \hline \hline
& & & C\seventeen O & & & \\ \hline \hline
(1-0)        & 0.03  & 0.02  & 0.02  & 0.01  & 0.01   & 0.03   \\
(2-1)        & 0.09  & 0.06  & 0.05  & 0.04  & 0.03   & 0.07   \\
(3-2)        & 0.12  & 0.09  & 0.08  & 0.06  & 0.06   & 0.14   \\ \hline \hline
& & & HCN & & & \\ \hline \hline
(1-0)              & 0.11  & 0.27  & 0.39  & 0.24  & 0.17   & 0.62   \\
(2-1)              & 3.42  & 3.27  & 3.13  & 0.41  & 0.21   & 2.44   \\
(3-2)              & 6.31  & 6.51  & 6.50   & 0.86  & 0.67   & 5.46   \\ 
(4-3)              & 7.35  & 8.58  & 9.05  & 0.65  & 0.63   & 8.1    \\\hline \hline
& & & H\thirteen CN & & & \\ \hline \hline
(1-0)       & 0.19  & 0.28  & 0.37  & 0.04  & 0.03   & 0.54   \\
(2-1)       & 1.23  & 1.19  & 1.29  & 0.07  & 0.03   & 1.16   \\
(3-2)       & 1.46  & 1.67  & 2.11  & 0.12  & 0.11   & 2.51   \\
(4-3)       & 0.73  & 1.03  & 1.60   & 0.07  & 0.08   & 2.77   \\ \hline \hline
& & & HC\fifteen N & & & \\ \hline \hline
(1-0)       & 0.07  & 0.10  & 0.11  & -     & -      & 0.23   \\
(2-1)       & 0.40  & 0.36  & 0.36  & -     & -      & 0.47   \\
(3-2)       & 0.34  & 0.36  & 0.41  & -     & -      & 0.88   \\
(4-3)       & 0.12  & 0.15  & 0.19  & -     & -      & 0.62   \\ \hline \hline
\end{tabular}
\end{table*}

\begin{table*}[]
\ContinuedFloat
\caption{continued.}
\centering
\begin{tabular}{ccccccc}
\hline
        Transition              & SSC-1*                        & SSC-4*                        & SSC-7*                         & SSC*-9                        & SSC-13*                       & SSC-14*                        \\ \hline
 & & & HCO$^{+}$ & & & \\ \hline
(1-0)        & 3.30   & 1.40   & 1.39  & 4.47  & 4.28   & 4.26   \\
(2-1)        & 13.49 & 8.23  & 8.35  & 17.43 & 16.79  & 17.75  \\
(3-2)        & 25.03 & 16.73 & 17.38 & 32.77 & 31.46  & 36.67  \\
(4-3)        & 31.8  & 24.72 & 26.86 & 42.83 & 40.88  & 57.43  \\ \hline \hline
 & & & H\thirteen CO$^{+}$ & & & \\ \hline \hline
 
(1-0) & 1.22  & 0.18  & 0.01  & 1.74  & 1.81   & 0.76   \\
(2-1) & 4.44  & 2.53  & 2.44  & 5.59  & 6.09   & 4.97   \\
(3-2) & 3.81  & 2.87  & 3.31  & 4.78  & 5.74   & 7.20    \\
(4-3) & 1.23  & 1.38  & 2.07  & 1.54  & 2.15   & 5.51   \\ \hline \hline
 & & & HC\eighteen O$^{+}$ & & & \\ \hline \hline
(1-0) & 0.67  & 0.07  & 0.04  & 0.64  & 0.67   & 0.34   \\
(2-1) & 2.17  & 1.30   & 1.33  & 1.58  & 1.74   & 2.64   \\
(3-2) & 1.13  & 1.02  & 1.31  & 0.59  & 0.70    & 2.52   \\ \hline \hline
\end{tabular}
\end{table*}

\newpage
\section{H39$\alpha$ Recombination Line Fitting}
\label{app:H39}
\begin{figure}
  \centering
  \includegraphics[width=0.5\textwidth]{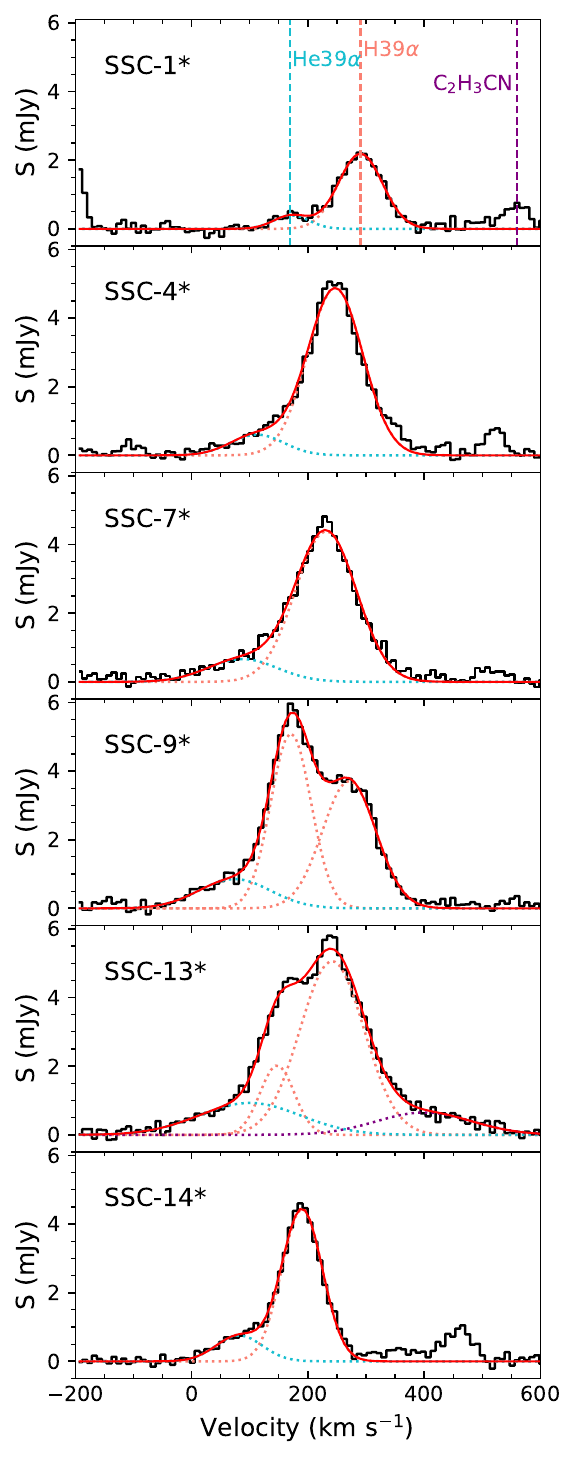}
  \caption{The spectra and resulting spectral fitting of H39$\alpha$.}
  \label{fig:H39_spectral_fitting}
\end{figure}

In Figure \ref{fig:H39_spectral_fitting}, we show spectra in heliocentric velocity with respect to the H39$\alpha$ transition at a rest frequency of 106.73738 GHz for each of the SSC*s. Other spectral emission that is present comes from He39$\alpha$ at a rest frequency of 106.78084 GHz which corresponds to $-$122.1 km~s$^{-1}$ offset in velocity from H39$\alpha$, and C$_2$H$_3$CN at a rest frequency of 106.64139 GHz or offset by -269.6 km~s$^{-1}$ with respect to H39$\alpha$. We’ve marked, with vertical dashed lines, the rest velocity velocities of He39$\alpha$ (cyan) and C$_2$H$_3$CN (purple) with respect to the fitted central velocity of H39$\alpha$ (pink) in the top panel.

We simultaneously fit Gaussian profiles to H39$\alpha$ and He39$\alpha$ emission, and for SSC-13* C$_2$H$_3$CN emission as well. Gaussian fits to the emission are shown with dotted lines for H39$\alpha$ (pink), He39$\alpha$ (cyan) and C$_2$H$_3$CN (purple). The sum of the Gaussian fits is shown by a solid red line.

By combining Equations A4 and A14 from \citet{Emig_2020}, the total ionizing photon rate expressed in terms of the flux density of the recombination line is

\begin{align*}
Q_0 = &\left(6.473\times 10^{51}~\mathrm{s}^{-1} \right) 
	\left(\frac{\int S_{\mathsf{n}}\,\mathrm{d}v}{100~\mathrm{mJy\, \text{km s}^{-1} }}\right)
	\left(\frac{1}{b_{\mathsf{n}+1}} \right) 
	\\ &
	\left( \frac{D}{4~\mathrm{Mpc}} \right)
	\left( \frac{\nu_{\mathsf{n}}}{100\,\mathrm{GHz}} \right)
	\left( \frac{ T_e }{ 10^4~\mathrm{K} } \right)^{0.667 - 0.034 \ln{(T_e\,/\,10^4\,\mathrm{K})} } 
\end{align*}
where $\int S_{\mathsf{n}}\,\mathrm{d}v $ is the integrated flux density of the RRL with principal quantum number $\mathsf{n}$, $b_{\mathsf{n}+1}$ is the LTE departure coefficient at $\mathsf{n}+1$,  $D$ is the distance to the emitting region, $T_e$ is the electron temperature, and $\nu_{\mathsf{n}}$ is the rest frequency of the RRL. 
To derive the total ionizing photon rate at each SSC*, we use the Gaussian area of the H39$\alpha$ emission or when necessary, the sum of the Gaussian areas of the H39$\alpha$ components.
We assume a temperature of $T_e = 6000$~K and an LTE departure coefficient of $b_{\mathsf{n}+1} = 0.8$ \citep{Bendo_2015, 2021_Mills}.

\end{document}